\documentclass[amsmath,amssymb,aps,revtex4-2]{revtex4-2}
\setcitestyle{square}
\usepackage{url}
\usepackage{hyperref}
\usepackage{lineno}
\usepackage{microtype}
\usepackage{subcaption}
\usepackage[onehalfspacing]{setspace}
\usepackage{chemist}
\usepackage{chemformula}
\usepackage{float}
\usepackage{caption}
\usepackage{multirow}
\usepackage{graphicx}
\usepackage{subcaption}

\begin{document}
\title {A comparison of different image analysis techniques for mapping spatiotemporal pH and carbon dissolution in density-driven convection of CO2 in water.} 

\author{Yao Xu$^{1}$}
\email{yaox@fys.uio.no}
\author{Marcel Moura$^{1}$} 
\author{Eirik Grude Flekk{\o}y$^{1,2}$}
\author{Knut J{\o}rgen  M{\aa}l{\o}y$^{1,3}$}
\affiliation{$^1$PoreLab, Department of Physics, The Njord Centre, University of Oslo, Oslo Norway}
\affiliation{$^2$PoreLab, Department of Chemistry, Norwegian University of Science and Technology, Trondheim, Norway
}
\affiliation{$^3$PoreLab, Department of Geoscience and Petroleum, Norwegian University of Science and Technology, Trondheim, Norway}

%\linenumbers
\begin{abstract}

%\textbf{Keywords:} Carbon capture and storage, experimental methods, porous media, convection %All article types: you may provide up to 8 keywords; at least 5 are mandatory.

\end{abstract}

\maketitle
\date{\today}

\section*{Abstract}
Density-driven convection enhances the carbon dissolution rate, which is significant for the geological carbon storage. This process will also influence the spatiotemporal pH and carbon concentrations of the underground fluid. To illuminate the convection mechanism, it is critical to understand the evolution of those properties within the porous media. However, determining the spatiotemporal pH and concentration within porous media is always challenging.  
This study employed a combination of three pH indicators that can track a wide range in pH from 4 to 9.5 in a convection experiment. Furthermore, we compared three image-processing techniques—Hue, gray-difference, and angular representation of RGB color space $\mathbf{(\phi,\theta)}$—for quantifying color changes from the universal $\text{pH}$ indicator arising from the carbon convection. The characterized colors were mapped into pH by calibrating against benchmark solutions. The comparative results demonstrate that the color quantified by the Hue technique is most robust, showing invariance to fluid thickness, camera settings, and LED luminance. In the convection experiments, it produces a continuous spatial distribution of pH and concentration level in the system. In contrast, the $\mathbf{(\phi,\theta)}$ and gray-difference techniques were more sensitive to environmental variations. They also have significant limitations for $\text{pH}$ interpolation in the critical range due to their non-monotonic calibration paths. Although all methods ultimately produced similar estimates of total dissolved carbon, the Hue technique offers greater stability and universality for high-resolution, dynamic measurements of pH and carbon concentration in the convection experiments.

\textbf{Keywords:} Density-driven convection, geological carbon storage, convection experiment, porous media, image processing, spatiotemporal carbon concentration.

\section{Introduction}

Geological carbon storage has been recognized as one of the most effective means to reduce anthropogenic carbon emissions in the atmosphere. Its achievements rely on four main mechanisms: 1) structure trapping, 2) dissolution trapping, 3) residual trapping, and 4) geochemical trapping\cite{al2022co2}\cite{bashir2024comprehensive}\cite{zhang2014mechanisms}. Among these, CO$_2$ dissolution is recognized as a stable means of permanently storing CO$_2$\cite{metz2005ipcc}\cite{soltanian2017dissolution}\cite{bakhshian2021dynamics}.
CO$_2$ dissolution is particularly linked to density convection, which is widely acknowledged as a driving force for enhancing geological carbon storage\cite{kim2019density}\cite{zhang2024permeability}\cite{xu2024effects}. The initiation of CO$_2$ dissolution triggers an increase in the solution’s density. Notably, once a considerable density gradient is established due to the dissolution process, the convective regime takes control of CO$_2$ transport over molecular diffusion. The denser fluid, now mixed with CO$_2$, descends, while the lighter, unmixed fluid tends to ascend. This natural convective phenomenon, known as density-driven convection, exerts a significant influence on the spatiotemporal mixing processes, thereby shaping the carbon distribution within porous media\cite{kim2019density}\cite{yang2006accelerated}\cite{chen2023density}.

Although there are many relevant experimental studies, the majority substitute fluids such as Methanol and Ethylene-Glycol (MEG), Propylene-Glycol (PPG), Potassium permanganate and water, which have a similar density difference as the system of the carbonic solution and water\cite{de2020non,de2023convective,de2023convectivediss,de2022experimental,tsai2013density,guo2021novel}. Those experiments provide insights into the convection and mixing process but blur many physicochemical details of the interaction between carbon and water. 
 For instance, when the gas carbon dioxide dissolves in water, a series of chemical reactions occurs between water molecules and carbonic species. Those interactions will influence the spatiotemporal pH and carbon concentration in the porous media\cite{xu2017effect}\cite{iglauer2011dissolution}\cite{wilkin2010geochemical}\cite{birgisson2025mapping}. 
 
 %Those interactions arising from the carbon convection will also influence the spatiotemporal pH and carbon concentration in the liquid phases. 
Attaining pH and carbon evolution is crucial in quantitatively unraveling the intricate convection processes. In addition, the two parameters are significant for the carbon mineralization, the most stable way to sequestrate CO$_2$ underground.
%The physical characterization of the growth rate and convection instability requires the knowledge of the spatialtemporal carbon concentration. Besides, pH and carbon concentration are also the paramount parameters determining the kinetics and equilibrium of chemical reactions. The understanding will serve as the foundational step for investigating the geochemical reactions of the CO$_2$-water-rocks system, the most stable way to sequestrate CO$_2$ underground.
Previous research has employed various methods to measure pH. The first and most common method is by the pH electrode. However, this method can only cope with equilibrium solutions and can not log its evolution in space. 
%To overcome the problems, studies also applied soluble fluorescence since the fluorescein light emission and adsorption are pH-independent\cite{brouzet2022co,faasen2023diffusive,kong2018dual}. This method is not intrusive and can offer quantitative instantaneous data in time and space. Based on the light intensity received by a camera, the method can directly visualize the mass and scaler transport and mixing in stationary or non-stationary flows, such as acid turbulent jets, CO$_2$ bubbles rising, and CO$_2$ dissolution in water\cite{lacassagne2018ratiometric}. 
To address these challenges, some studies have applied soluble fluorescence, as fluorescein emission and absorption are pH-independent \cite{brouzet2022co,faasen2023diffusive,kong2018dual}. This non-intrusive method provides quantitative, instantaneous data in both time and space. By measuring the light intensity captured by a camera, it enables direct visualization of mass and scalar transport, as well as mixing processes, in both stationary and non-stationary flows, such as acid turbulent jets, rising CO$_2$ bubbles, and CO$_2$ dissolution in water \cite{lacassagne2018ratiometric}.
%However, the limitations are also evident. For example, the fluorescein emission could be weak and require a longer exposure time from the camera. That means some fast turbulence or mass transfer could not be captured. Besides, the overlapping between fluorescein’s adsorption and emission could lead to the fluorescence re-absorption\cite{lacassagne2018ratiometric}. In addition, using intensity as the parameter to determine pH could be inaccurate since the dissolution of CO$_2$ is normally low. 
Nevertheless, the limitations of this approach are evident. For example, fluorescein emission can be relatively weak, demanding longer camera exposure times. This constraint may result in the inability to capture rapid phenomena such as mixing and reaction kinetics or mass transfer events. 
In addition, using fluorescence intensity as a parameter for determining pH may introduce inaccuracies, particularly because the dissolution of CO$_2$ under typical conditions is generally low\cite{lacassagne2018ratiometric}. %Furthermore, the overlap between fluorescein’s absorption and emission spectra can cause fluorescence re-absorption, as noted by Lacassagne et al.\cite{lacassagne2018ratiometric}.
To overcome the restrictions, a growing trend involves adding pH indicators to solutions. These indicator solutions exhibit color changes based on its pH levels. Most pH indicators are accompanied by a calibrated color palette that correlates colors with known pH values, allowing for visual assessment of samples' pH. However, unlike physical quantities such as length or temperature, color perception is inherently subjective and can vary depending on the observer's interpretation\cite{cantrell2010use}.
To combat that, the wavelength of a color was utilized. 
Transmitting white light through liquids and measuring its absorbance at various wavelengths can be used to quantify the colors of solutions. This spectral information can then be precisely calibrated to the solution’s pH. However, such experiments require large sample volumes, significant time, and expensive instruments to measure absorbance across different wavelengths\cite{koschan2008digital}. %While the wavelength spectrum can be characterized as a physical quantity of electromagnetic radiation, this approach is not convenient for quantifying color and image processing.
Fortunately, colors can also be quantified using the common digital photography devices and color spaces. A color space is a mathematical representation of perceived colors. The most common space in modern digital photography is red-green-blue (RGB) primaries. In this framework, each pixel of an image can be represented digitally as a combination of red, green, and blue values.
An appropriate form of color representation is essential for accurate data storage, consistent image display across different media, and efficient mathematical processing. While the RGB color space is the most widely used format for digitally representing color in devices such as CRT monitors, scanners, and digital cameras, it is not capable of simultaneously fulfilling all these requirements.
%When white lights at different intensities shine through the same bulk of liquid, the absorbance of each channel of RGB should be the same as per the Beer-Lambert law. However, most digital cameras have non-linear responses of RGB values to light illumination. That is because the camera changes its perception of colors and behaves more sensitively in a dark environment than in very bright conditions. If the RGB method is applied for image processing, this will lead to different color perceptions of the same fluid under different illumination. 
Theoretically, the absorbance across the RGB channels remains invariant to changes in incident light intensity. Since absorbance is defined by the ratio of transmitted to incident light intensity, it remains a function of the medium's properties—concentration and path length—rather than the absolute magnitude of the light source, provided the camera sensor operates within its linear range\cite{lothian1963beer}\cite{mayerhofer2019beer}. %.according to Beer's law, which states that the concentration, path length, and molar absorptivity are all directly proportional to the absorbance. 
However, most digital cameras exhibit non-linear responses to light illumination, resulting in deviations from this principle. This non-linearity arises because cameras tend to adjust their perception of colors, becoming more sensitive in low-light conditions and less sensitive under bright illumination. Consequently, if the RGB color space is used for image processing, it can result in inconsistent color interpretations for the same liquid under different lighting conditions.
Therefore, other color spaces have been considered for color visualization and image processing. Previous research showed there are 38 major color spaces\cite{kahu2019review}, among which, the HSV color space is highly aligned with human perception, and the Hue parameter is barely influenced by the brightness\cite{cantrell2010use,koschan2008digital}. 
HSV color spaces are the abbreviations for ‘Hue,’ ‘Saturation,’ and ‘Values.’ In broad terms, H is a numerical representation of the dominant wavelength of a color, S gives a basic measure of apparent grayness of the Hue, and V represents the intensity of light shining through the object\cite{cantrell2010use}\cite{yabusaki2014novel}\cite{macdonald1999using}. Compared to the RGB color space, the HSV color space offers distinct advantages, particularly in its treatment of color intensity variations. Additionally, previous studies verified that the Hue (H) parameter remains independent of variations in light intensity caused by changes in indicators' concentration and optical path length\cite{cantrell2010use}. Alternatively, the saturation (S) and value (V) parameters will incorporate that information with the Hue parameter unaffected. This makes Hue a robust analytical parameter for quantifying colors and a reliable method for determining the properties of fluids. HSV color space has been utilized in analytical chemistry and biology to determine the concentration of serum creatinine and the pH of the solution and to detect flames\cite{tarim2024colorimetric}. To the best of our knowledge, there is little work measuring pH by HSV color space within the context of CO$_2$ convection, and there is no work systematically comparing different techniques by colors in measuring pH and carbon concentration in convection experiments. 

Although the Hue method was proven to be useful for determining pH, its abilities also depended greatly on the choices of the indicators. A single pH indicator, such as bromocresol green (yellow at pH 4 to blue at pH 5.6) or bromothymol blue (yellow at pH 5.2 to purple at pH 6.8), can only show color variations in a narrow range, unable to satisfy the needs in the carbon convection experiments\cite{de2021bi,de2022two}.
To address such constraints, combining multiple pH indicators to cover a broader measurable pH range is desirable. Universal indicators, which are mixtures of several pH indicators and additives, are widely used for this purpose. Different recipes for universal indicators exist, with variations in the type, number, and concentration of the constituent pH indicators\cite{thomas2015experimental}. An optimal combination of pH indicators should not only cover the intended pH range but also exhibit distinguishable color changes in response to subtle variations in pH (e.g., changes as small as 0.1 pH units). Furthermore, considerations such as the toxicity and cost of the indicators must also be accounted for.

This study employed the recipe of Birgisson et al\cite{birgisson2025mapping}. The combination of three safe indicators, along with some other additives, has been applied to carbon convection experiments and achieved good resolution of pH change in the range from 4 (red) to 9.5 (blue). %Since carbon dioxide is a weak acid, its injection is less likely to lower the fluid’s pH below 4.4, but provides a higher capacity to measure the solutions initially under basic conditions, which is common in saline aquifers or underground water. 
Moreover, we compared three imaging analysis techniques for measuring pH and carbon concentration in carbon convection experiments. The results indicate that Hue can effectively determine both pH and carbon concentration. The calibration curve derived from Hue demonstrates broad applicability, being minimally affected by light intensity, camera settings, or other environmental factors. 
In addition, in the convection experiments, it produces a more complete, continuous spatial distribution of pH and concentration level in the system. In contrast, the other two techniques are more sensitive to environmental variations and have more errors in pH and concentration determination in the convection experiments.
Owing to its robustness, this Hue technique can be reliably applied across diverse experimental conditions, with enhanced flexibility and accuracy.

%In this study, we utilized novel experimental techniques for CO$_2$ convection experiments. Combined with the application of the customized pH indicators, we employed hue values in the HSV color space to determine the spatiotemporal pH in the CO$_2$ convection experiment. This approach enables both quantitative and qualitative analysis of carbon convection in porous models.
%The experimental technique can effectively determine subtle pH variations in the pH range from 4.4 to 9.6. The calibration curve developed in this study exhibits broad applicability, as it is minimally affected by light intensity and camera settings. Given its robustness, this calibration curve can be reliably applied to measure pH under diverse experimental conditions, providing greater flexibility and accuracy. 

\section{Methodology}

In this section, we describe the experimental procedures for constructing calibration curves using the Hue, gray-difference, and $\mathbf{(\phi,\theta)}$ techniques.  This process was conducted using transparent multi-well microplates, which facilitated high-throughput, batch imaging of the fluid samples. These calibration curves were subsequently applied for the spatiotemporal interpolation of pH in the density-driven convection (DDC) experiments conducted within the milled PMMA model later described in Section \ref{exp procedure}.

\subsection{Construction of calibration curves}
\begin{figure}[htbp]
    \centering
    % First figure
        \includegraphics[width=\textwidth]{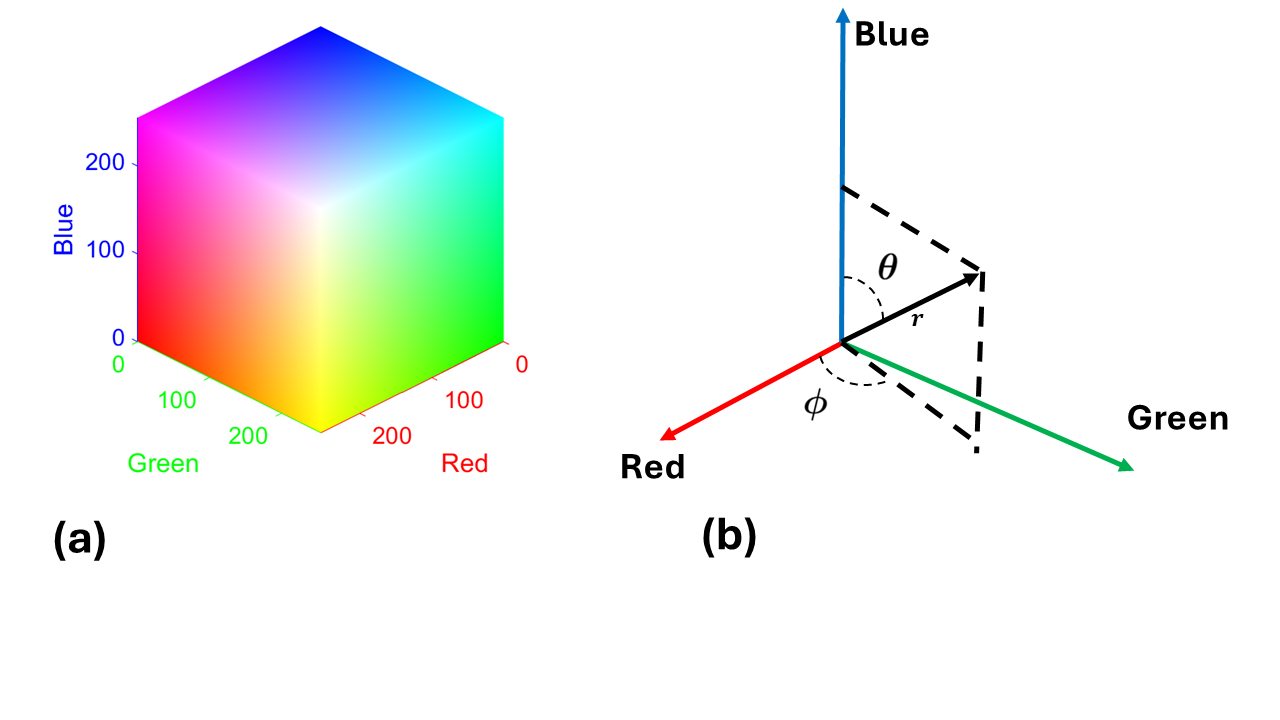}
    \caption{RGB color space and graphical illustration of space $\mathbf{(\phi,\theta)}$. (a) In the RGB color space, every color is co-determined by the three elements, namely, red, green, and blue, as indicated by the colors in the Cartesian coordinates. (b) The$\mathbf{(\phi,\theta)}$ are the azimuth and polar angles in the spherical coordinate system. Those two angles can be calculated and converted from the RGB values in the Cartesian coordinate, as indicated in the subplot (b). It can be seen as the angular representation of the RGB color space.}
    \label{rgb_polar}
\end{figure}

To correlate colors numerically with pH, the colors need to be quantified. 
There are various ways to convert the colors numerically. The most well-known is by the standard color space (RGB), where in 8-bit images the color is composed of the three elements: red, green, and blue, denoted as $\mathbf{(R,G,B)}$ with each color channel ranging from [0-255]. The color space is graphically illustrated in Figure \ref{rgb_polar}. 

\begin{figure}[H]
    \centering
    \includegraphics[width=0.9\linewidth]{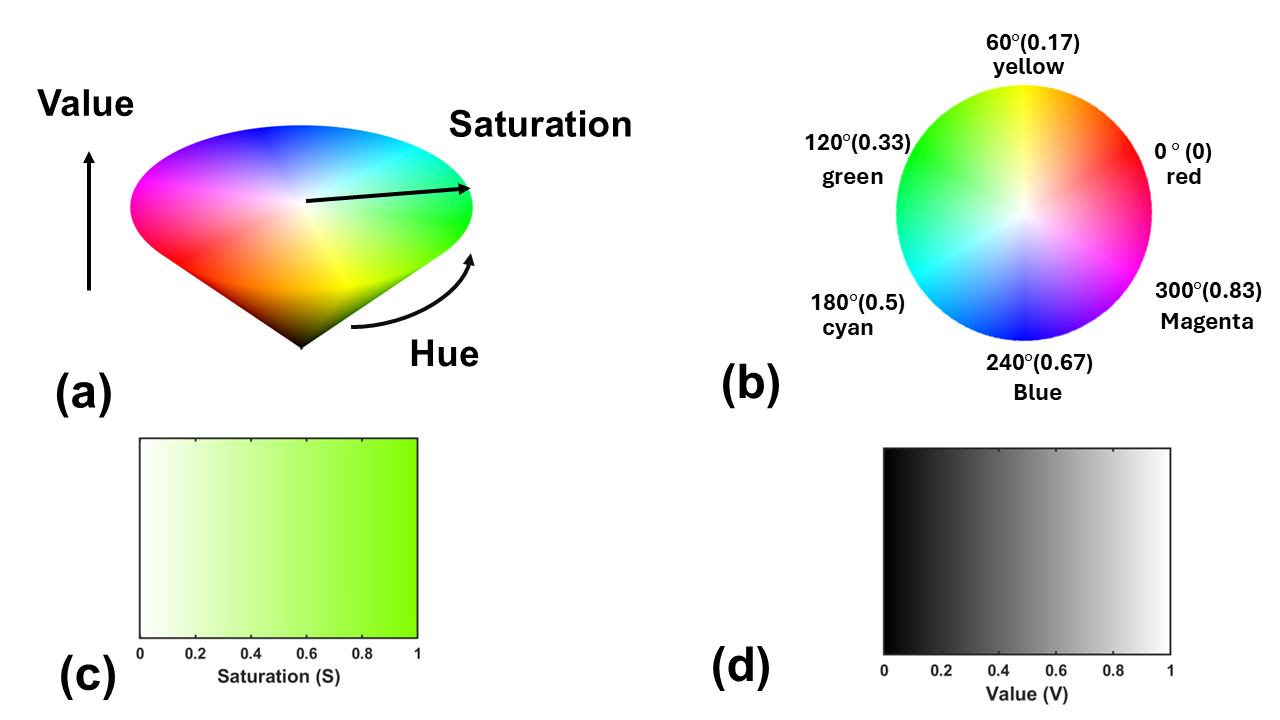}
    \caption{Graphical illustration of the HSV (Hue, Saturation, Value) color space. (a) The 3D color cone model illustrating the three parameters: Hue (H) (angular position on the color wheel), Saturation (S) (radial distance from the center, quantifying colorfulness/grayness), and Value (V) (vertical axis, representing brightness or lightness). (b) Key Hue values are marked on the color wheel. For example, Red is at 0$^{\circ}$. Numbers in parentheses represent the normalized [0, 1] scale. (c) Saturation modulates the color transition from white to pure color when it varies from 0 to 1 (at V=1 and H=0.25). (d) Value modulates the color, changing from black to white as V varies from 0 to 1 (at S=0, H=0.25).}   
    \label{Hsv_color space}
\end{figure}

However, using all three color elements to map the pH is not a numerically trivial task. Instead, an alternative single quantity, the gray value, calculated by:
\begin{equation}
    \text{Gray}=(0.299R+0.587G+0.114B)/255,
    \label{cal_gray}
\end{equation}
is sometimes employed. In addition, the HSV (Hue, Saturation, and Value) color space has also been widely utilized for visualization. In this way, color has been characterized by three independent parameters, with hue representing a coordinate along the rainbow (spectral sequence). By mapping hue to a normalized interval $[0, 1]$, we can represent the transition of dominant wavelengths found in the visible spectrum: beginning with spectral red (0.00), progressing through the yellow-green regions (0.17–0.33), and continuing with spectral blue (0.67). This normalized scale effectively bridges the linear physical spectrum into a perceptually continuous cycle; saturation specifies a measure of the grayness, and value characterizes the brightness of the color\cite{yabusaki2014novel}. The HSV parameters are graphically illustrated in Figure \ref{Hsv_color space} and can be numerically calculated using:
\begin{equation}
    \begin{split}
                        \max &= \max\{R, G, B\}; \\
                        \min &= \min\{R, G, B\}; \\
        H &= 
        \begin{cases}
            (\frac{G - B}{\max - \min} + 0) / 6; & \text{if } \max = R \\
            (\frac{B - R}{\max - \min} + 2 )/ 6; & \text{if } \max = G \\
            (\frac{R - G}{\max - \min} + 4 )/ 6; & \text{if } \max = B
        \end{cases}\\
                        S &=\dfrac{\max - \min}{\max};\\
                        V &=\max,
    \end{split}
    \label{cal_hue}
\end{equation}
where {$R, G, B$} are normalized to the interval [0, 1].   

A previous study also employed both polar and azimuth angles in the polar coordinates to separate colors\cite{birgisson2025mapping} and showed high efficiency. The parameters can be formulated as: 
\begin{equation}
\begin{split}
    \theta=\arccos{\frac{B}{\sqrt{R^2+G^2+B^2}}};\\
    \phi=\arctan{\frac{G}{R}}.
\end{split}
\label{cal_angles}
\end{equation}
The graphical illustration is presented in Figure \ref{rgb_polar}(b).

% plot: illustration of the color spaces: sRGB and HSV.
To facilitate color separation in a wide pH range, a broad mixture of pH indicators was used\cite{birgisson2025mapping}. It can display color differentiation with subtle pH changes. The components of the indicator mixture are shown in the Table \ref {receipe}.

\begin{table}[H]
\centering
\begin{tabular}{c |c | c} 
 \hline
  & Stock & 1:50 Dilution \\ [1.0ex] 
 \hline
 Thymol blue [mg/L] & 1250 & 25 \\
 Bromothymol blue [mg/L] & 2500 & 50 \\
 Methyl red [mg/L] & 1600 & 32 \\
\hline
 2-propanol \%vol & 12.5 & 0.25\\
 NaOH [mol/L] & $2.5 \cdot 10^{-3}$ & $5.0 \cdot 10^{-5}$\\[1ex] 
 \hline
\end{tabular}
\caption{ The wide-range pH indicator developed for \ch{CO2} DDC experiments.}
\label{receipe}
\end{table}

\begin{figure}[H]
   \hspace{4cm} % adjust the length as needed
    \includegraphics[width=0.7\linewidth]{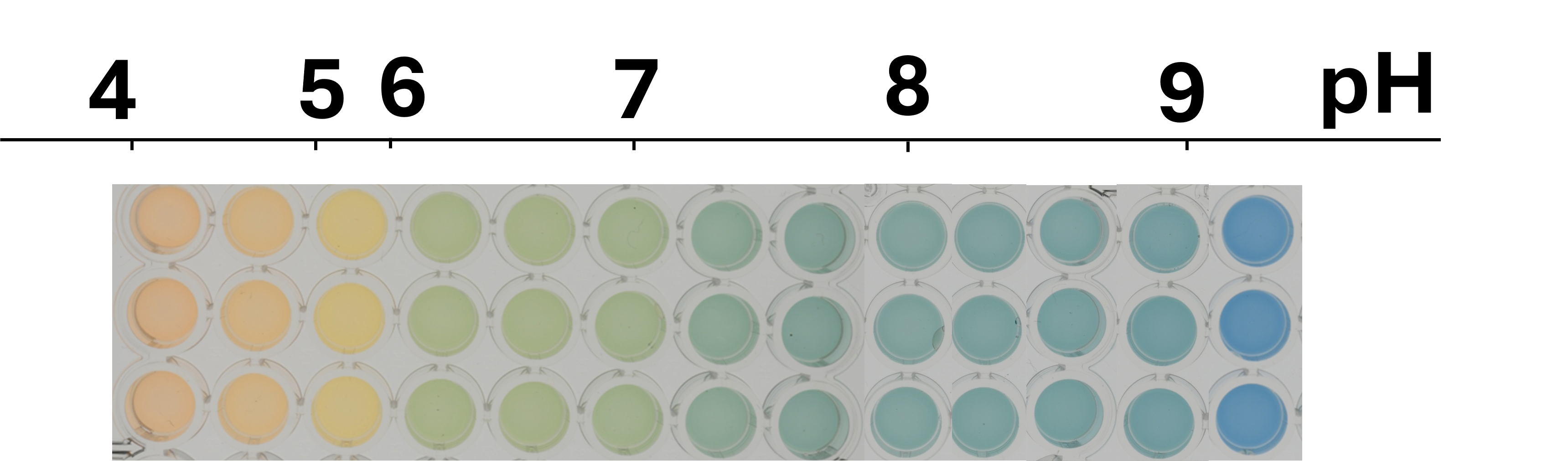}
    \caption{The 13 samples in the microplate. Every column has the fluid from the same sample. Hue, gray value and $\mathbf{(\phi,\theta)}$ were calculated as the the average of the three wells in a column. The fluid depth in each sample is 2 mm.}
    \label{microplate}
\end{figure}

%One of the aims of this study is to compare the versatility of the three color quantification techniques with application to interpolate pH. 
The pH indicator solution can be titrated to various known pH levels. In this study, there are 13 samples (benchmark solutions) with various levels.  As present in Figure \ref{microplate}, colors transit from red to green and end at blue when the pH changes from 4 to 9.5. Each column in Figure \ref{microplate} shows 3 repetitions from the same sample, with the purpose of averaging measurements of colors and reducing errors.

The pH of each sample was determined as the average of three measurements using the pH meter (VWR pHenomenal pH 1100) with an ideal resolution of 0.001. However, in practice, the resolution is lower, and measurement variation can normally reach 0.05. 
The $\mathbf{(R,G,B)}$ values were obtained by dispensing the mixture into three wells of a transparent microplate and imaging them. For each well, the gray, Hue, and $\mathbf{(\phi,\theta)}$ values were calculated according to Equations \ref{cal_gray}, \ref{cal_hue}, and \ref{cal_angles}, and the mean across the three wells was used for calibration analysis. Following this procedure, the calibration path was established and subsequently applied to interpolate pH in the carbon convection experiment.

\subsection{Experimental procedure}\label{exp procedure}

\begin{figure}[H]
    \centering
    \includegraphics[width=0.85\linewidth]{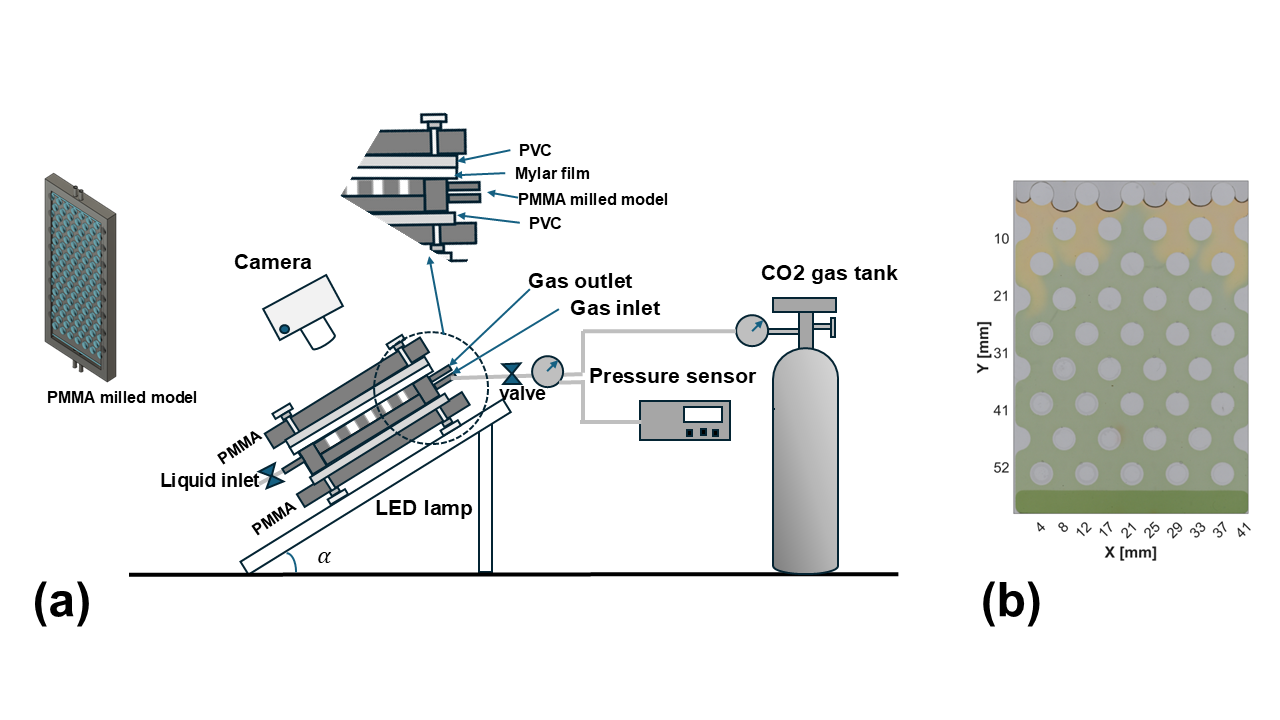}
    \caption{(a) The sketch of the experimental setup. The 3D rendering of the PMMA-milled model is also present. (b) An example of the ongoing DDC experiment after 10 minutes of carbon injection. This image will be further analyzed in the results section.}
    \label{setup-sketch}
\end{figure}

The sketch of the experimental setup of carbon convection is shown in Figure \ref{setup-sketch}.
A porous medium model milled from Polymethyl methacrylate (PMMA) was tightly sandwiched between polyvinyl chloride (PVC), Mylar film, and two other flat PMMA sheets, ensuring a leak-free environment. The high optical transparency of these materials ensures minimal light absorbance, facilitating clear imaging. Mylar film was utilized for its low chemical affinity for stains, preventing residual dye adhesion between experiments. 
The model is then placed on the illumination LED lamp, with the entire setup tilted at an angle of 30$^{\circ}$. Positioned in alignment with it, a high-resolution camera captured images from the top every 5 seconds. The model has a size of 65 mm × 41 mm, with uniformly distributed cylinders that mimic the grains of the rocks, allowing the fluid to flow between them. The cylinders have both a diameter and a height (thickness of the fluid) of 2mm. The model was first saturated with the pH indicator solution with a pH of 7.00, and then the gas carbon dioxide was injected on top with an injection pressure of 2mbar.

The spatial-temporal pH in the porous medium model can be interpolated by the built calibration curves of the three techniques.
Afterwards, the corresponding carbon concentration is calculated using:

\begin{equation}
  C_T=\frac{1}{1+2K_{a2}/[H^+]}\left([Na]^++[H]^+-\frac{10^{-14}M^2}{[H^+]}\right)\left(\frac{K_{a2}}{[H^+]}+1+\frac{[H^+]}{K_{a1}}+\frac{[H^+]}{K_{a1}K_H}\right),
  \label{carb-con}
\end{equation}   
where K$_H$, K$_{a1}$ and K$_{a2}$ are the equilibrium constants for the hydration of \ch{CO2} and dissociation steps of carbonic acid respectively\cite{birgisson2025mapping}.
Their values are given in Table \ref{table:1}. The expression above, along with the fact that \ch{[H^+]}$=10^{-\ch{pH}}$, gives an estimation of the total carbon. The concentration of \ch{Na$^+$} is incorporated into the calculations because \ch{NaOH} was included in the formulation as shown in Table \ref{receipe}, to neutralize the initial indicator solution to pH 7.
\begin{table}[h!]
\centering
\begin{tabular}{c |c | c | c} 
 \hline
  Equilibrium & Definition & Eq. constant & $pKa$ \\ [1.0ex] 
 \hline
 Gas dissolution & $K_h=\frac{[CO_2]}{p_{CO_2}}$ & $3.4 \cdot 10^{-2}$ M/atm & - \\
 Hydration  & $K_H=\frac{[H_2CO_3]}{[CO_2]} $ & $1.66 \cdot 10^{-3}$  & - \\
 First dissociation & $K_{a1}=\frac{[H^+][HCO_3^-]}{[H_2CO_3]}$ & $2.67 \cdot 10^{-4}$ M & $3.6$\\
 Second dissociation & $K_{a2}=\frac{[H^+][CO_3^{2-}]}{[HCO_3^-]}$ & $4.47 \cdot 10^{-11}$M  & $10.3$\\[1ex] 
 \hline
\end{tabular}
\caption{ Equilibrium constants relevant to the carbonic acid equilibrium in fresh water at temperature 25$^{\circ}$C and atmospheric pressure\cite{greenwood2012chemistry,sander2015compilation} Acid $pKa$ values are presented for the dissociation equilibrium.}
\label{table:1}
\end{table}

%% graphically show the color space
%%% their representation by one of the pictures in the experiment (or can also be put in the results part)
%\subsection{Mapping colors to pH}
%%The receipe of the pH indicators solution
%%% how to make the solutions
%The color of each sample is quantified by injecting the mixture into a microplate and imaging it.
%The colors were determined by averaging three measurements per sample, obtained using a Nikon Z7II camera. 
%The construction of the calibration path is achieved by injecting the mixture with known pH levels into a microplate and imaging them. 
%%% using the Microplate as the cell to conduct the calibration experiments: transparent and refracts little light when the light passes through, demands fewer fluid samples since the wells are small in radius, and is easy to carry out the experiments/repeat: has many wells
%%% illustrated how to set the things in the microplate and its comparison with the fluid in the 3D-printed model/plixel glasses model.  
%%% how to measure it/ three times measurement of pH

%For the standard of a robust calibration curve, it is supposed to meet the requirements below: 1) It is reproducible, 2) it is applicable in various scenarios, 3) it is handy to apply or interpolate pH, 4) it has relatively high resolution and accuracy, 5) it should be easy to establish without the utilization of the expensive facilities or difficult operations and procedures.  
A previous study reported that the error in pH will significantly propagate to the determination of total dissolved carbon\cite{birgisson2025mapping}.
Therefore, obtaining a robust calibration curve is crucial for illuminating pH and carbon evolution during the convection experiments.  
In the following sections, we will compare the three techniques from various perspectives to assess their robustness for accurate pH and concentration measurements.

\section{Results and discussion}
\subsection{The impact of time}

\begin{figure}[htbp]
    \centering
    % First figure
    \begin{subfigure}[b]{0.32\textwidth}
        \includegraphics[width=\textwidth]{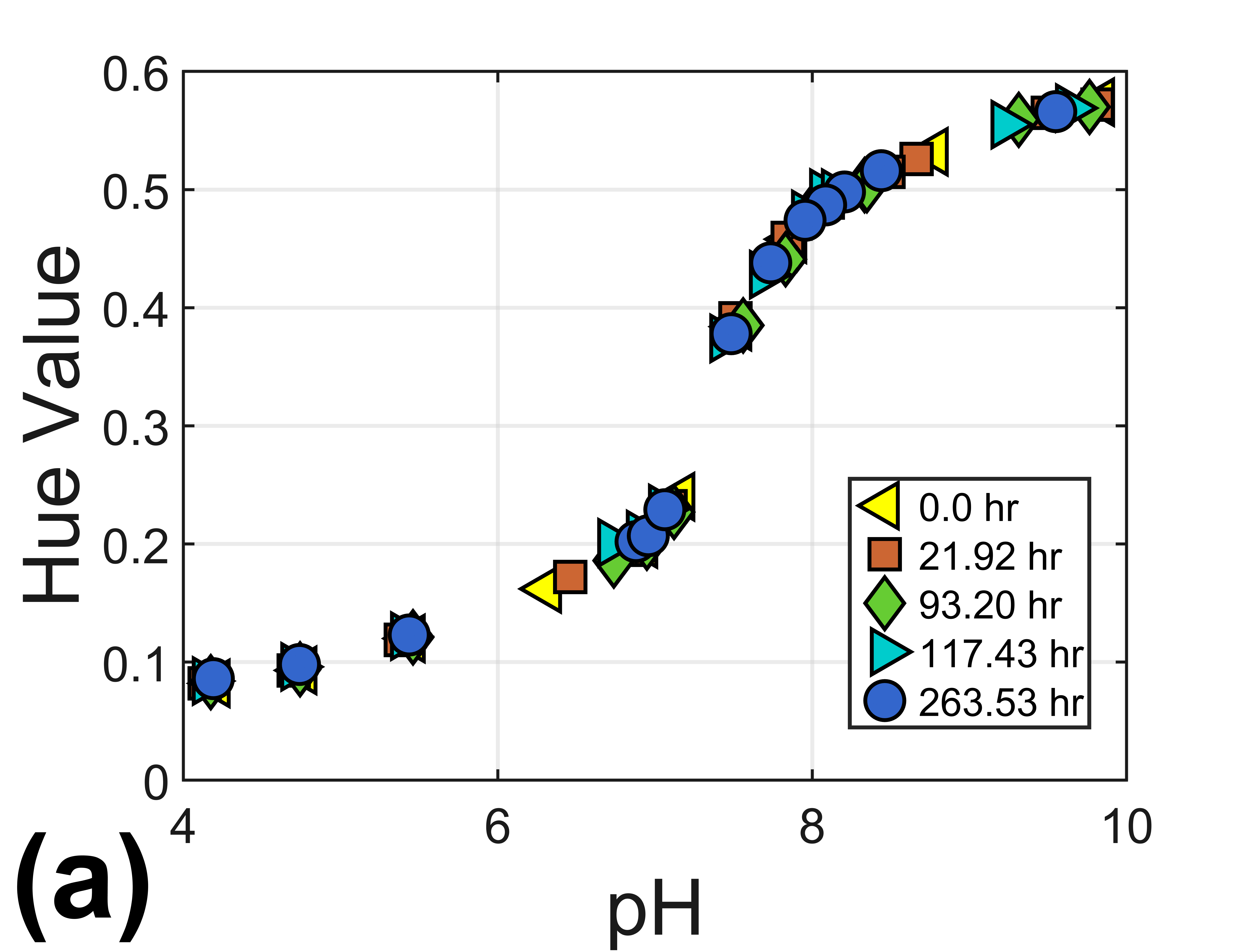}
        %\caption{First}
        \label{fig:first}
    \end{subfigure}
    \hfill
    % Second figure
    \begin{subfigure}[b]{0.32\textwidth}
        \includegraphics[width=\textwidth]{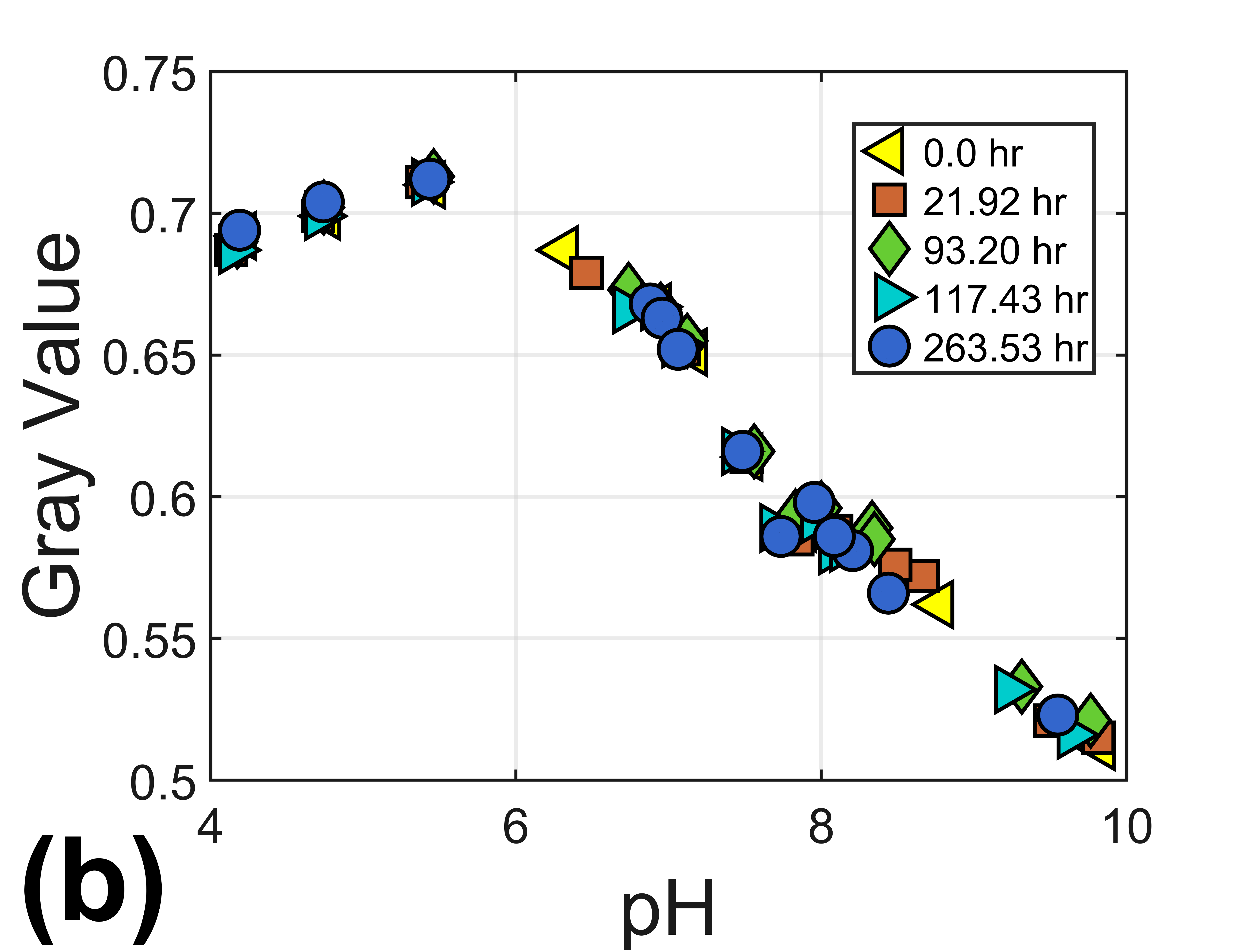}
        %\caption{Second}
        \label{fig:second}
    \end{subfigure}
    \hfill
    % Third figure
    \begin{subfigure}[b]{0.34\textwidth}
        \includegraphics[width=\textwidth]{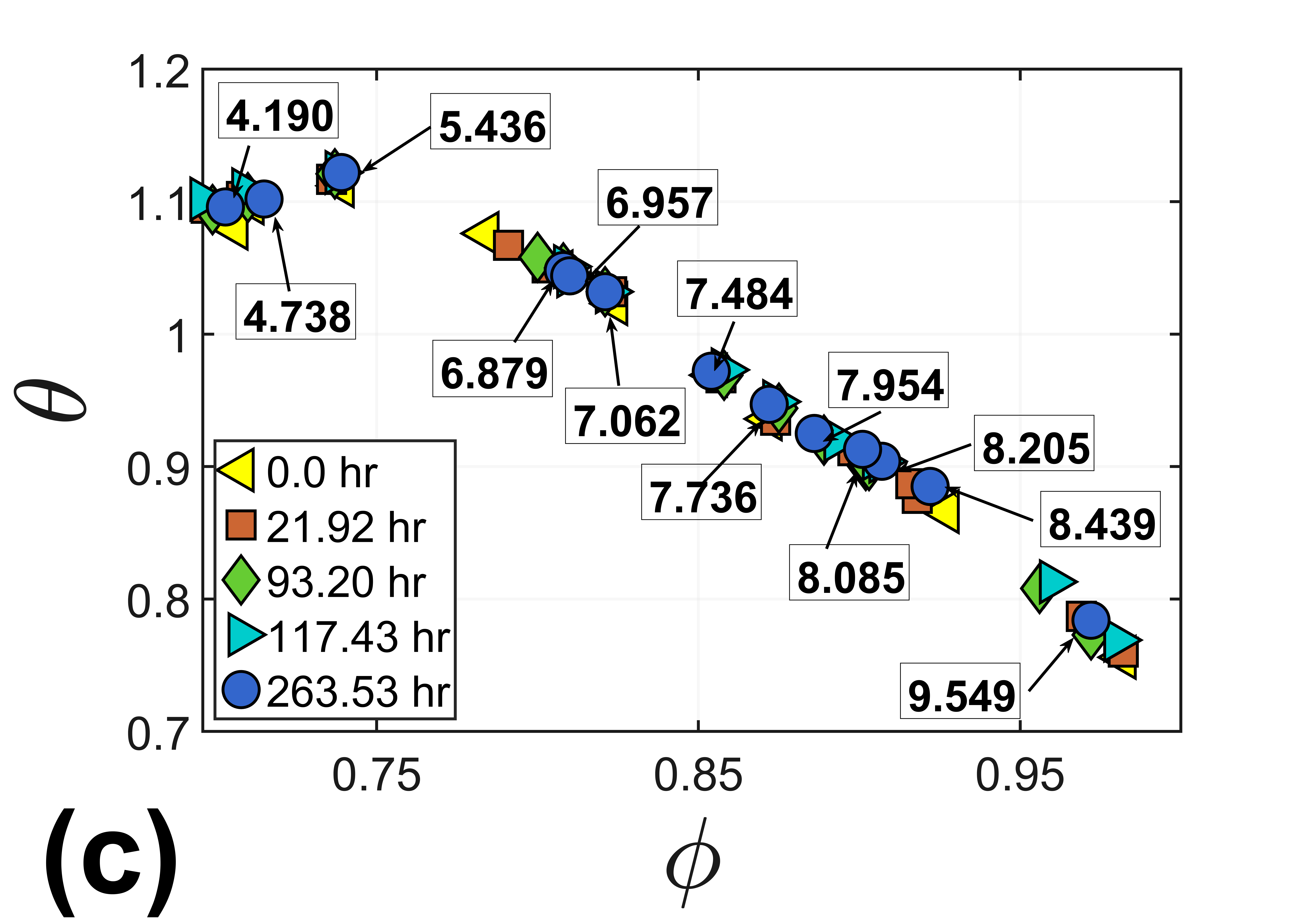}
        %\caption{Third}
        \label{fig:third}
    \end{subfigure}
    \caption{The calibration curves with various time: (a) the calibration curve of the Hue-based method; (b) the calibration curve of the gray-based method; (c) the calibration curve of the $\mathbf{(\phi,\theta)}$-based method where the text in the inset corresponds to the pH values of 263.53 hr.}
    \label{Fig_Time}
\end{figure}

%The calibration curve should ensure reproducibility and allow straightforward interpolation.
%The color and pH of each sample may vary over time, potentially as a result of the instability of the LED light luminance, the shift of equilibrium reactions due to the temperature change.
A robust calibration procedure needs to be repeatable over time.
The pH and colors were remeasured for repeated times following the steps above. Figure \ref{Fig_Time} shows the impacts of time on the calibration path by the three techniques. Every data point is the average of three measurements. 
In general, slight pH shifts were observed in some samples over time, most likely due to environmental fluctuations such as temperature.
%Correspondingly, the Hue, gray value, and $\mathbf{(\phi,\theta)}$ were able to capture these pH variations, ensuring that the calibration paths consistently followed a defined trend and remained reproducible, as shown in Figure.\ref{Fig_Time}. 
We notice, however, that the curve of the gray-based method is still a little dispersed within the range of pH 7-9, while the curves for the Hue and $\mathbf{(\phi,\theta)}$ methods were mostly unaffected. 

%Moreover, the calibration curves employed by the hue method are monotonic, facilitating a more efficient and straightforward pH interpolation compared with the other two methods. 

\subsection{The impact of camera settings}
\begin{figure}[H]
    \centering
    % First figure
    \begin{subfigure}[b]{0.32\textwidth}
        \includegraphics[width=\textwidth]{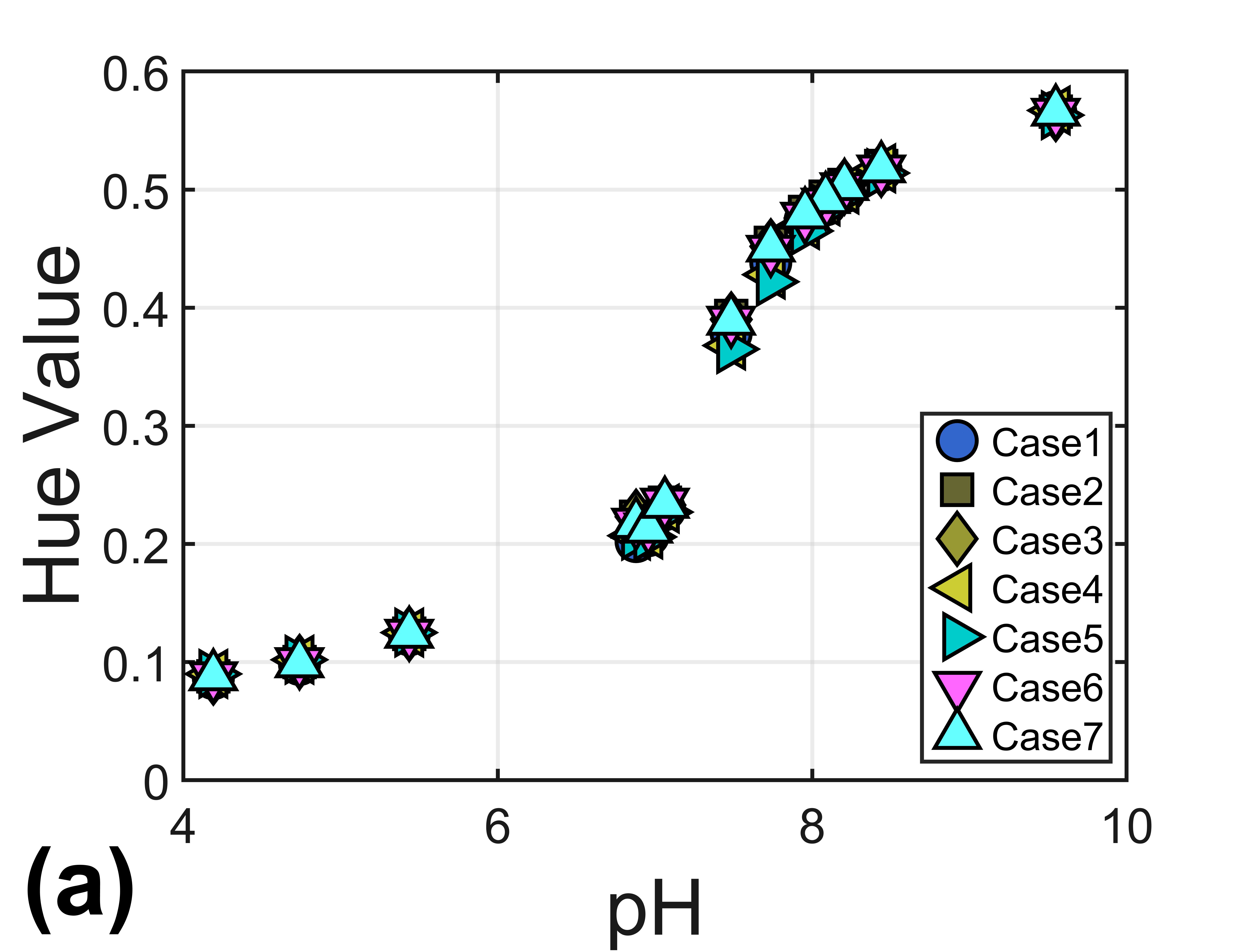}
        %\caption{First}
        \label{fig:first}
    \end{subfigure}
    \hfill
    % Second figure
    \begin{subfigure}[b]{0.32\textwidth}
        \includegraphics[width=\textwidth]{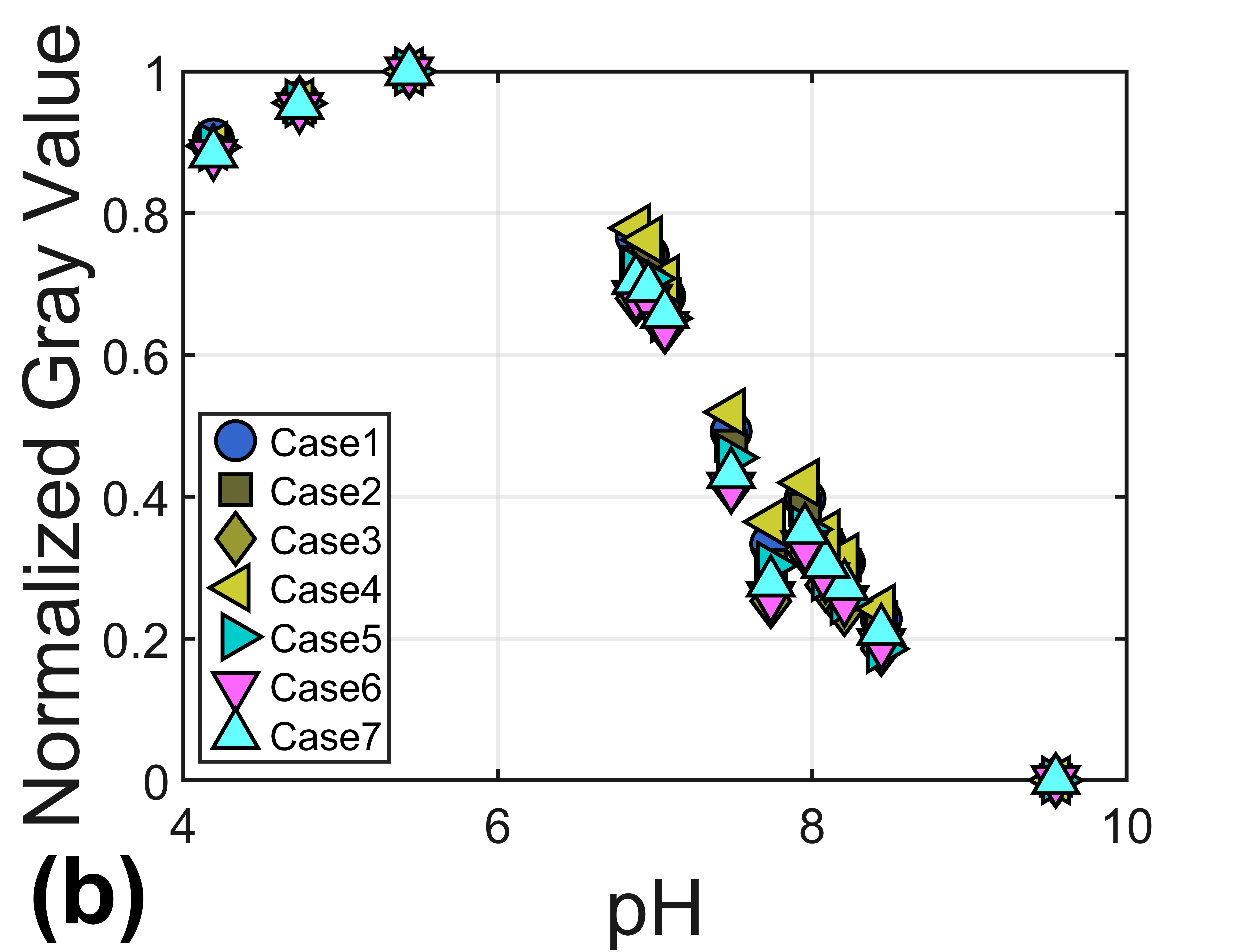}
        %\caption{Second}
        \label{fig:second}
    \end{subfigure}
    \hfill
    % Third figure
    \begin{subfigure}[b]{0.34\textwidth}
        \includegraphics[width=\textwidth]{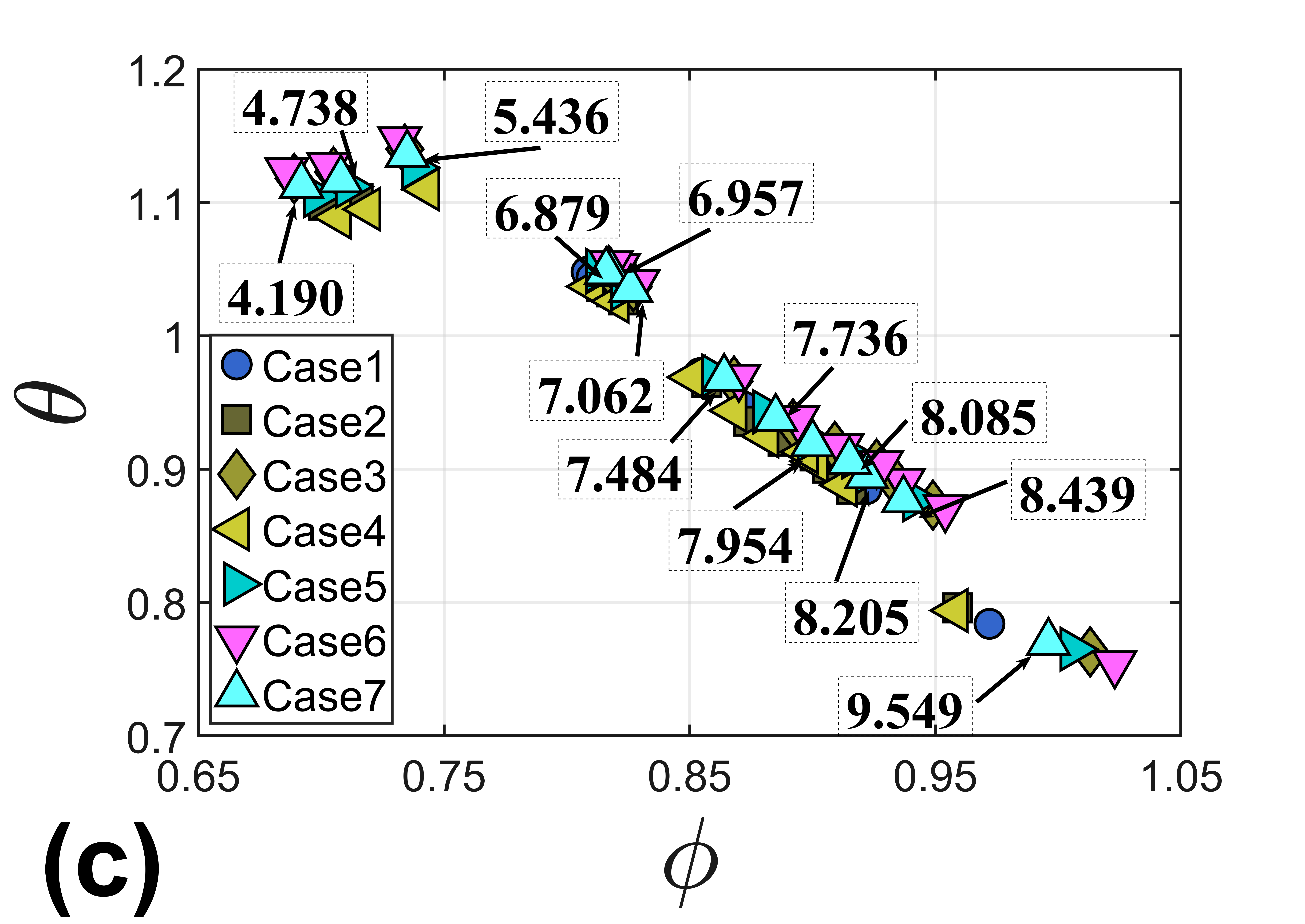}
        %\caption{Third}
        \label{fig:third}
    \end{subfigure}
     \caption{The calibration curves with various camera settings. The calibration test was conducted at the same time as 263.53 hr, but under different camera settings. (a) the calibration curve of the Hue-based method; (b) the calibration curve of the gray-based method; (c) the calibration curve of the $\mathbf{(\phi,\theta)}$-based method where the text in the inset corresponds to the case 7.}
    \label{Fig_camera}
\end{figure}

Camera settings such as the aperture, the shutter speed, and ISO could exert a large influence on the color perception by cameras. 
%Even for the same settings, the different qualities among cameras can also lead to discrepancies in the imaging. 
Due to that, calibration procedures may need to be rebuilt once the settings are varied, causing those calibration curves to be less applicable.
A robust calibration technique should always allow for a stable calibration path, less dependent on the camera settings. This ensures a broad and universal applicability of the calibration curve under different scenarios. 
In this section, the impacts of various camera settings were investigated, and the relevant parameters in each case are shown in the Table \ref {table:2}. 
As noticed in Figure \ref{Fig_camera}, the Hue-pH correlation curve almost follows the same trend with a nearly negligible drift, regardless of the camera settings. However, the $\mathbf{(\phi,\theta)}$ experienced a small degree of drift in both  $\mathbf{(\phi)}$ and $\mathbf{(\theta)}$. 
We noticed that the gray value levels across different camera parameters differed mostly in overall scale rather than in the shape of the calibration curve. In order to account for that, we introduced a normalized gray value level given by:

%Since the gray values are similar in shape, with variations only in scale, the normalized gray value was utilized to evaluate the variation and calculated by: 
\begin{equation}
    \text{Normalized gray value} = \frac{gray - \min\{gray\}}{\max\{gray\} - \min\{gray\}}.
    \label{normalized_g}
\end{equation}
Moreover, even after normalization, we observed that the gray value shifted more significantly than for the other two techniques, with values fluctuating around 0.2 from pH 7 to pH 8. This would result in significant errors in the pH interpolation if the camera settings vary.  

%This demonstrates that the hue method is a potentially robust technique for calibrating pH due to its stability against variations of the camera settings.

%where the $\max_g$ and $\min_g$ are the maximum and minimum gray values in each case, respectively

% \usepackage{multirow}
\begin{table}[H]
\centering
\begin{tabular}{l| l| l| l| l}
\hline
Case & The variations                      & Shutter Speed {[}s{]} & ISO      & Aperture \\ \hline
1    & Basic Case                          & 1/100                 & ISO 100  & f/13     \\ \hline
2    & \multirow{2}{*}{Speed variation} & 1/80                  & ISO 100  & f/13     \\ %\cline{1-1} \cline{3-5} 
3    &                                     & 1/160                 & ISO 100  & f/13     \\ \hline
4    & \multirow{2}{*}{Aperture variation} & 1/100                 & ISO 100  & f/11     \\ %\cline{1-1} \cline{3-5} 
5    &                                     & 1/100                 & ISO 100  & f/16     \\ \hline
6    & \multirow{2}{*}{ISO variation}      & 1/100                 & ISO 64   & f/13     \\ %\cline{1-1} \cline{3-5} 
7    &                                     & 1/100                 & ISO 80   & f/13     \\ \hline
\end{tabular} 
\caption{The camera setting parameters in various cases. The parameters in the basic case are what utilized in the convection experiments.}
\label{table:2} 
\end{table}

\subsection{The impact of LED light illuminance}

\begin{figure}[H]
    \centering
    % First figure
    \begin{subfigure}[b]{0.32\textwidth}
        \includegraphics[width=\textwidth]{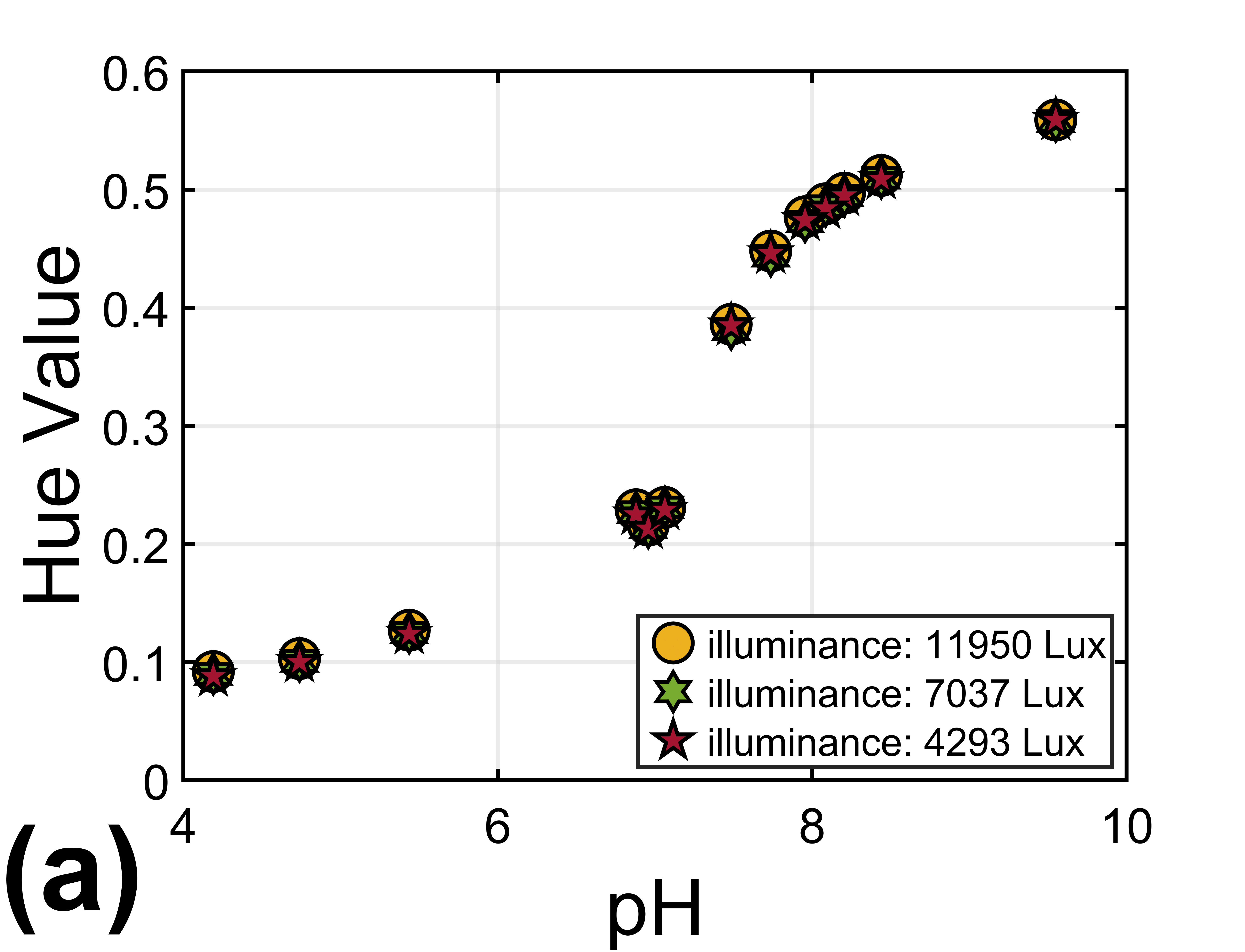}
        %\caption{First}
        \label{fig:first}
    \end{subfigure}
    \hfill
    % Second figure
    \begin{subfigure}[b]{0.32\textwidth}
        \includegraphics[width=\textwidth]{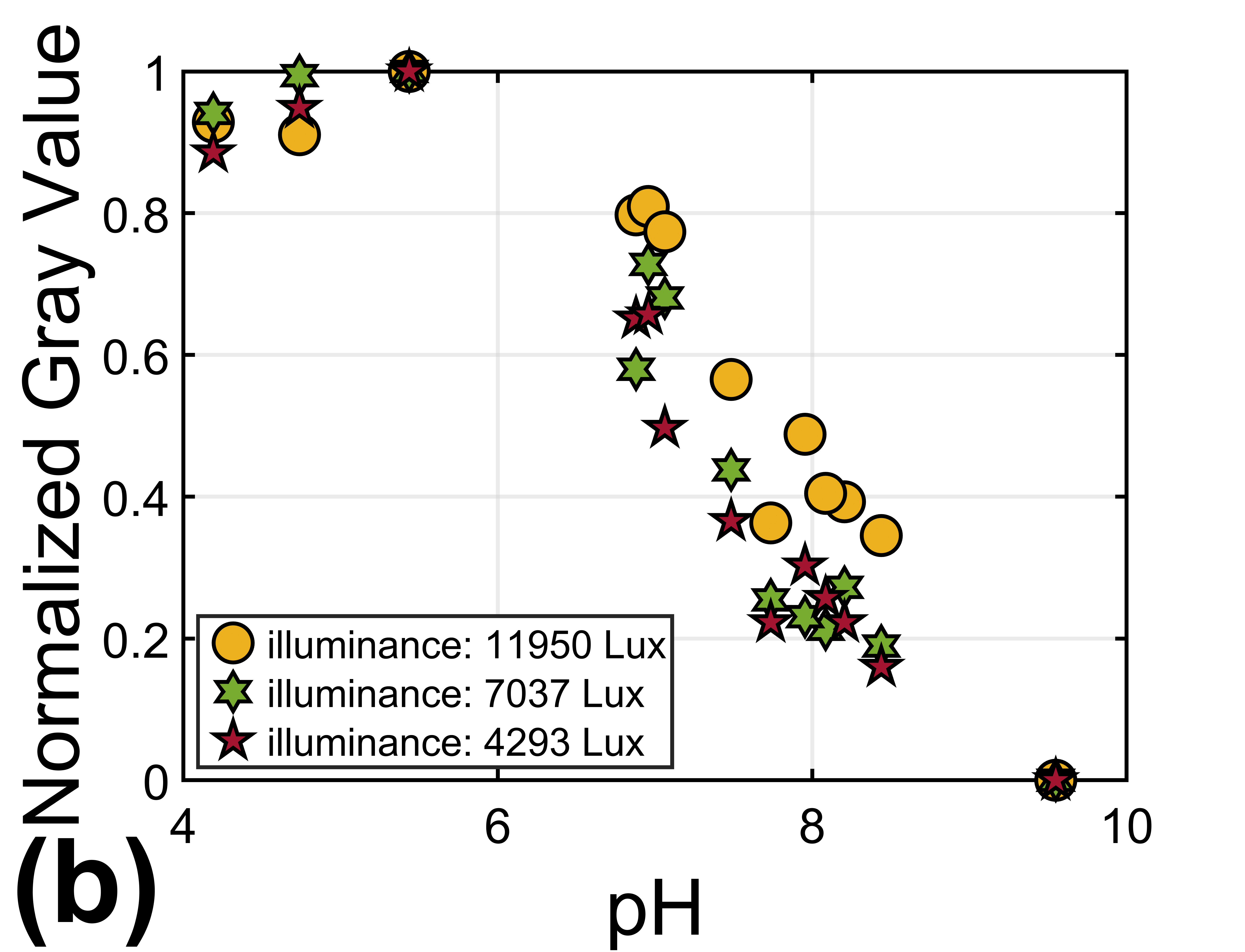}
        %\caption{Second}
        \label{fig:second}
    \end{subfigure}
    \hfill
    % Third figure
    \begin{subfigure}[b]{0.34\textwidth}
        \includegraphics[width=\textwidth]{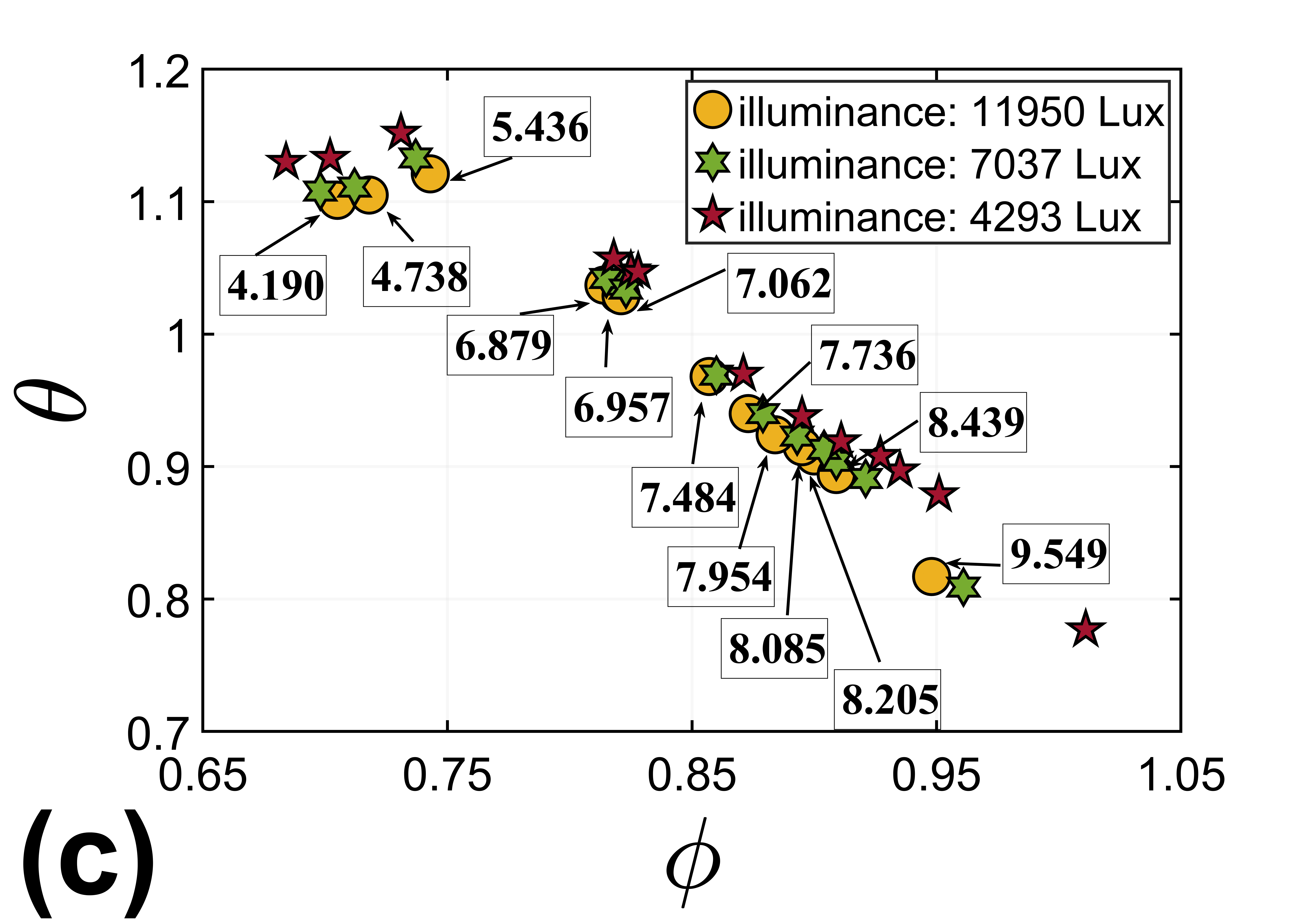}
        %\caption{Third}
        \label{fig:third}
    \end{subfigure}
    \caption{The calibration curves with various LED illuminance. The calibration test was conducted at the same time: (a) the calibration curve of the Hue-based method; (b) the calibration curve of the gray-based method; (c) the calibration curve of the $\mathbf{(\phi,\theta)}$-based method where the text in the inset corresponds to pH values of illuminance: 11950 lux.}
    \label{Fig_LED}
\end{figure}

Moreover, the heterogeneity of LED illuminance in time and space could also have an impact on color imaging. This is another uncertainty that a robust calibration should overcome. A good technique should be robust against color changes caused by LED illuminance drift or environmental factors, while remaining sensitive enough to shift noticeably even with small pH variations. In this case, the normalized gray values experienced a significant change, even for the same samples as shown in Figure \ref{Fig_LED}. This may be because the digital camera's sensor is not linearly responsive to ambient illuminance. The illuminance of the light box was quantified using a digital lux meter (UNI-T, UT383). To ensure consistency, the meter was positioned at the geometric center of the light box, oriented parallel to the surface at a fixed vertical distance of 3 cm. The reported values in Figure \ref{Fig_LED} represent the averages of three independent measurements.

\subsection{The impact of fluid thickness }\label{thickness_sec}
The attenuation of light is dependent on the light's passage through the fluid.
%The greater distance the light passes through the fluid, the more attenuation it will experience. 
In other words, the color perceived by the camera may also rely on the fluid thickness inside the pores.  % should have a sketch for the experiment set up to indicate this h.
This presents another challenge in building a robust calibration curve.
%The fluid thickness can also influence the colors captured by a camera. 
As per Beer's law:
\begin{equation}
    A=\epsilon lc,
\end{equation}
where $A$ is the absorbance, $\epsilon$ is the molar absorptivity, $l$ is the length of the solution the light passes through, and $c$ is the concentration of the solution.

\begin{figure}[H]
    \centering
    % First figure
    \begin{subfigure}[b]{0.32\textwidth}
        \includegraphics[width=\textwidth]{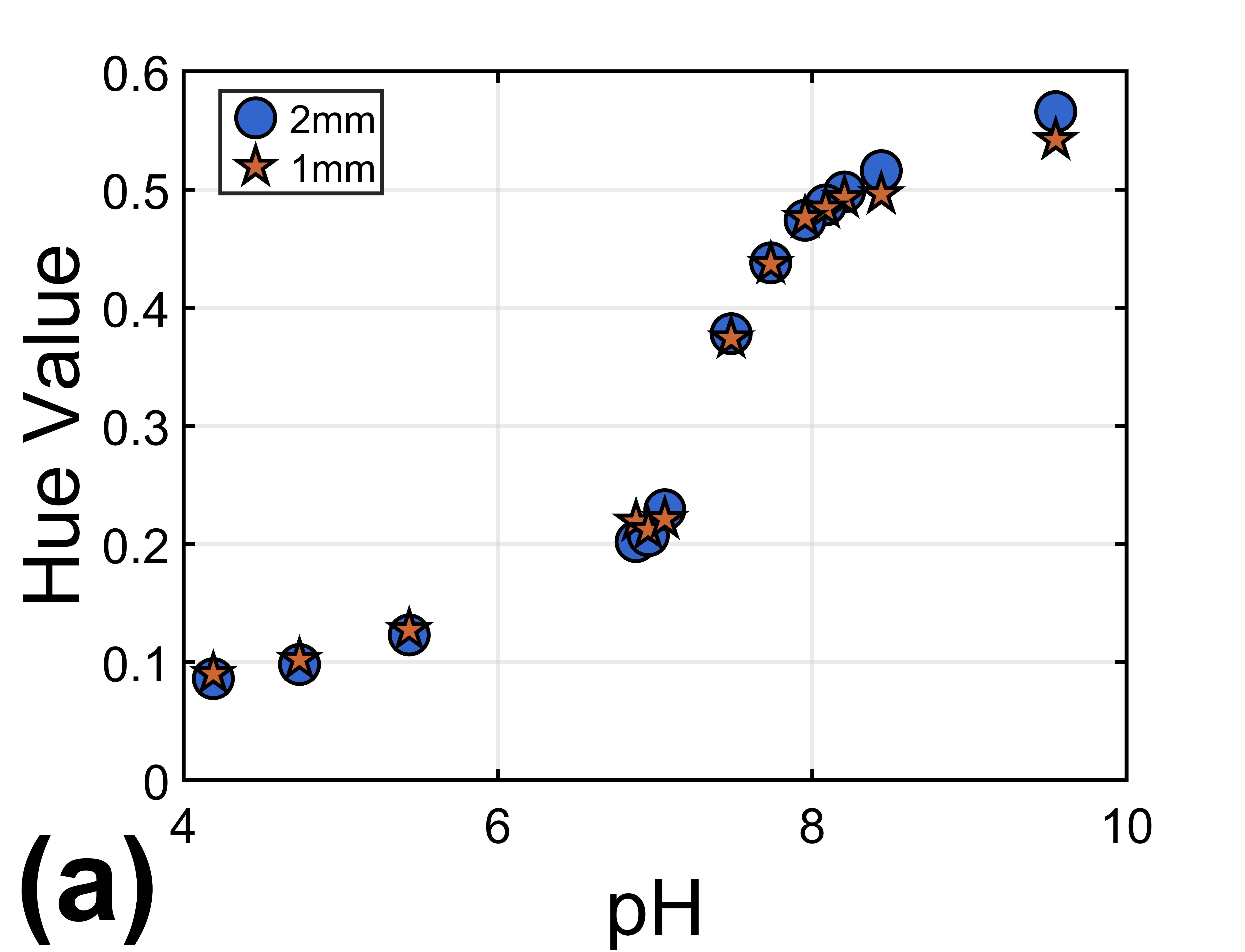}
        %\caption{First}
        \label{fig:first}
    \end{subfigure}
    \hfill
    % Second figure
    \begin{subfigure}[b]{0.32\textwidth}
        \includegraphics[width=\textwidth]{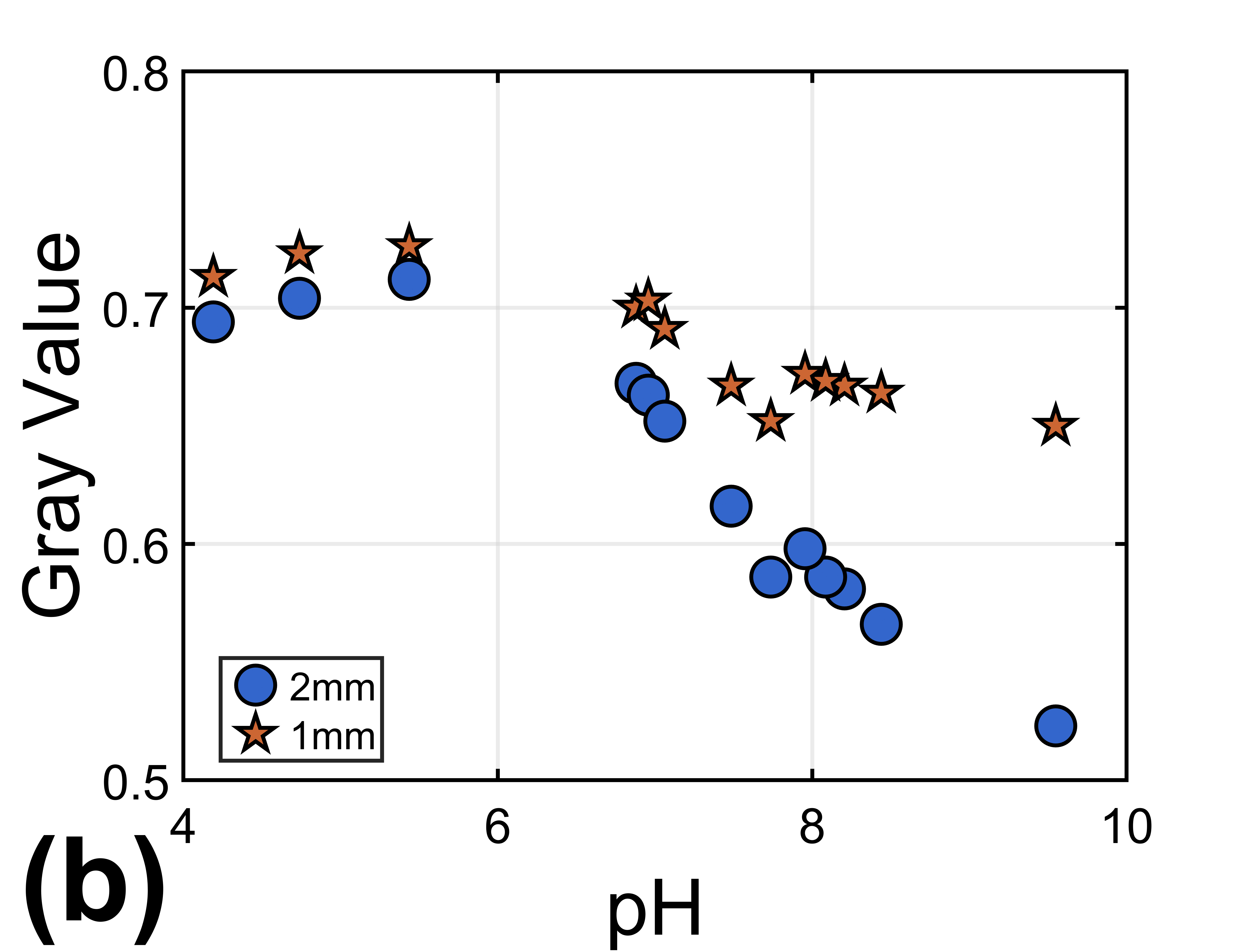}
        %\caption{Second}
        \label{fig:second}
    \end{subfigure}
    \hfill
    % Third figure
    \begin{subfigure}[b]{0.34\textwidth}
        \includegraphics[width=\textwidth]{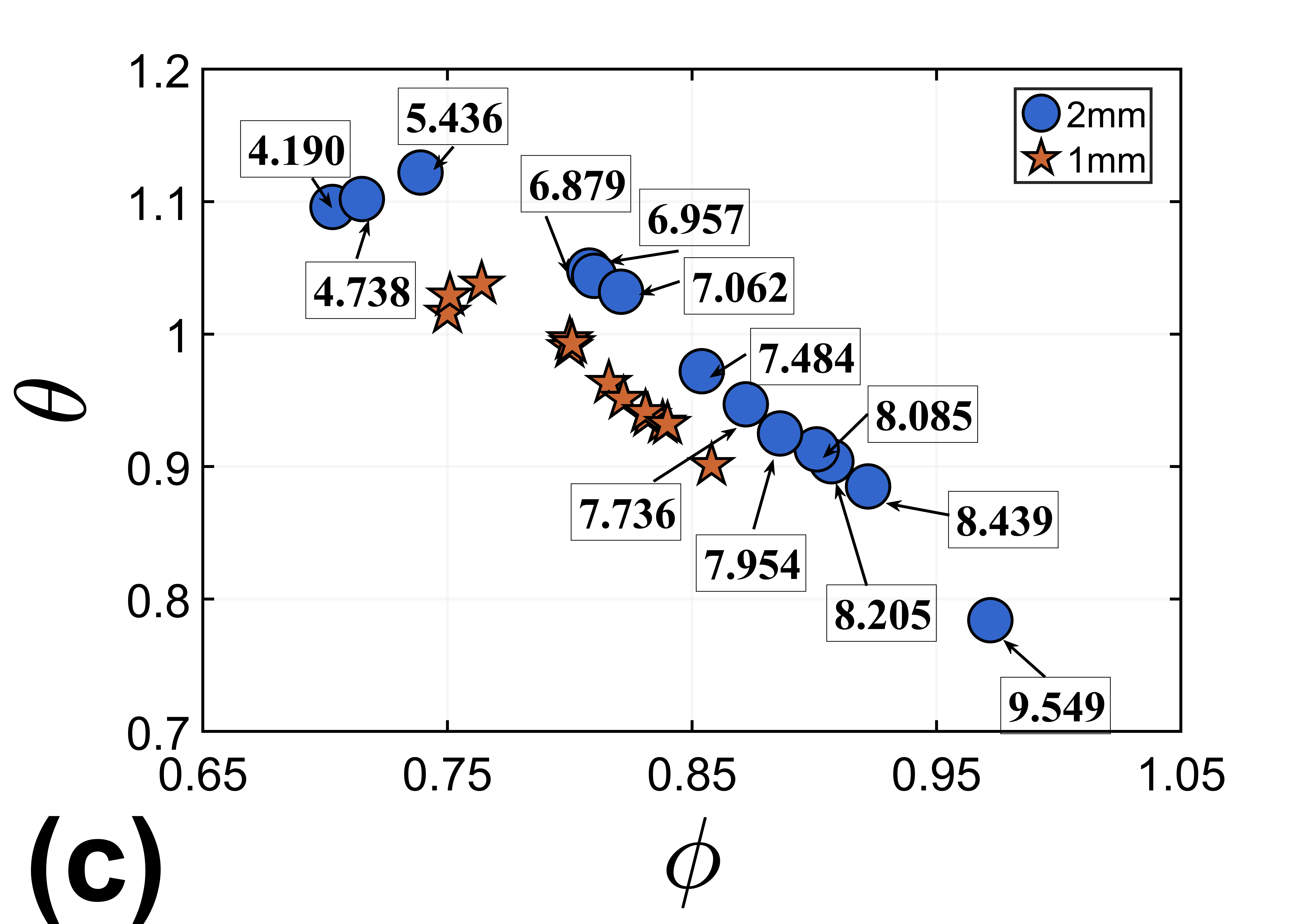}
        %\caption{Third}
        \label{fig:third}
    \end{subfigure}
    \caption{The calibration curves with various fluid thickness: (a) the calibration curve of the Hue-based method; (b) the calibration curve of the gray-based method; (c) the calibration curve by $\mathbf{(\phi,\theta)}$, where the text in the inset corresponds to pH values of fluid height of 2 mm.}
    \label{Fig_height}
\end{figure}

\begin{figure}[H]
    \centering
    \includegraphics[width=0.7\linewidth]{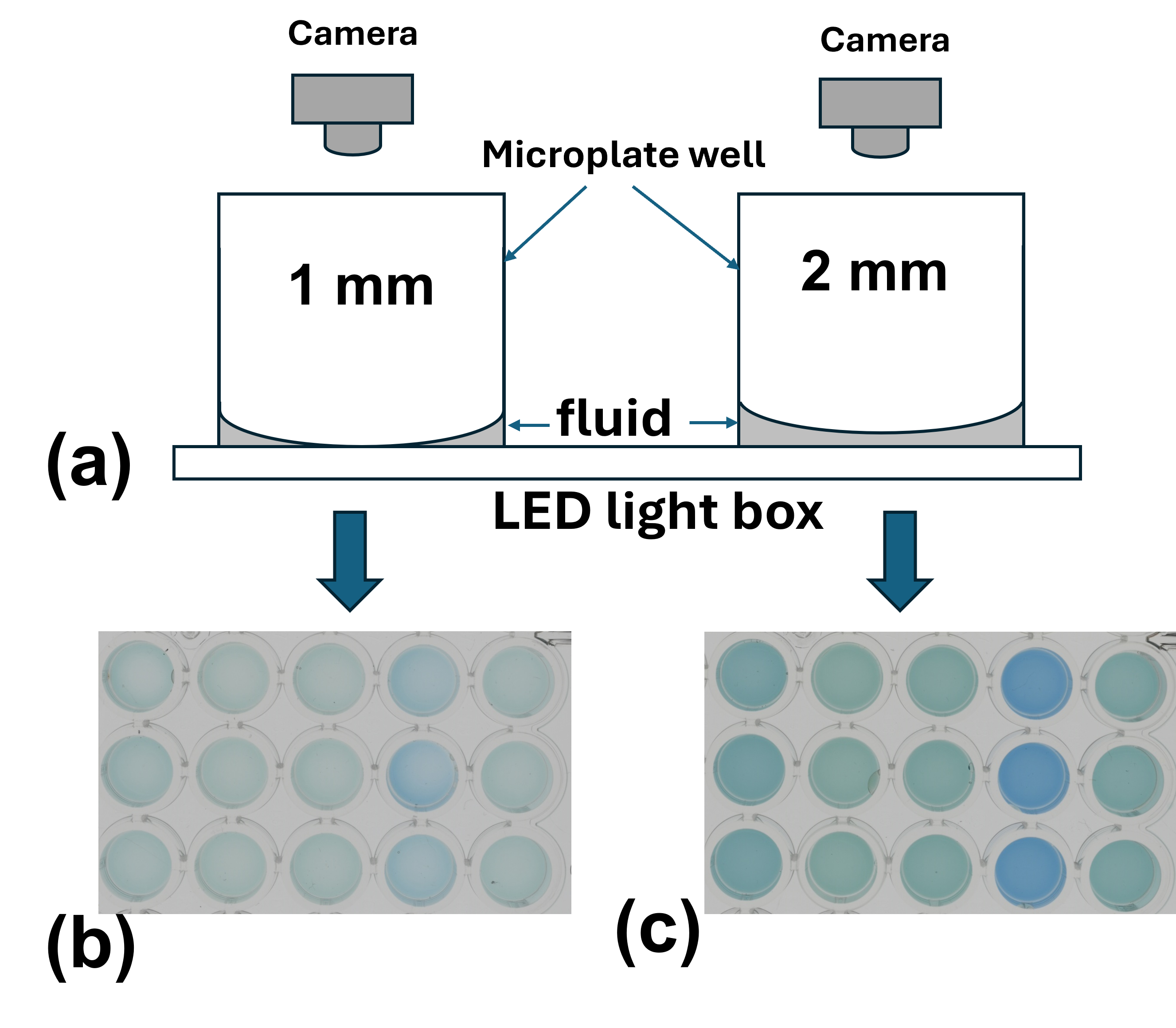}
    \caption{(a) Schematic illustration of varying fluid depths within the microplate wells. (b) Representative top-view images of samples with a $1\text{ mm}$ fluid depth; each column displays triplicate replicates of a single sample, with different samples presented across the columns. (c) Corresponding images of the same fluid samples at a $2\text{ mm}$ depth. The fluid depth was calculated based on the fluid volume and the cross-section of the wells.}
    \label{thickness}
\end{figure}

As shown in Figure \ref{Fig_height}, there is an apparent shift in the gray values when the height of the fluid has changed. In addition, $\mathbf{(\phi,\theta)}$ noticed more changes than in other scenarios where we changed the luminance and camera settings. Nevertheless, the Hue values exhibit greater variability in the $1\text{ mm}$ case, which is likely attributable to a more pronounced meniscus effect. As illustrated in Figure \ref{thickness}, the low fluid volume prevents the formation of a uniform layer across the well. Instead, surface tension causes the fluid to accumulate near the walls, significantly reducing the liquid depth at the center.  This non-uniformity  leads to fluctuations in the calculated Hue at 1 mm case.
Apart from that, Hue values still seem to be more invariant, subject to the different fluid thickness.

%where most solutes accumulate at the edge but present little in the center of the wells. 

\subsection{pH interpolation}\label{interpolation}

In this section, we perform mathematical fitting of the calibration data for the three proposed methods. These fitted models are subsequently applied to interpolate the spatiotemporal pH distributions within the density-driven convection (DDC) experiments.

\begin{figure}[H]
    \centering
    % First figure
    \begin{subfigure}[t]{0.48\textwidth}
        \includegraphics[width=\textwidth]{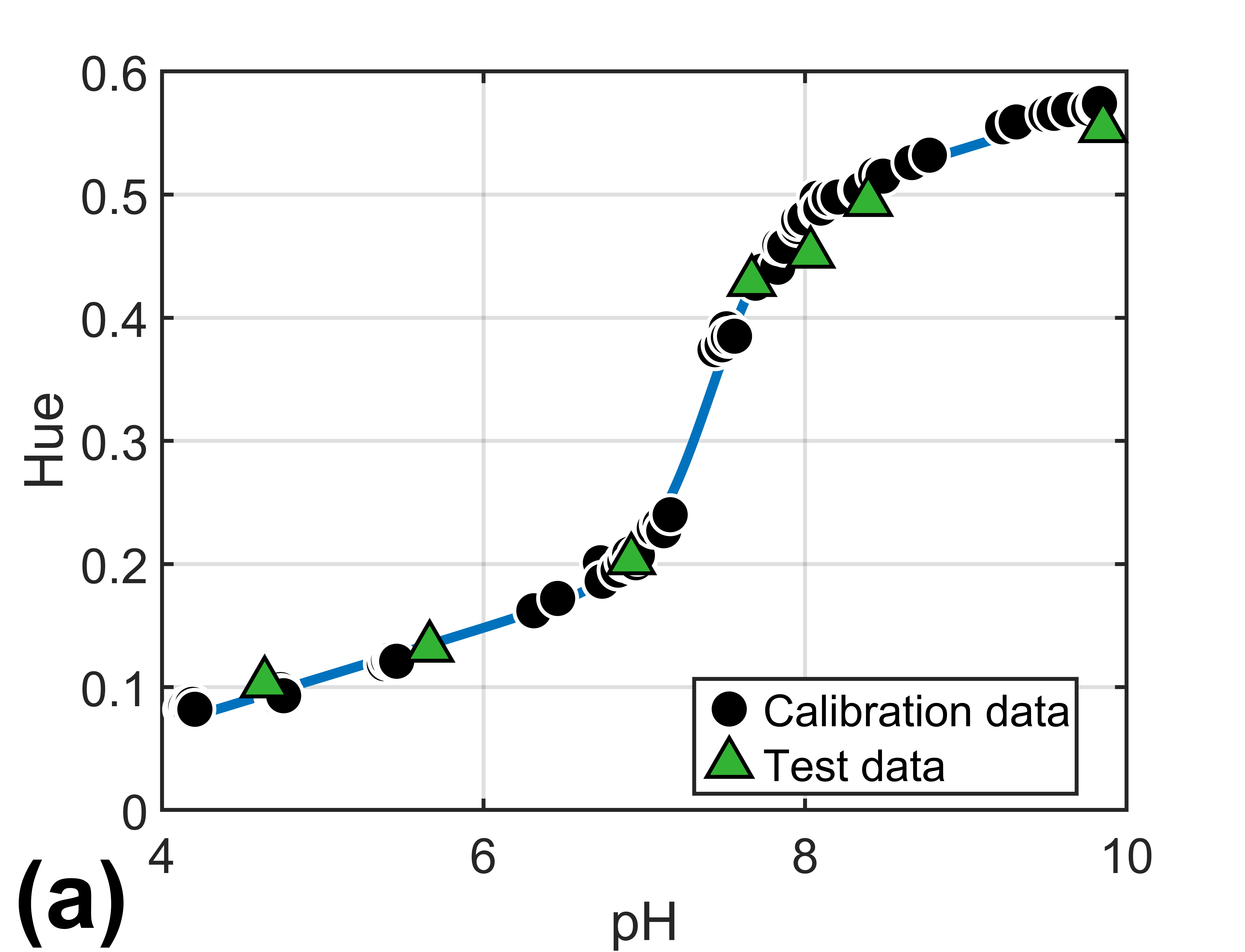}
        %\caption{First}
        \label{fig:first}
    \end{subfigure}
    \hfill
    % Second figure
    \begin{subfigure}[t]{0.48\textwidth}
        \includegraphics[width=\textwidth]{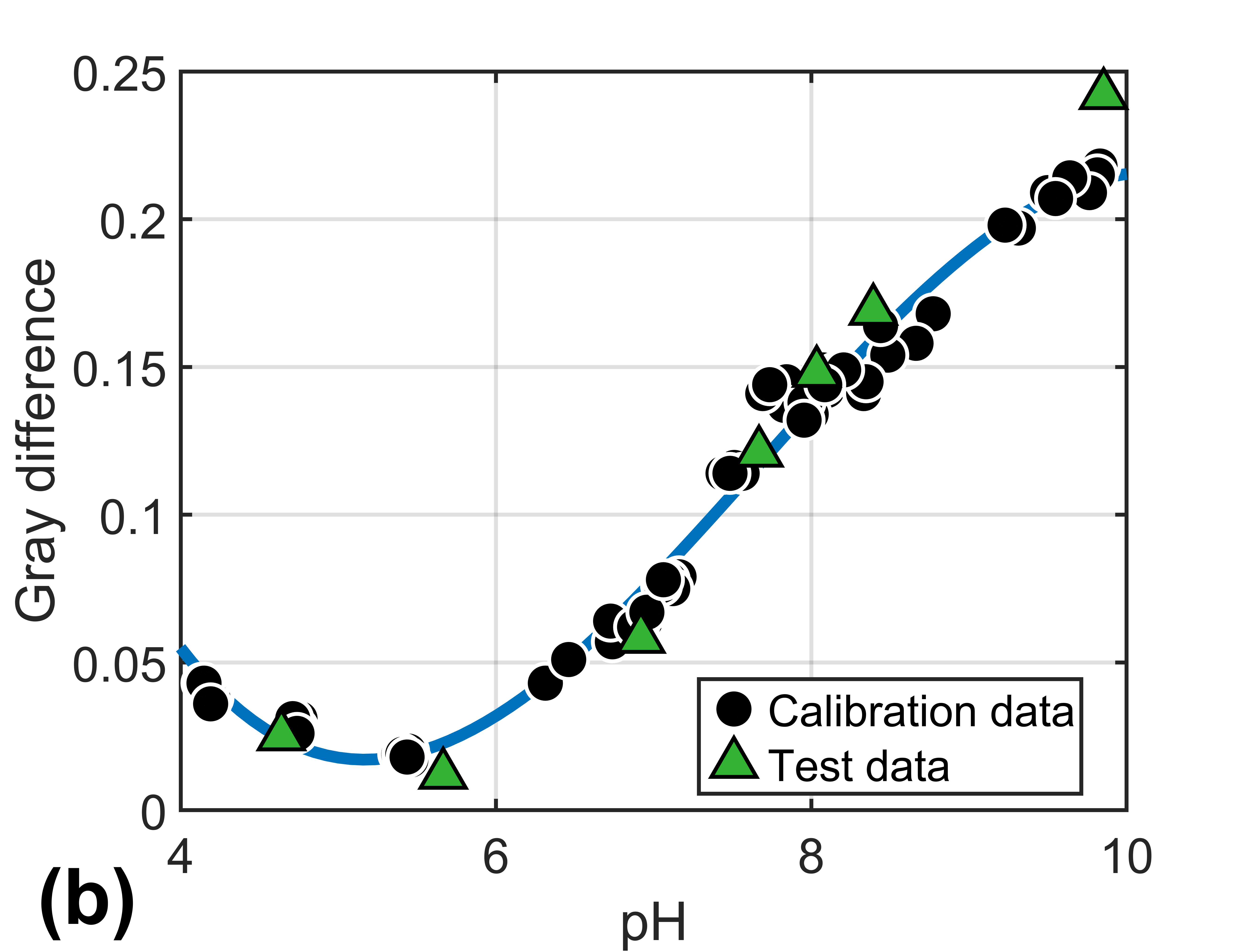}
        %\caption{Second}
    \end{subfigure}   
    \caption{The fitting curves. (a): The Hue fitting curve by the sigmoidal relationship. (b): The gray-difference value fitted by the cubic polynomial. The calibration data were measured from the 13 samples at five different times. The fitting was based on these data and showed an excellent agreement. The test data were from different sample solutions to avoid bias or overfitting. This type of data will be utilized to test the fitting quality.}
     \label{fitting}
   \hfill
    \end{figure}

The pH-Hue data were fitted using: 
\begin{equation}
    Hue=b_1+\frac{(b_2-b_1)}{1+10^{b_3(pH-b_4)}}+b_5(pH-b_4),
    \label{sigmoid}
\end{equation}
which was adapted from a sigmoidal relationship reported in the literature to calculate Hue\cite {cantrell2010use}. The change was the addition of a linear part in our fitting: $b_5(pH-b_4)$, which is critical to fitting the tilted trend in the range of pH 4-6 and pH 8-10.  Without it, the curves will remain flat in both ranges. 
The calibration data are exactly the same as shown in Figure \ref {Fig_Time}(a). 
The fitting curve is shown in Figure \ref{fitting}(a).

%where the calibration data set fit well, showing an excellent correlation ($R^2=0.998$).
%As shown in Fig.\ref{fitting}, the test data set was also well fitted by the curve with $RMSE=0.0171$.% and $MAE=0.0128$.

\begin{figure}[H]
    \centering
    \includegraphics[width=0.95\linewidth]{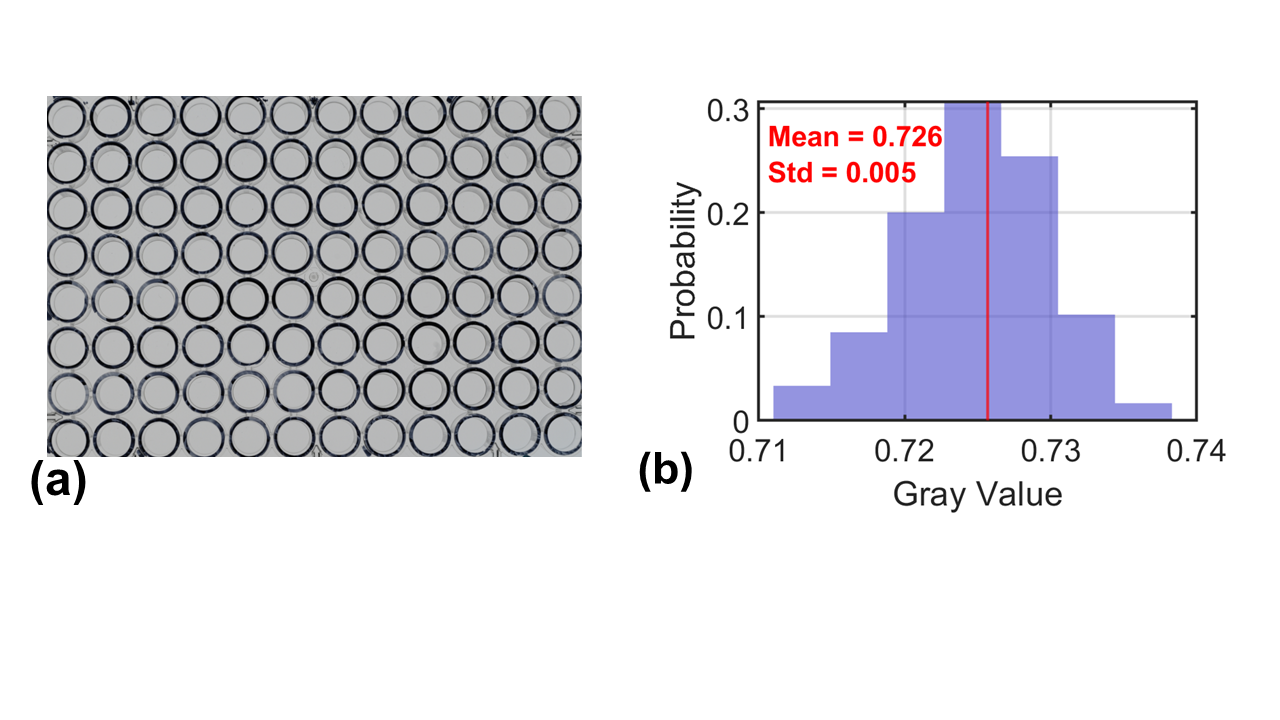}
    \caption{(a): Visualization of the empty microplate model. (b): Corresponding grayscale value distribution for pixels in all the empty wells ($\mu = 0.726, \sigma = 0.005$).}
    \label{microplate_background}
\end{figure}

Although the microplates are transparent, our analysis shows that they can still absorb some light and attenuate the intensity. Therefore, the gray-value difference was considered for fitting. This quantity is computed as:

\begin{equation}
    gray_{diff}=gray_{0}-gray,
\end{equation}
where $gray_0$ denotes the gray value of the microplate without fluid, and $gray$ corresponds to the data shown in Figure \ref{Fig_Time}(b). This represents the fluid-induced gray attenuation without interference from the plate.
Although $gray_0$ may vary slightly from well to well in the background (empty model), the standard deviation is only 0.005 (Figure \ref{microplate_background}(b)), indicating a highly uniform gray-value distribution. To simplify the procedure and calculations, we assume a uniform background and adopt the mean value, 0.726, for all wells. Using this approximation, the calibration data $gray_{diff}$ are obtained for fitting, as shown in Figure \ref{fitting}(b).

The gray-difference value was fitted by the cubic polynomial:
\begin{equation}
    gray_{diff}=a_1pH^3+a_2pH^2+a_3pH+a_4
    \label{polynomial}
\end{equation}
and the fitting graph was shown in Figure \ref{fitting}(b). 

%The consistency between the calibration data and the fitting was evaluated by $r^2$, calculated by:  
%\begin{equation}
%r^2 = 1 - \frac{\sum_{i=1}^n (pH_i - \hat{pH}_i)^2}{\sum_{i=1}^n (pH_i - \bar{pH})^2}
%\end{equation}
The error between the fitted curve and the experimental data was quantified using the root mean square error (RMSE), calculated as follows:
\begin{equation}
\mathrm{RMSE} = \sqrt{\frac{1}{n} \sum_{i=1}^n (pH_i - \hat{pH}_i)^2}.
\end{equation}
where $pH_i$ denotes the $i$-th observed pH value in the calibration dataset, $\hat{pH}_i$ represents the corresponding value predicted by the fitted model. The subscript $i$ indicates the individual data point index, and $n$ ($n=62$) is the total number of observations.

Since both cases involve various fitting parameters, several new test solutions (distinct from the 13 samples used to construct the calibration curves) were prepared to evaluate the fitting quality and to prevent bias and noise in the calibration dataset from being overfitted.
As per Table \ref{fitting table}, the curves show excellent agreement with the calibration data, with RMSE values of 0.0076 and 0.0082 for Hue and gray-difference fitting, respectively. The RMSEs of the test data in both cases were also small (0.0171 for Hue and 0.0139 for the gray-difference method), indicating good fitting quality. 

 \begin{table}[H]
\centering
\begin{tabular}{lll}
\hline
                         & Hue method                                                                                  & Gray method                                                                             \\ \hline
Fitting                  & Modified sigmoidal                                                                                   & Cubic polynomial                                                                        \\ %\hline
Parameters               & \begin{tabular}[c]{@{}l@{}}$b_1$=0.20, $b_2$=0.48, $b_3$=-2.2,\\  $b_4$=7.41, $b_5$=0.04.\end{tabular} & \begin{tabular}[c]{@{}l@{}}$a_1$=-0.0032, $a_2$=0.074, \\ $a_3$=-0.50, $a_4$=1.10.\end{tabular} \\ %\hline
 
RMSE of calibration data & 0.0076                                                                                     & 0.0082                                                                                \\ %\hline
RMSE of test data        & 0.0171                                                                                      & 0.0139                                                                                  \\ \hline
\end{tabular}
\caption{The summary of the fitting conditions by Hue-based and gray-based techniques.}
\label{fitting table}
\end{table}

In both the Hue and gray-difference techniques, interpolation of pH requires only a single variable, which offers mathematical simplicity. Nevertheless, for the gray difference technique, within the pH range of 4 to 6, a given gray value may correspond to two distinct pH values, as per Figure \ref{fitting}(b). This non-uniqueness introduces considerable uncertainty into the gray-based method and can result in erroneous interpolation.
These uncertainties primarily arise from the coloring properties within the pH range of 4–6, where the indicators exhibit only minor color changes. As a result, the gray-based technique failed to distinguish these subtle variations, leading to overlaps in gray values across different pH levels. %In addition, pH interpolation is very sensitive to changes in gray difference. %As per Figure.\ref{fitting}(b), a change of 0.02 will lead to the pH shift from 6 to 6.5. 

Despite the limitations of the indicators, the Hue technique demonstrated superior sensitivity, effectively distinguishing these minor color transitions and maintaining a unique, one-to-one mapping across the target range.
However, in the Hue-based fitting curve, the slope is flatter in the pH ranges of 4–6 and 8–10 compared with 6–8, indicating that the color variation per pH unit differs across ranges. This curve shape is primarily governed by the coloring characteristics of the pH indicators, with steeper slopes corresponding to more distinct and pronounced color changes, and flatter slopes reflecting weaker color responses. Thus, the reduced slope in the Hue curve reflects the inherent limitations of the indicators rather than the Hue method itself.

%These uncertainties can be mainly attributed to the coloring properties within the pH range (from 4-6), in which the changes of the indicators' colors are minor. This brings challenges for gray techniques to separate those colors and cause the overlap for the different pH values.

%Accordingly, the pH can be interpolated from Hue values in an easy and efficient manner.  

\begin{figure}[H]
    \centering
    % First figure
    \begin{subfigure}[t]{0.48\textwidth}
        \includegraphics[width=\textwidth]{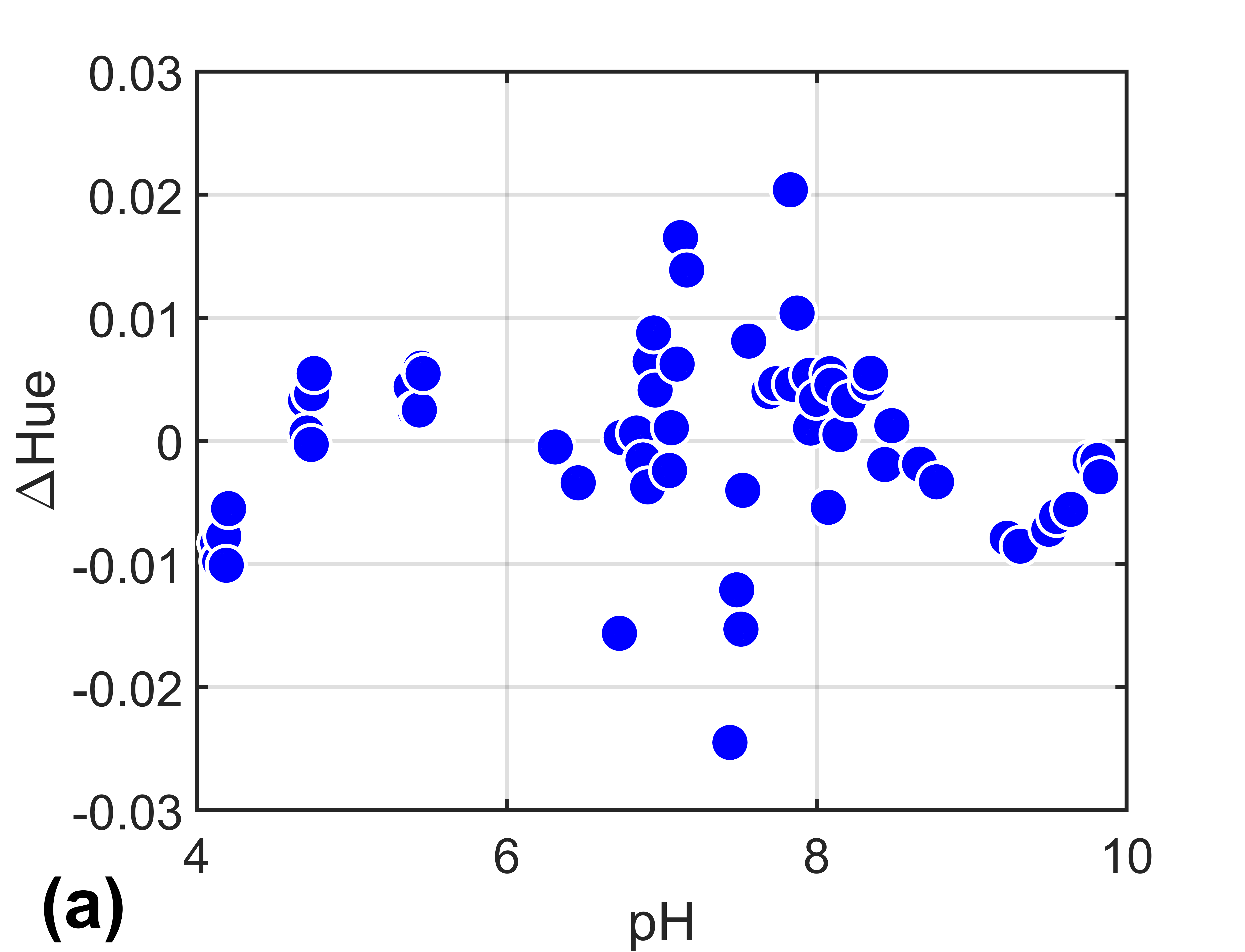}
        %\caption{First}
        \label{fig:first}
    \end{subfigure}
    \hfill
    % Second figure
    \begin{subfigure}[t]{0.48\textwidth}
        \includegraphics[width=\textwidth]{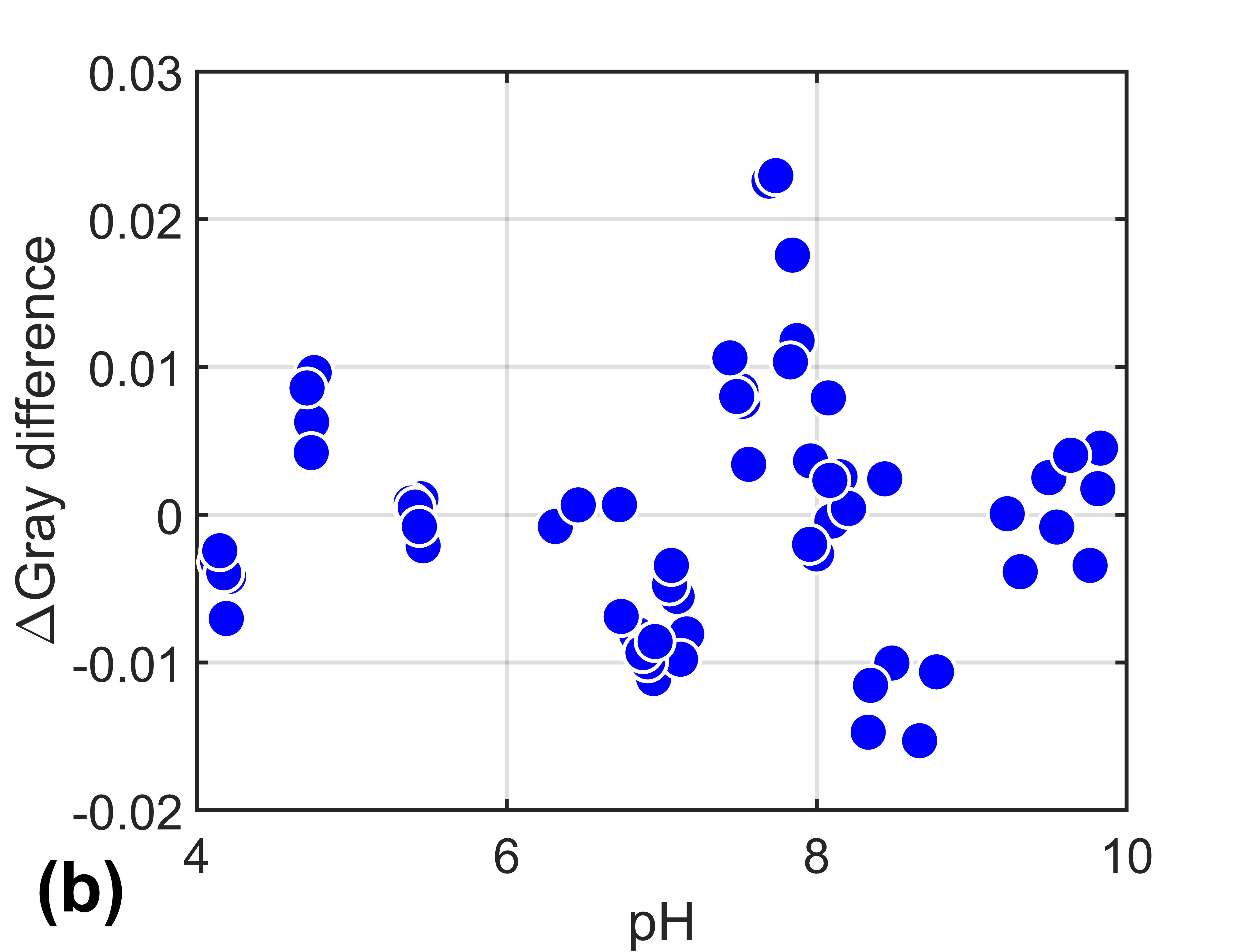}
        %\caption{Second}
    \end{subfigure}   
    \caption{The deviation of calibrated data from the fitting curve. (a)(b): The case of Hue and gray difference, respectively.}
     \label{deviation}
   \hfill
    \end{figure}

In addition, the deviation of the calibration data from the fitted model is shown in Figure \ref{deviation}.
    
As per Figure \ref{deviation}, the residuals from both methods generally fluctuate around 0 without a regular trend, indicating a good fitting quality and no systematic error.

Furthermore, the pH errors propagated from the uncertainties of measuring Hue and gray difference values are quantified below: 

\begin{equation}\label{coeff_hue}
    \sigma_{pH}=\frac{1}{{\left| \frac{\mathrm{d}(\mathrm{Hue})}{\mathrm{d}(\mathrm{pH})}\right|}}\sigma_{Hue},
\end{equation}

\begin{equation}\label{coeff_gray}
    \sigma_{pH}=\frac{1}{\left | \frac{\mathrm{d}(\mathrm{gray\ difference})}{\mathrm{d}(\mathrm{pH})}\right |}\sigma_{\mathrm{gray\ difference}},
\end{equation}
where $\sigma$ represents measurement errors, with the subscript identifying the specific parameter to which it pertains.

The sensitivity coefficients $|d(\text{gray difference}) / d(\text{pH})|^{-1}$ and $\left| d(\text{Hue})/d(\text{pH}) \right|^{-1}$ serve as critical parameters for quantifying the propagation of errors. Specifically, they determine the extent to which measurement noise in the Hue and grayscale-difference propagates into the final pH determination. As per Equations \ref{coeff_hue} and \ref{coeff_gray}, the pH determination becomes increasingly sensitive to experimental error as these coefficients rise, leading to reduced measurement precision.
The plot of the two coefficients with time is shown below:

    \begin{figure}[H]
    \centering
    % First figure
    \begin{subfigure}[t]{0.48\textwidth}
        \includegraphics[width=\textwidth]{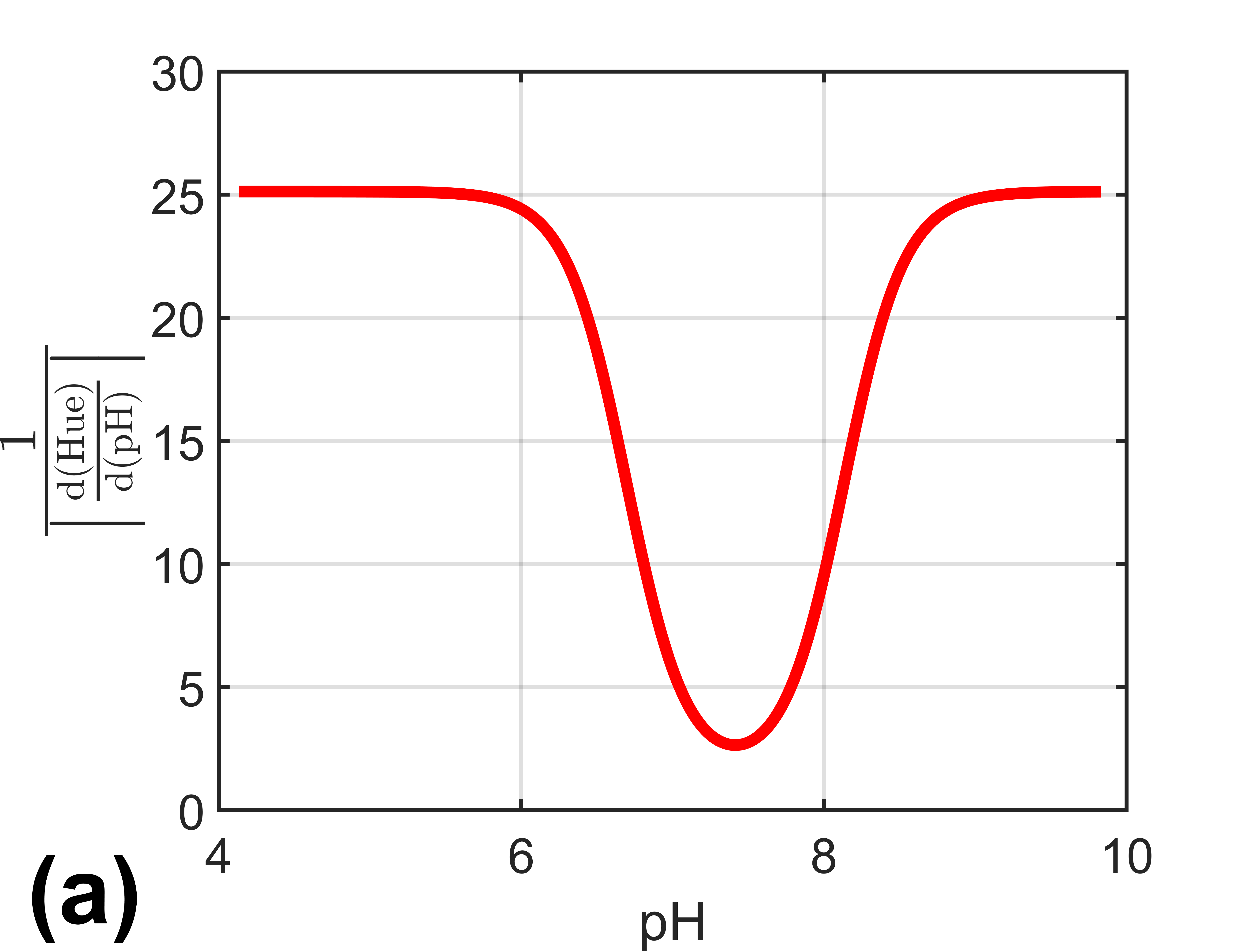}
        %\caption{First}
        \label{fig:firs}
    \end{subfigure}
    \hfill
    % Second figure
    \begin{subfigure}[t]{0.48\textwidth}
        \includegraphics[width=\textwidth]{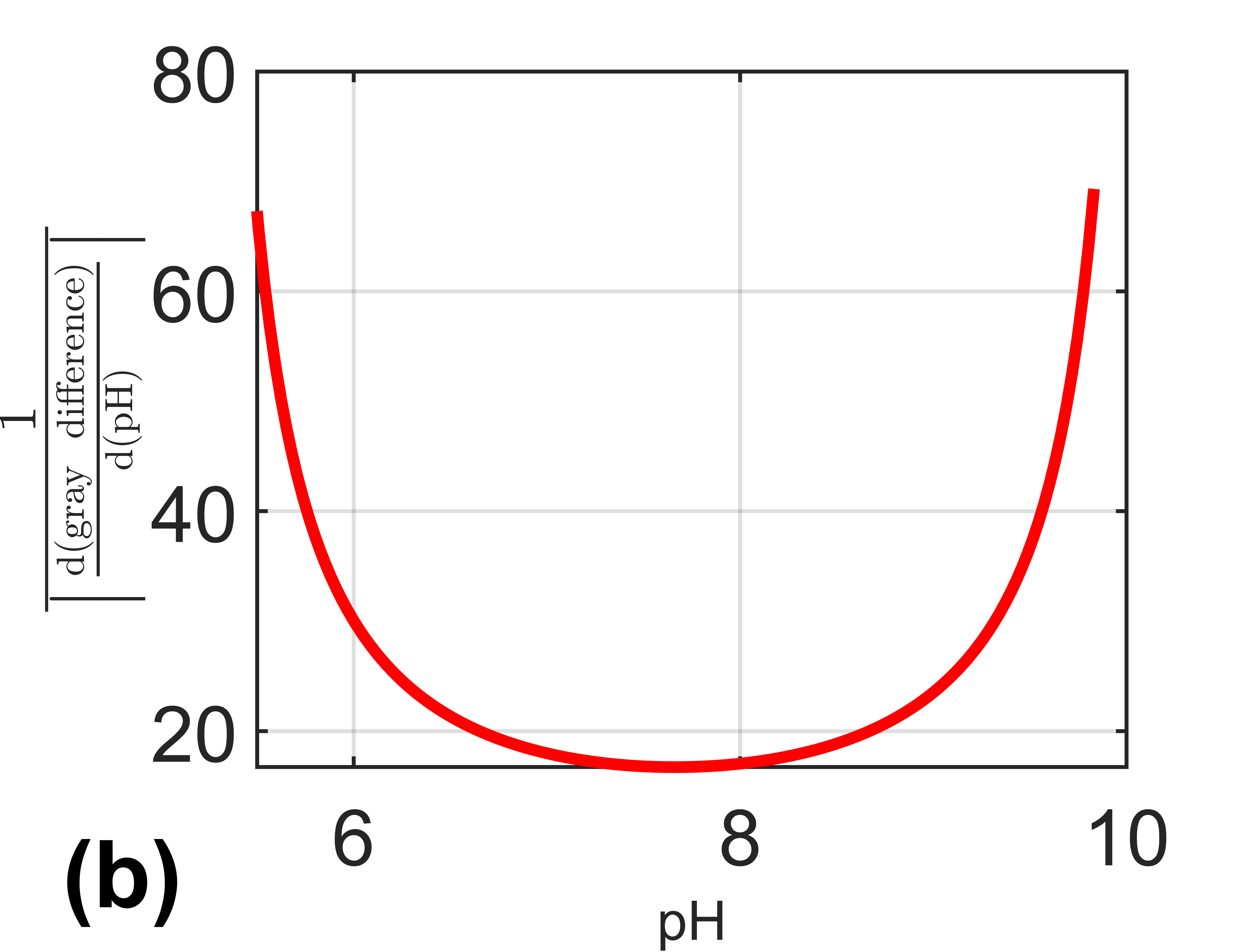}
        %\caption{Second}
    \end{subfigure}   
    \caption{(a) The inverse derivative of Hue with respect to pH across the pH range from 4 to 10. (b) The inverse derivative of gray difference with respect to pH across the pH range from 5.5 to 10.}
     \label{deriviative}
   \hfill
    \end{figure}

Notably, both methods exhibit high precision within the pH range of 7–8, characterized by low sensitivity to measurement errors. This enhanced performance is due to the sharp color change of pH indicator in this pH range. 
However, typical numerical values of $|d(\text{gray difference}) / d(\text{pH})|^{-1}$ are locally larger compared to $\left| d(\text{Hue})/d(\text{pH}) \right|^{-1}$, resulting in the pH determinations interpolated by the grayscale difference method being less precise than those determined by Hue.  
%However, the grayscale difference method is consistently more sensitive compared to the Hue method across the entire range, resulting in the pH determinations interpolated by the grayscale difference method being less precise than those determined by Hue. 

In the angular technique, there were two parameters $\mathbf{(\phi,\theta)}$ to map pH levels. The process is more mathematically demanding compared with the two methods above. There could be various interpolation techniques. In this study, projection interpolation was used, as explained below \cite{birgisson2025mapping}. 

\begin{figure}[H]
    \centering
    % First figure
    \begin{subfigure}[t]{0.48\textwidth}
        \includegraphics[width=\textwidth]{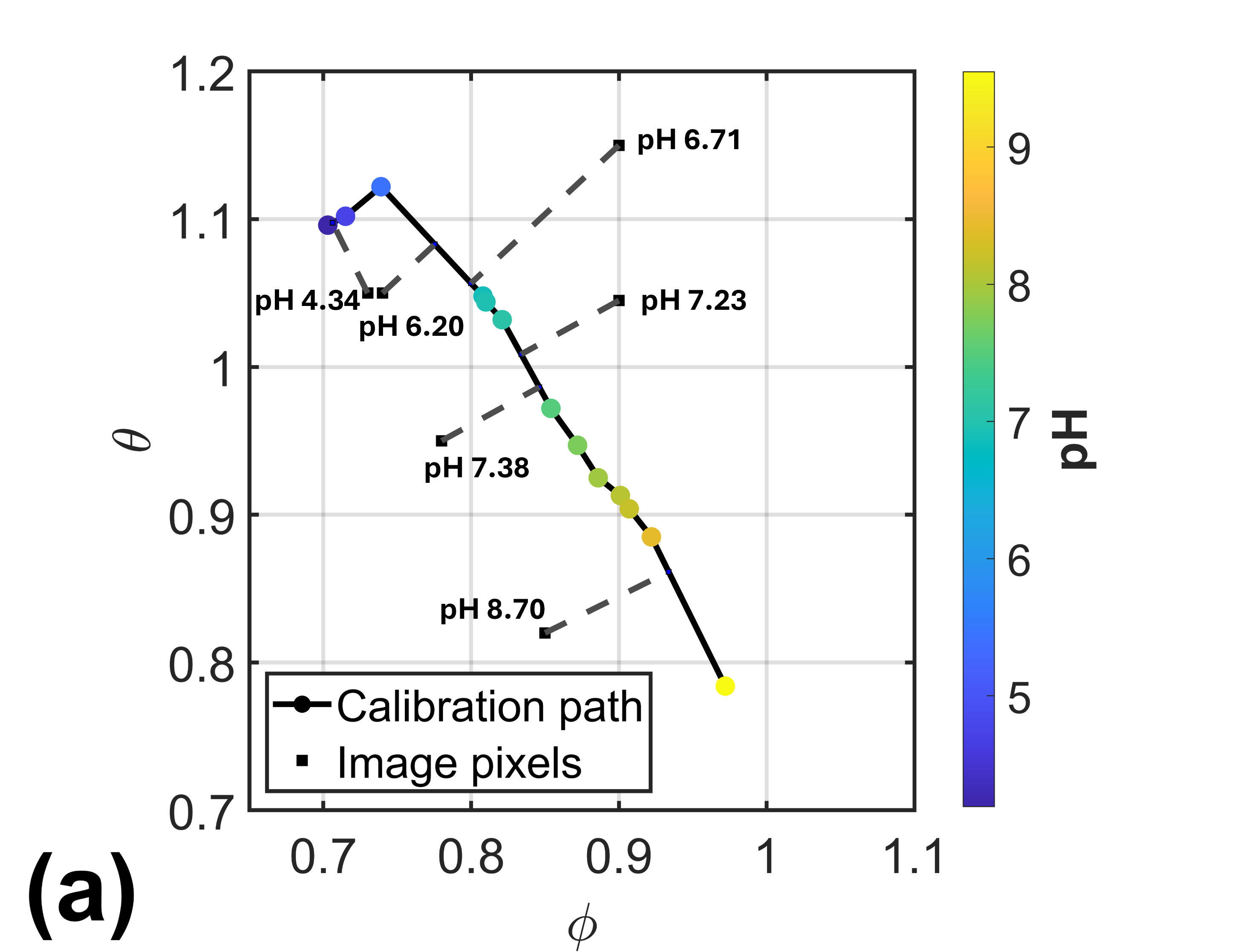}
        %\caption{First}
        \label{fig:first}
    \end{subfigure}
    \hfill
    % Second figure
    \begin{subfigure}[t]{0.48\textwidth}
        \includegraphics[width=\textwidth]{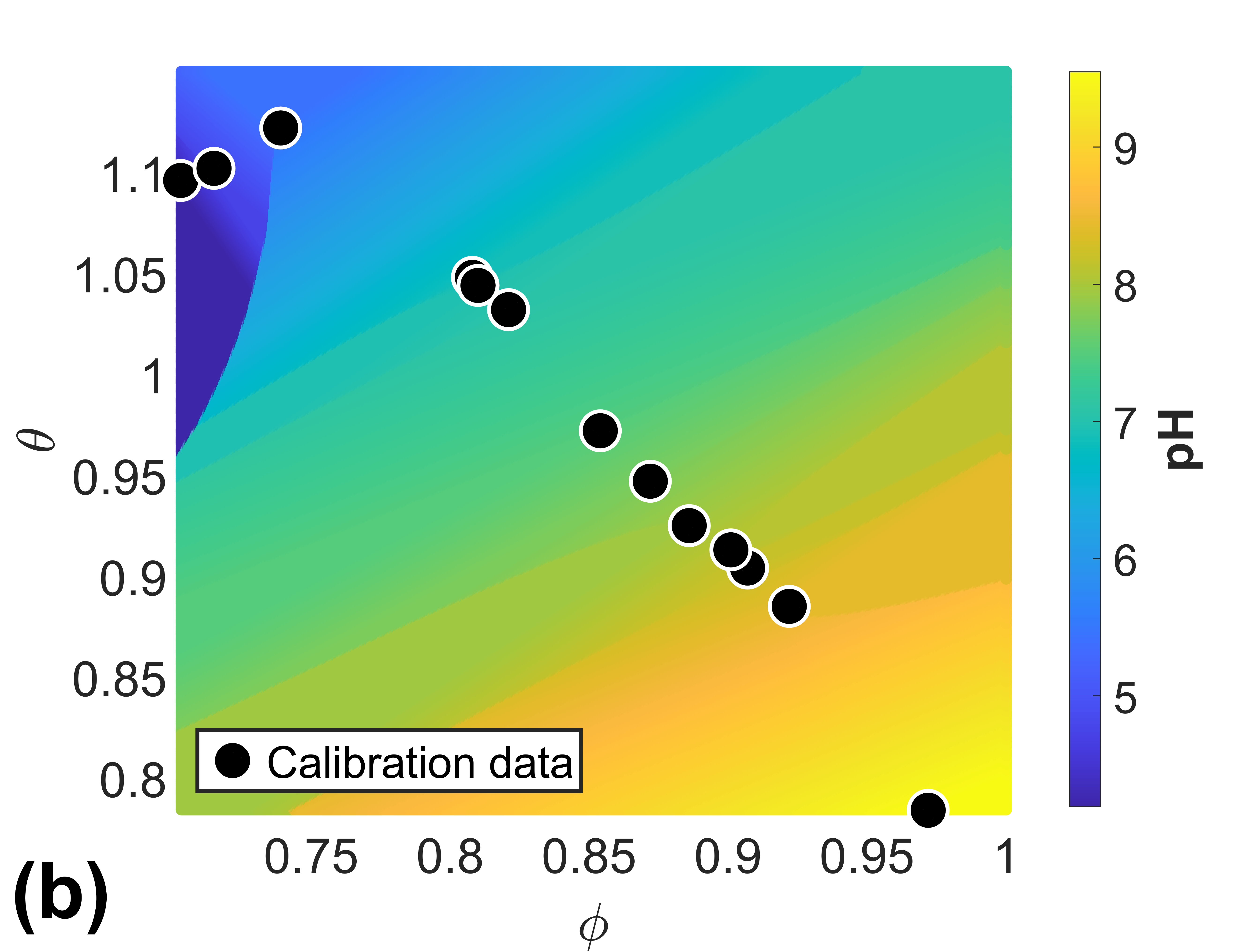}
        %\caption{Second}
    \end{subfigure}   
    \caption{The projection method of interpolating pH from $\phi$ and $\theta$. (a) Examples of interpreting image pixels from $\phi$ and $\theta$ into pH levels. A pixel was projected onto the calibration path using its shortest distance to the path. The pH was linearly interpolated by the two neighboring calibrated points of the projection point. (b) The pH map, ranging from $\phi$ $\in$ [0.6, 1] and $\theta$ $\in$ [0.7, 1.15], by the projection method. The map contains 1000*1000 linearly spaced points.}
     \label{projection method}
   \hfill
    \end{figure}

To deduce the pH of any unknown pixel, one then finds the shortest distance from it to this calibrated path in the $\mathbf{(\phi,\theta)}$ coordinate. The shortest distance from any point to the calibration path can be found analytically as shown in Figure \ref{projection method}(a), and said point is then assigned to the line segment to which the distance is shortest. On this line segment, the distance between the two neighboring calibration colors can be used to interpolate the pH value, which will then be assigned to the pixel in question.

A pH map generated using a 1000×1000-pixel interpolation is shown in Figure \ref{projection method}(b). In a previous study, the angular method demonstrated relatively good accuracy for pH measurement. However, its non-monotonic calibration path can lead to erroneous interpolation. For example, two pixels with nearly identical values of $\phi$ and $\theta$—(0.73, 1.05) and (0.74, 1.05), as shown in Figure \ref{projection method}(a)—are assigned markedly different pH values by this projection method: pH 4.34 and pH 6.20, respectively. This is also verified by the pH map shown in Figure \ref {projection method}(b). There is a pH jump near the third point of the calibration data. %It could arise from non-monotonic calibration data, caused by minor color changes of the pH indicators across pH 4-6.  

\subsection{Error analysis} \label{Error}% is possible to move to the supplementary material

Many factors can cause errors in pH determination. For example, the intensity of LED illumination can be nonuniform in space, and the milled PMMA model used for the convection experiment can present imperfections that affect its transparency, such as scratches and stains. 
%is not fully transparent due to stains from previous experiments or scratches on the model. %Those effects could lead to errors at different pH ranges.

\begin{figure}[H]
    \centering
    \includegraphics[width=1\linewidth]{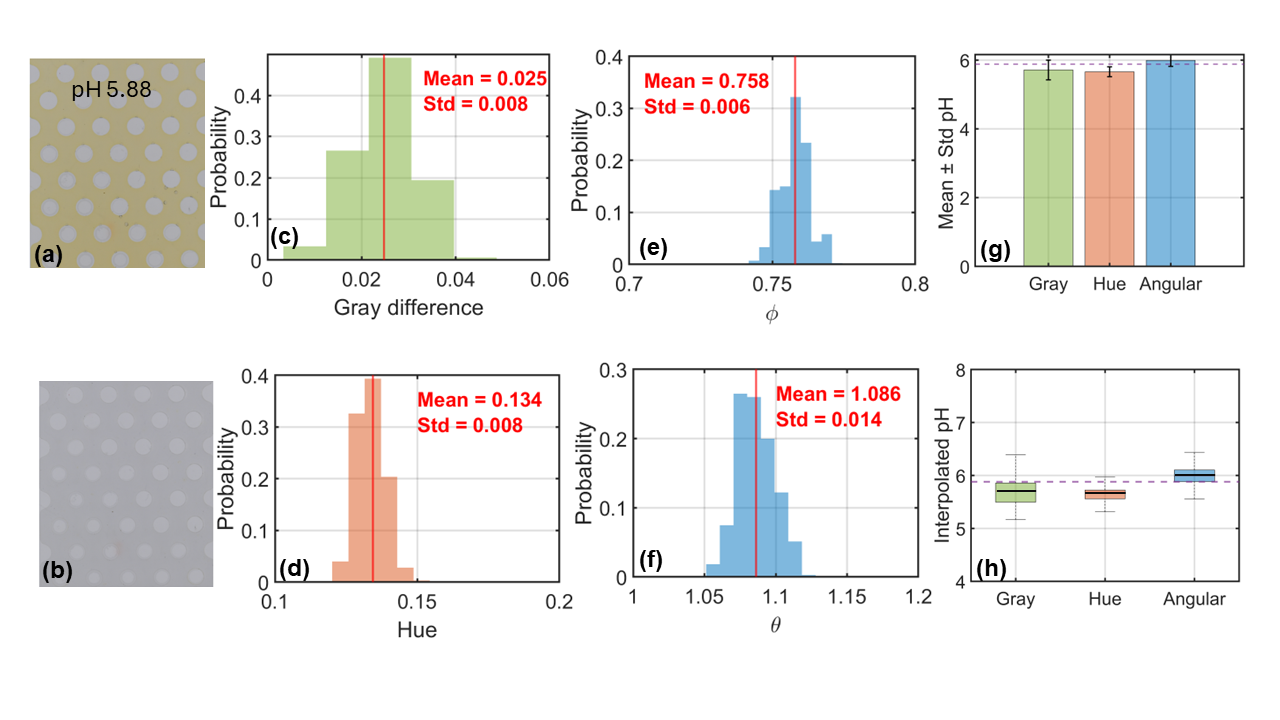}
    \caption{Error analysis of the fluid with pH 5.88 $\pm$ 0.09 in the PMMA model. The red vertical line shown in each histogram denotes the mean. The mean value and its corresponding standard deviation are additionally reported in the figure. (a) The model containing the fluid; the circular features represent particles (non-pore space), while the remaining regions correspond to pore space. The measured pH by the pH meter is additionally reported in the subplot. (b) The empty model. (c)(d)(e)(f): Distributions of pixel-color properties within the pore space, including gray-value difference, Hue, $\phi$, and $\theta$. Hue, $\phi$, and $\theta$ are evaluated from image (a), whereas the gray-value difference is obtained from the pixel-wise gray-level difference between images (b) and (a). (g): The mean pH and its standard deviation for pixels in the pore space obtained using the three techniques. The horizontal dashed line is the measured pH (5.88) by the pH meter. (h): The box plot of the interpolated pH where the line inside the box represents the median, the lower boundary of the box represents the 25th percentile, and the upper part of the box is the 75th percentile. The whiskers extend to the most extreme data points not considered outliers.}% The whiskers are drawn within 1.5 times of the distance between the upper and lower quartiles.}
    \label{E:5.860}
\end{figure}

\begin{figure}[H]
    \centering
    \includegraphics[width=1\linewidth]{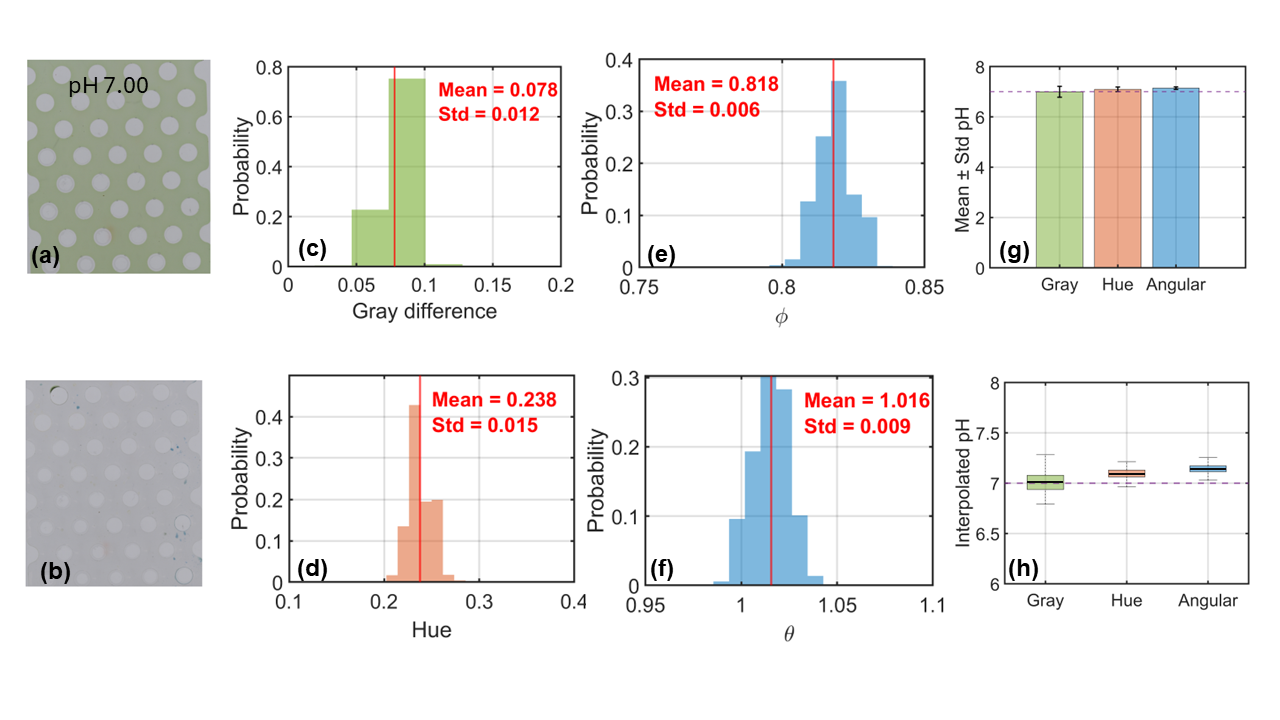}
    \caption{Error analysis of the fluid with pH 7.00  $\pm$ 0.03 in the PMMA model.The red vertical line shown in each histogram denotes the mean. The mean value and its corresponding standard deviation are additionally reported in the figure. (a) The model containing the fluid; the circular features represent particles (non-pore space), while the remaining regions correspond to pore space. The measured pH by the pH meter is additionally reported in the subplot. (b) The empty model. (c)(d)(e)(f): Distributions of pixel-color properties within the pore space, including gray-value difference, Hue, $\phi$, and $\theta$. Hue, $\phi$, and $\theta$ are evaluated from image (a), whereas the gray-value difference is obtained from the pixel-wise gray-level difference between images (b) and (a). (g):The mean pH and its standard deviation for pixels in the pore space obtained using the three techniques. The horizontal dashed line is the measured pH (7.00) by the pH meter. (h): The box plot of the interpolated pH where the line inside the box represents the median, the lower boundary of the box represents the 25th percentile, and the upper part of the box is the 75th percentile. The whiskers extend to the most extreme data points not considered outliers.} %The whiskers are drawn within 1.5 times of the distance between the upper and lower quartiles.}
    \label{E:7.010}
\end{figure}

\begin{figure}[H]
    \centering
    \includegraphics[width=1\linewidth]{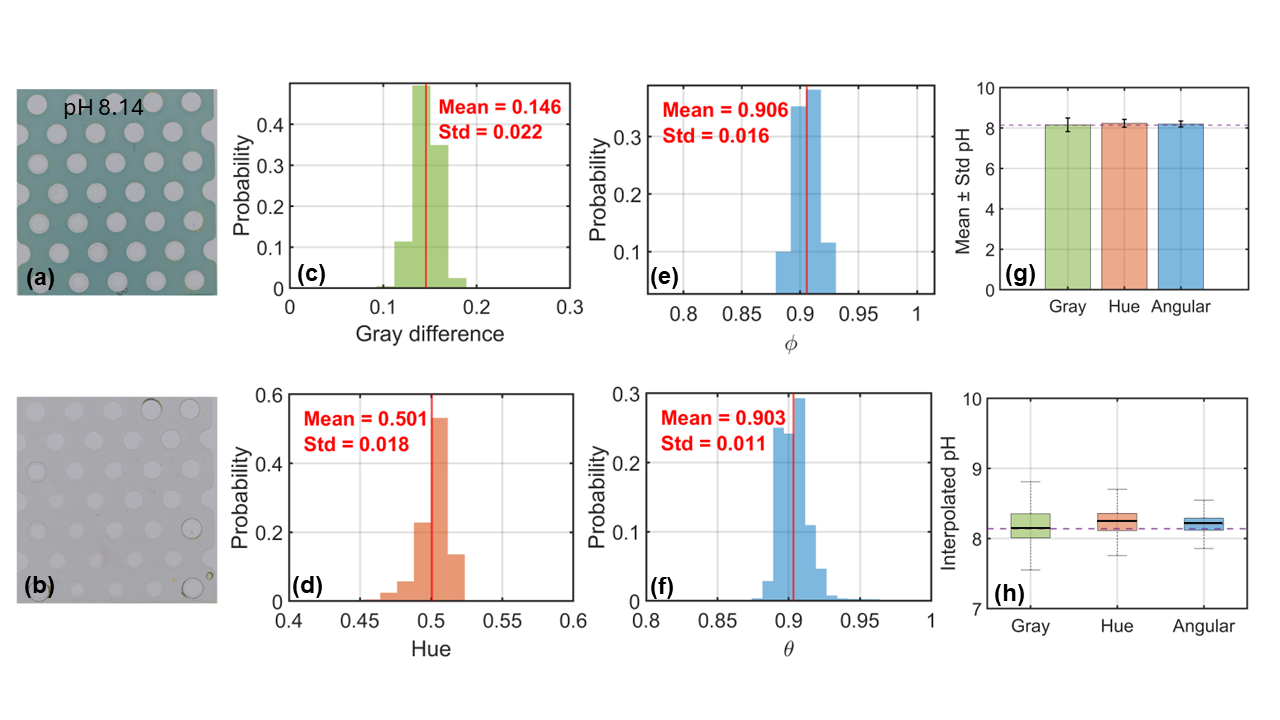}
    \caption{Error analysis of the fluid with pH 8.14 $\pm$ 0.021 in the PMMA model. The red vertical line shown in each histogram denotes the mean. The mean value and its corresponding standard deviation are additionally reported in the figure. (a) The model containing the fluid; the circular features represent particles (non-pore space), while the remaining regions correspond to pore space. The measured pH by the pH meter is additionally reported in the subplot. (b) The empty model. (c)(d)(e)(f): Distributions of pixel-color properties within the pore space, including gray-value difference, Hue, $\phi$, and $\theta$. Hue, $\phi$, and $\theta$ are evaluated from image (a), whereas the gray-value difference is obtained from the pixel-wise gray-level difference between images (b) and (a). (g): The mean pH and its standard deviation for pixels in the pore space obtained using the three techniques. The horizontal dashed line is the measured pH (8.14) by the pH meter. (h): The box plot of the interpolated pH where the line inside the box represents the median, the lower boundary of the box represents the 25th percentile, and the upper part of the box is the 75th percentile. The whiskers extend to the most extreme data points not considered outliers.} %The whiskers are drawn within 1.5 times of the distance between the upper and lower quartiles.}
    \label{E:8.154}
\end{figure}

In this section, three fluids with different pH levels were used to investigate errors in pH determination.  The pH of the fluids was measured by the pH meter (VWR pHenomenal
pH 1100) three times, and the average value was taken as the final benchmark value. Subsequently, the clean, dry, and empty PMMA model was placed on the LED light box, and a photo was taken as the background, as shown in Figure \ref {E:5.860}(a). This photo was used to compute the gray-difference in pixels.  
Afterward, a fluid was injected into the model without moving the PMMA model, and another photo was taken as shown in Figure \ref{E:5.860}(b). The pixel colors in the pore space were analyzed by the three techniques, and their histogram plots are shown in Figure \ref {E:5.860}(c)(d)(e)(f). The pixels' pH were interpolated by the curves from Section \ref{interpolation}, and the mean pH and its standard deviation was shown in Figure \ref{E:5.860}(g). The box plot showing the pH's distribution was displayed in Figure \ref{E:5.860}(h). 
The procedure was repeated for other fluids, and the results were shown in Figure \ref{E:7.010} and \ref{E:8.154}.

%The light is not homogeneous
%The stain of the PMMA model is not totally transparent, which may cause the error
%Why three models: pH may be different in different pH.

\begin{table}[H]
\centering
\begin{tabular}{l|l|l|l|l|l|l|l|l}
\hline
    Fluid No.    & pH & $\phi$     & $\theta$   & pH\_ Angular & Hue         & pH\_Hue     & Gray difference & pH\_Gray    \\ \hline
 1 & 5.88 ($\pm$ 0.09)       & 0.758 ($\pm$ 0.006) & 1.086 ($\pm$0.014) & 5.99 ($\pm$0.17)  & 0.134($\pm$0.008)  & 5.66 ($\pm$0.14) & 0.025 ($\pm$0.008)   & 5.71($\pm$0.28)  \\ \hline
 2 & 7.00 ($\pm$ 0.03)        & 0.818 ($\pm$0.006) & 1.016 ($\pm$0.009) & 7.14 ($\pm$0.05)  & 0.238 ($\pm$0.015) & 7.10 ($\pm$0.09) & 0.078 ($\pm$0.012)   & 6.98($\pm$0.28)  \\ \hline
 3 & 8.143($\pm$ 0.021)       & 0.906 ($\pm$0.016) & 0.903 ($\pm$0.011) & 8.20 ($\pm$0.15)  & 0.501($\pm$0.018) & 8.23 ($\pm$0.20) & 0.146 ($\pm$0.022)   & 8.2 ($\pm$0.3) \\ \hline
\end{tabular}
\caption{The summary of error analysis for three solutions with various pH. Each column represents the average value, and the number in parentheses represents its corresponding standard deviation. The second column lists the pH measured by the pH meter, while the remaining columns present the pH values interpreted by the respective techniques, as indicated by their subscripts.}
\label{E:table}
\end{table}

\begin{figure} [H]
    \centering
    \includegraphics[width=0.8\linewidth]{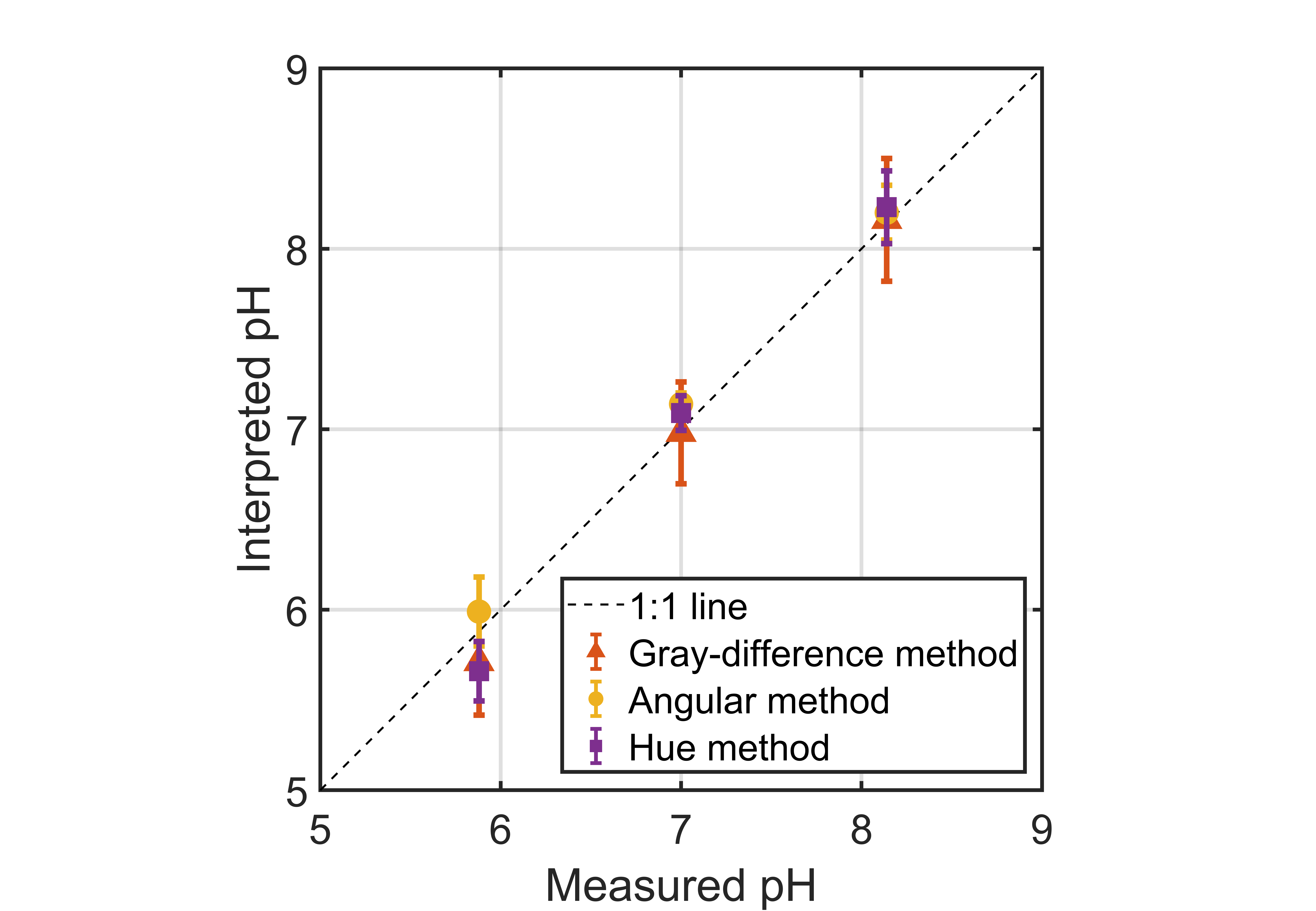}
    \caption{The comparison of the measured pH and interpreted pH by the three techniques.}
    \label{E:plot_11}
\end{figure}

Overall, the pH values derived from all three techniques agree well with the true values measured by the pH meter across the three pH levels, as shown in Figure \ref{E:plot_11}. The angular method is the least sensitive to image noise because it incorporates two parameters rather than interpolating pH from a single variable. However, this dual-parameter approach is mathematically more involved and computationally slower. In contrast, the gray-value method exhibits the largest error (standard deviation) but provides a faster pH estimate than the angular technique. The Hue method combines the above advantages, providing both good accuracy and computational efficiency.

% should have had a pH-pH measured map. 

\subsection{Qualitative Visualization}

\begin{figure}[H]
    \centering
    \begin{subfigure}[t]{0.25\textwidth}
        \includegraphics[width=\textwidth]{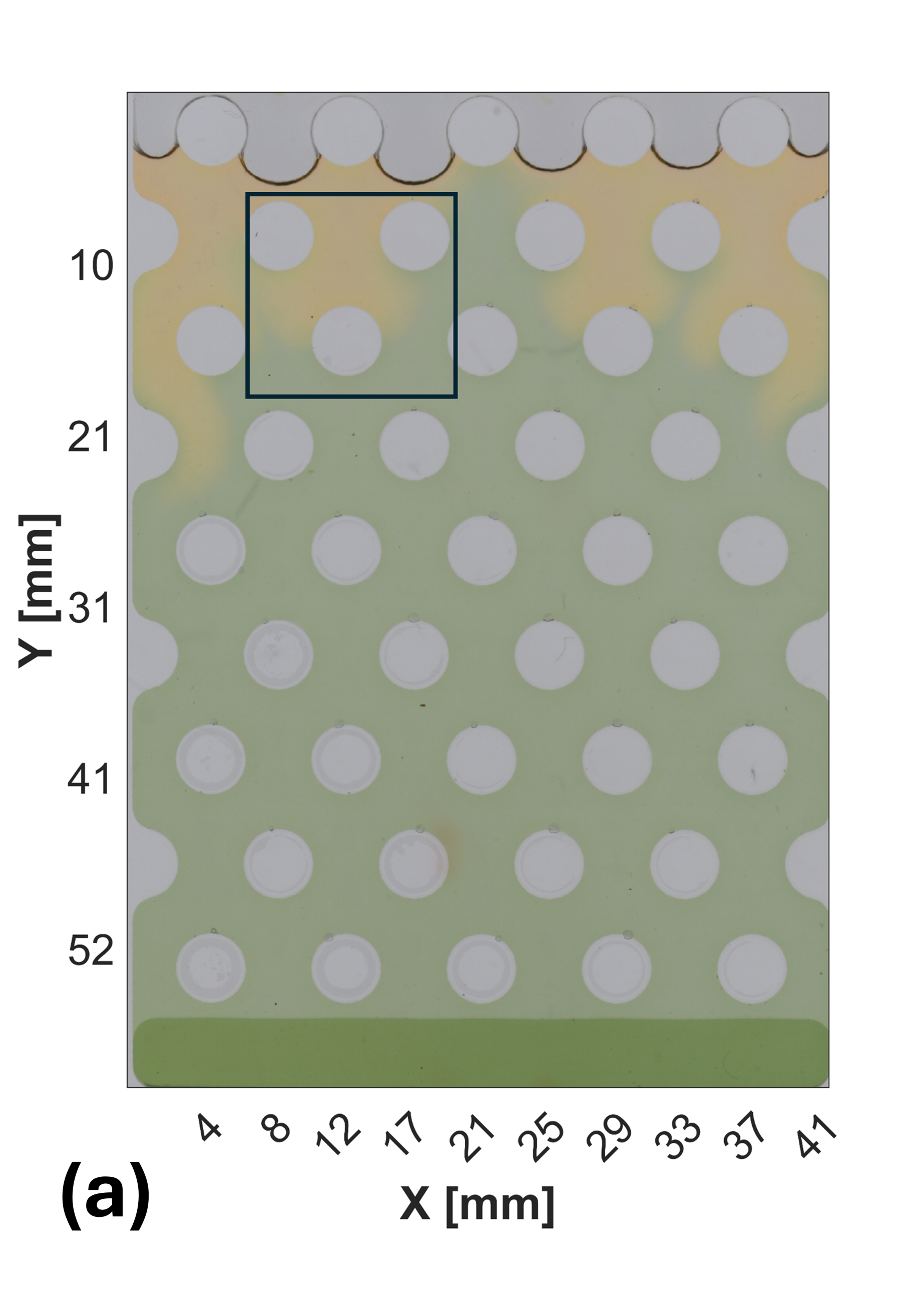}%{plots/raw_convection3.png}
        %\caption{First}
        \label{fig:first}
    \end{subfigure}
    % Second figure
    \begin{subfigure}[t]{0.3\textwidth}
        \includegraphics[width=\textwidth]{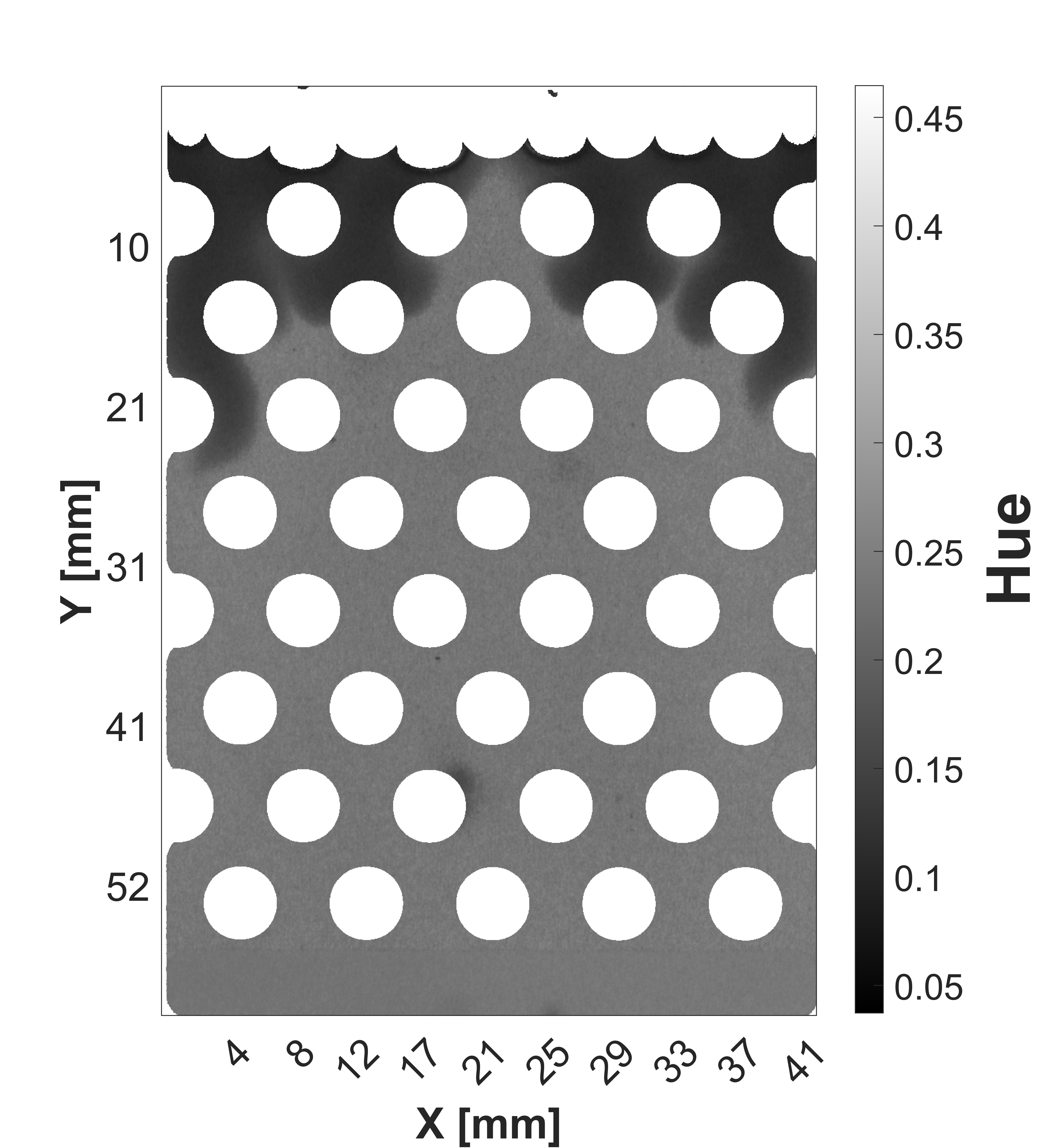}%{plots/hue_convection3.png}
        %\caption{Second}
    \end{subfigure}  
   \begin{subfigure}[t]{0.29\textwidth}
        \includegraphics[width=\textwidth]{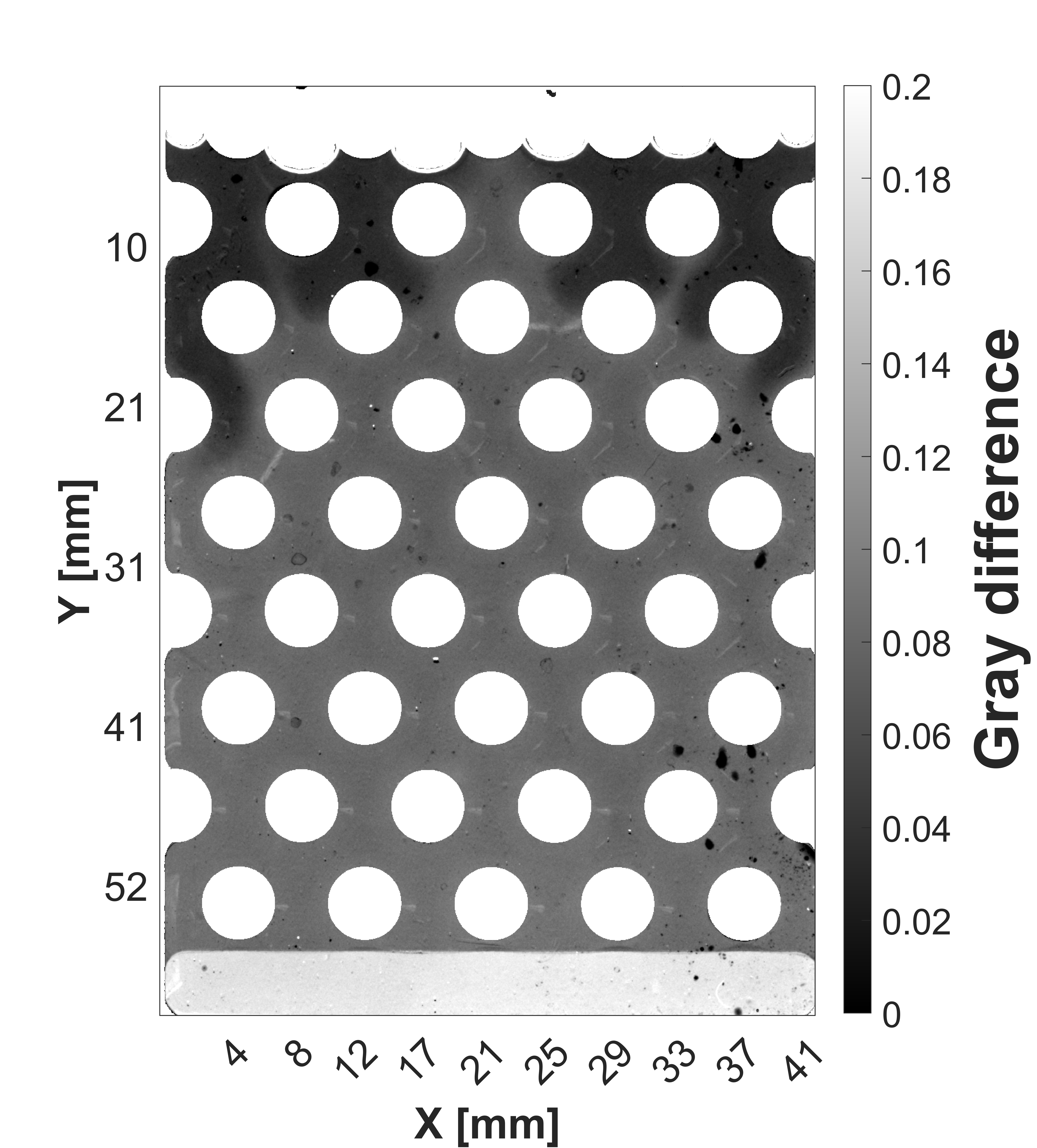}%{plots/gray_convection.png}
        %\caption{Second}
    \end{subfigure}   \\
    \begin{subfigure}[t]{0.3\textwidth}
        \includegraphics[width=\textwidth]{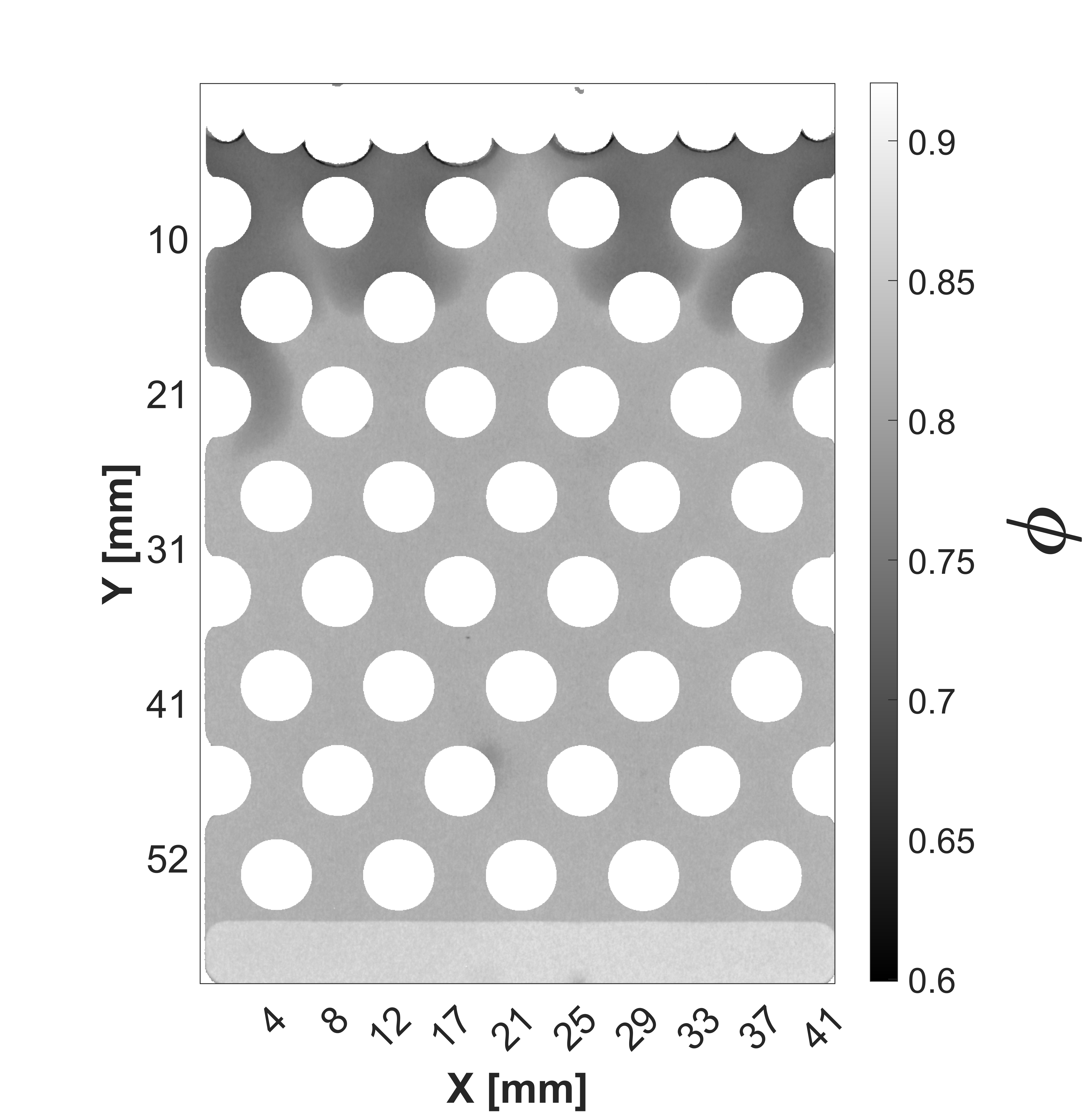}%{plots/phi_convection.png}
        %\caption{Second}
    \end{subfigure}   
   \begin{subfigure}[t]{0.3\textwidth}
        \includegraphics[width=\textwidth]{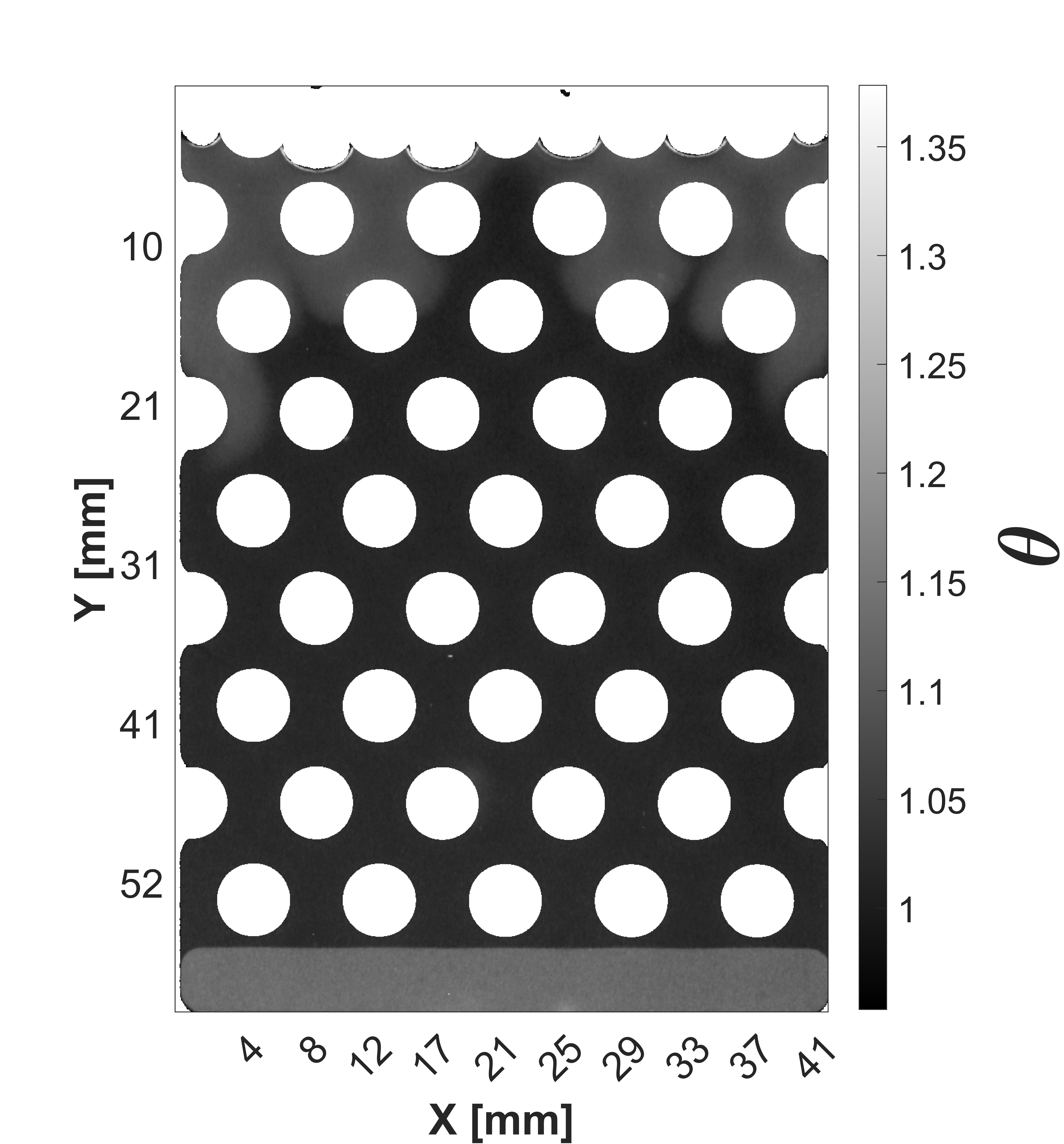}%{plots/theta_convection.png}
        %\caption{Second}
    \end{subfigure}   
     \caption{The qualitative visualization of the carbon convection experiment in the porous medium made from PMMA. They showed the carbon plumes after 10 minutes of injection. (a)(b)(c)(d)(e): The raw image, and the images of Hue, Gray-difference, $\phi$, and $\theta$, respectively. The pixels in the square marked in (a) will be further analyzed and presented in Figure.\ref{single plume}.}
     \label{quanlitative_visulization}
    \end{figure}

Following the analysis above, the established techniques were applied to a carbon dioxide ($\text{CO}_2$) injection experiment conducted within the milled PMMA model as seen in Figure \ref{setup-sketch}.
The raw image of carbon convection and its corresponding maps of Hue, gray difference value, and  $\mathbf{\phi}$, $\mathbf{\theta}$ were shown in Figure \ref{quanlitative_visulization}.

The porous medium fabricated from PMMA materials enable clear visualization of carbon convection, from which several qualitative features can be observed, such as finger morphologies and preferential convection pathways, as shown in Figure \ref{quanlitative_visulization}. For instance, an intensity gradient is visible across each finger, reflecting variations in carbon concentration: concentrations are typically higher at the core of a finger and gradually decrease toward the edges. Beyond qualitative observations, some fundamental features can also be quantified. The onset time of instability—defined as the moment when fingers appear—can be identified, providing an important metric for comparing conditions under which convection begins to dominate over diffusion. Furthermore, other key parameters, such as the number and width of fingers, the extent of the mixing zone, and related measures, can also be extracted from the visualization for analysis of convection behavior. 

It is worth noting that the model included a channel at the bottom containing fluids with a thickness of 4 mm, compared to 2 mm in the porous medium. As shown in Figure \ref{quanlitative_visulization}, the gray, $\mathbf{(\phi,\theta)}$, and raw images all displayed noticeable color variations between fluids of different thicknesses, despite being the same fluid. In contrast, Hue remained consistently unaffected by fluid thickness, validating the calibration results presented in Section \ref{thickness_sec}.

\subsection{pH and carbon concentration}

%However, the information above is not sufficient to physically characterize the growth rate of the convection instability and to identify the dominant wavenumber in the instability. In addition, the influence of the instability on the dissolution rate remains insufficiently understood.
%However, the qualitative characterization is insufficient to illuminate the convection mechanism. As carbon convection proceeds, the spatial pH and carbon concentration in the porous media will also be affected. Those parameters are crucial for unwinding the convection mechanism. However, their determination is challenging. 
While qualitative characterization provides initial insights, it is insufficient to fully elucidate the underlying convection mechanisms. For example, as carbon convection proceeds, the spatial distribution of pH and carbon concentration within the porous medium is significantly altered. Although these parameters are critical for revealing the convective dynamics, their precise determination remains technically challenging\cite{de2021bi}. As indicated in previous literature\cite{birgisson2025mapping}, a 0.2-unit measurement error in pH can lead to a 10\% error in the calculation of total dissolved carbon. 
%To calculate those quantities, a precise measurement of the spatial-temporal pH and carbon concentration is demanded. 
The accuracy of the results is highly dependent on the experimental techniques employed. This section provides a comparative analysis of the $\text{pH}$ and concentration distributions interpolated by the Hue, gray-difference, and $(\phi,\theta)$ methods within the porous medium.

\begin{figure}[H]
    \centering
    \begin{subfigure}[t]{0.25\textwidth}
        \includegraphics[width=\textwidth]{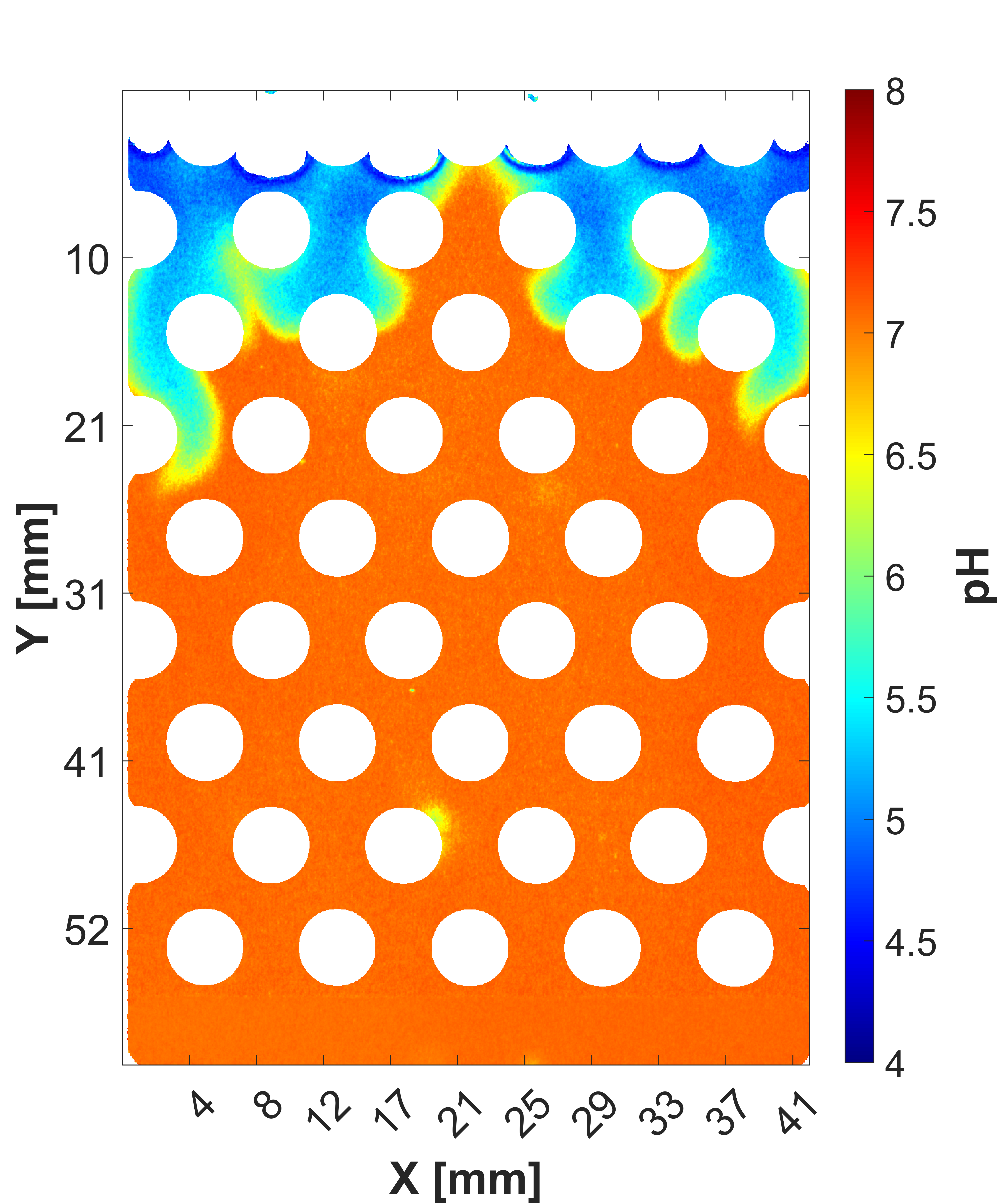}%{plots/pH_map.png}
        %\caption{First}
        \label{fig:first}
    \end{subfigure}
    % Second figure
    \begin{subfigure}[t]{0.25\textwidth}
        \includegraphics[width=\textwidth]{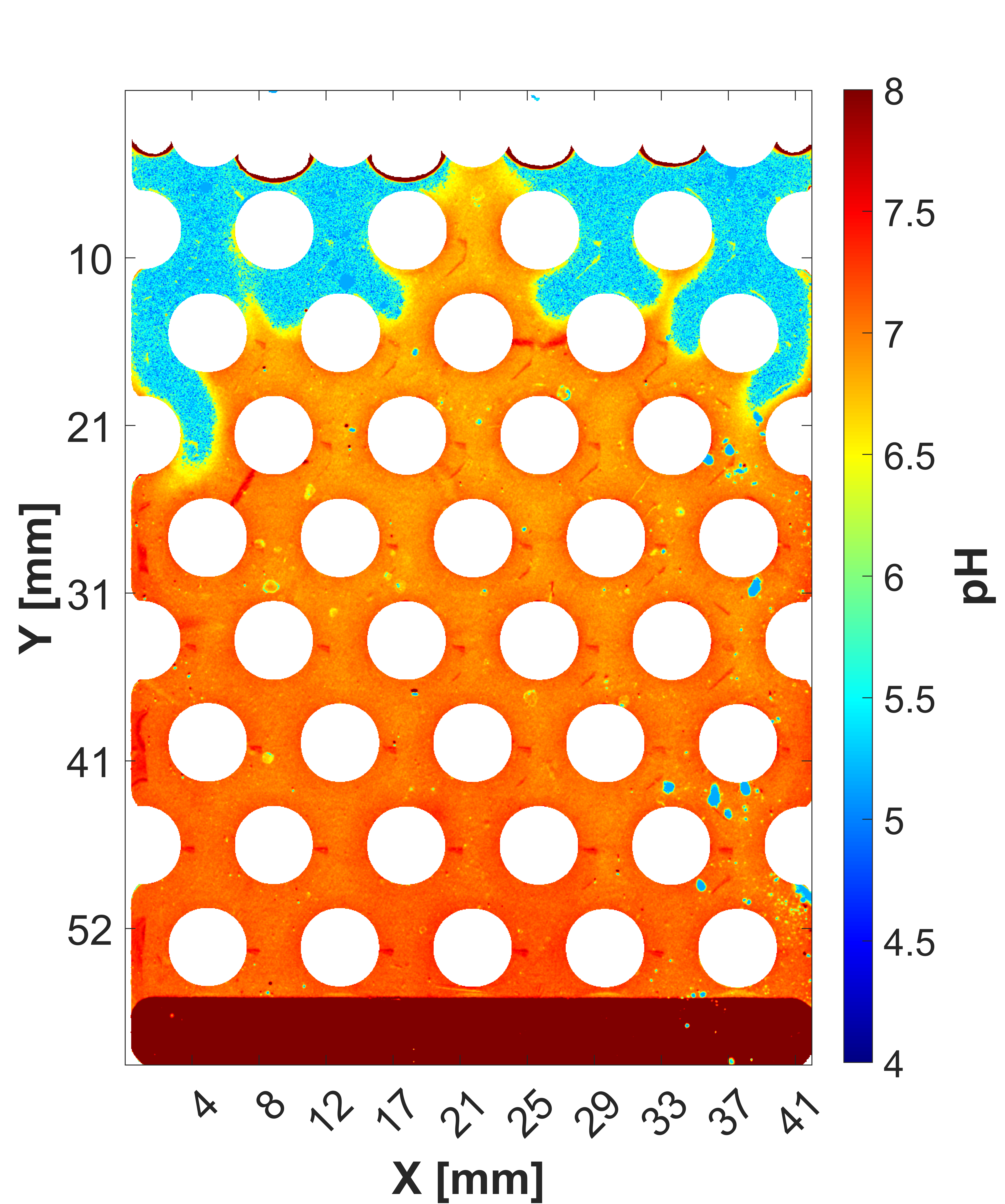}%{plots/pH_map_gray.png}
        %\caption{Second}
    \end{subfigure}   
   \begin{subfigure}[t]{0.25\textwidth}
        \includegraphics[width=\textwidth]{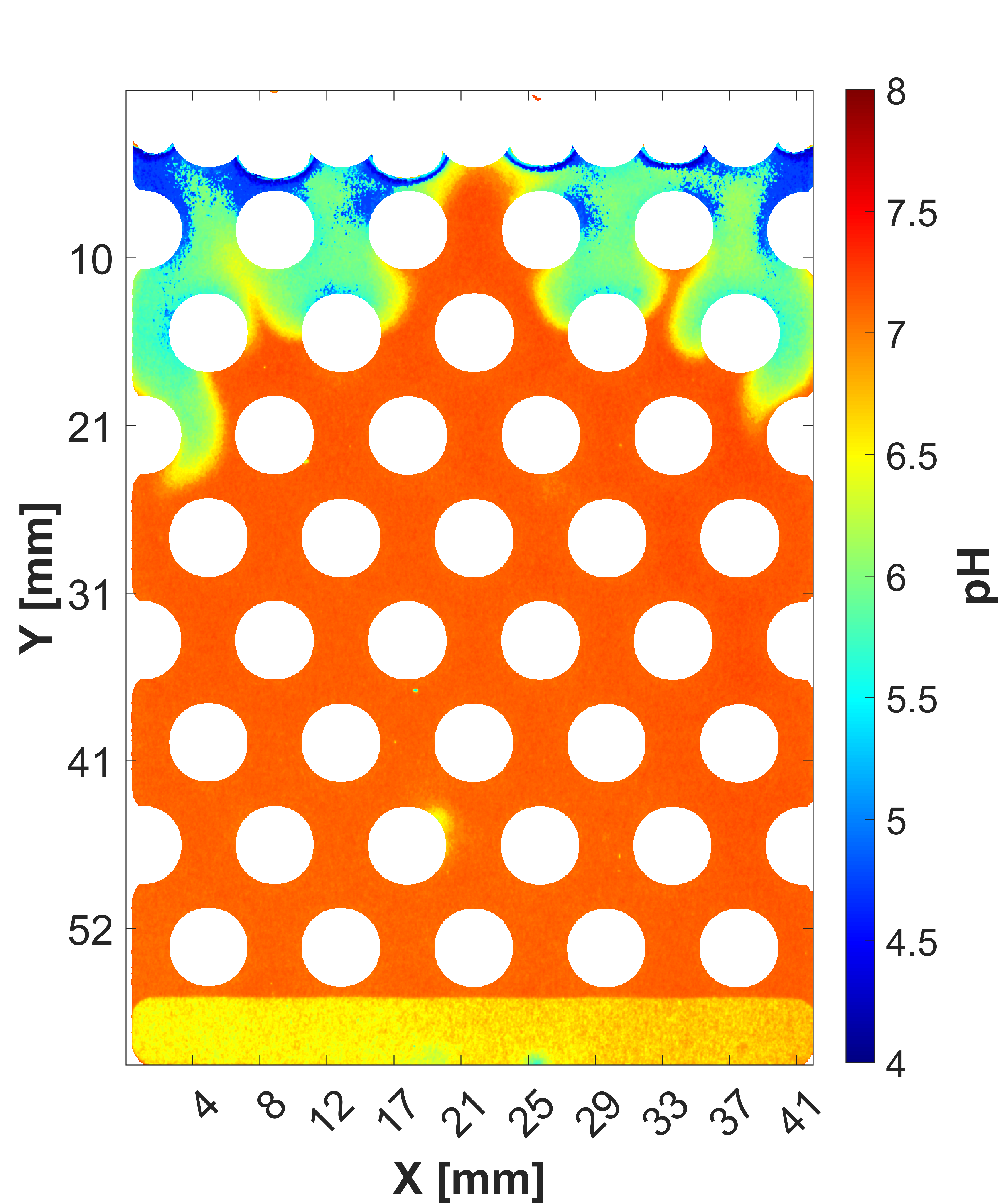}%{plots/pH_map_polar.png}
        %\caption{Second}
    \end{subfigure}   
    \caption{The spatial pH in the porous media after 10 minutes of carbon injection. From left to right are the pH map by Hue, gray difference, and $\mathbf{(\phi,\theta)}$ respectively.}
     \label{pH in porous media}
    \end{figure}

\begin{figure}[H]
    \centering
    % First figure
    \begin{subfigure}[t]{0.45\textwidth}
    \centering
        \includegraphics[width=\textwidth]{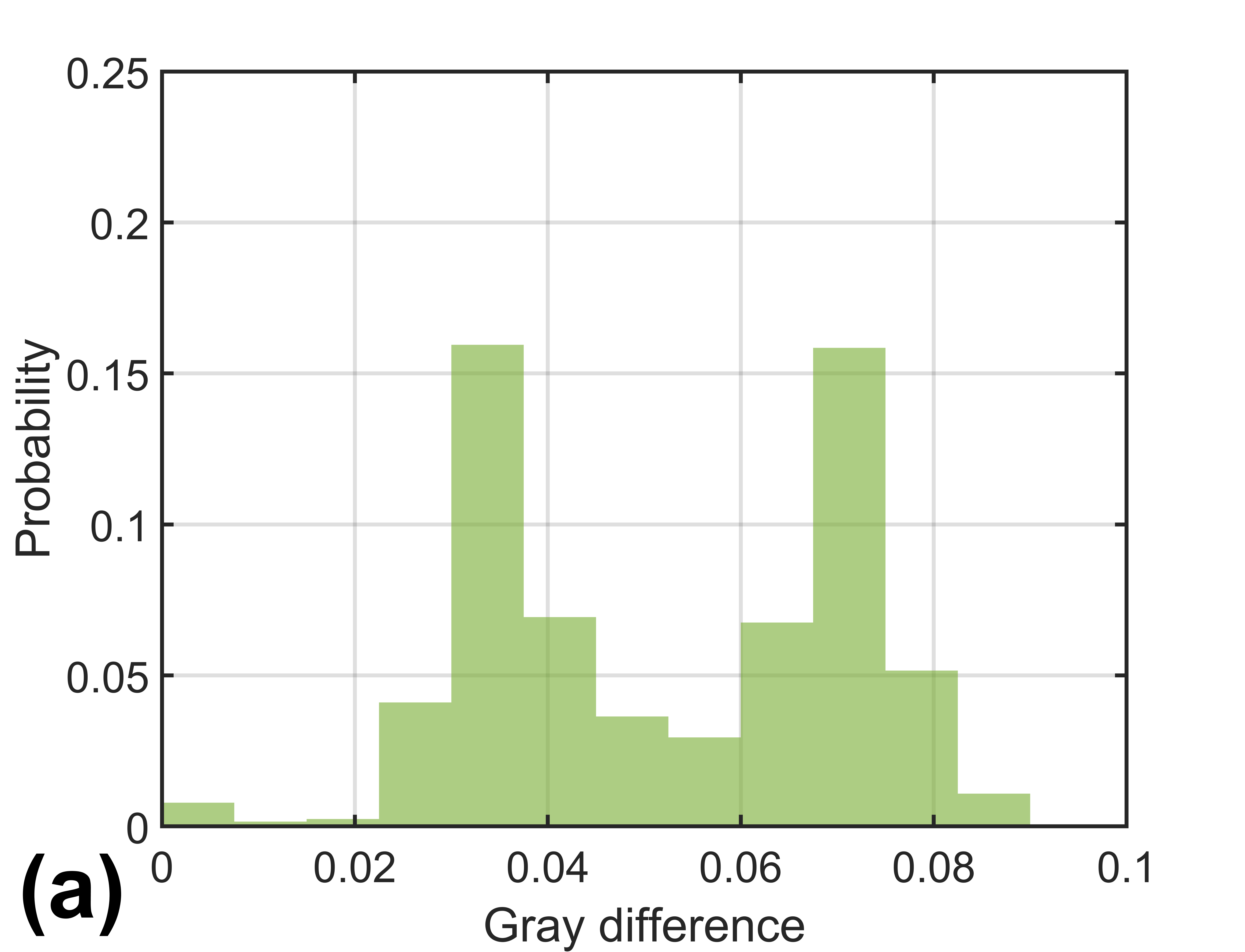}
        %\caption{First}
        \label{fig:first}
    \end{subfigure}
    \hfill
    % Second figure
    \begin{subfigure}[t]{0.45\textwidth}
    \centering
        \includegraphics[width=\textwidth]{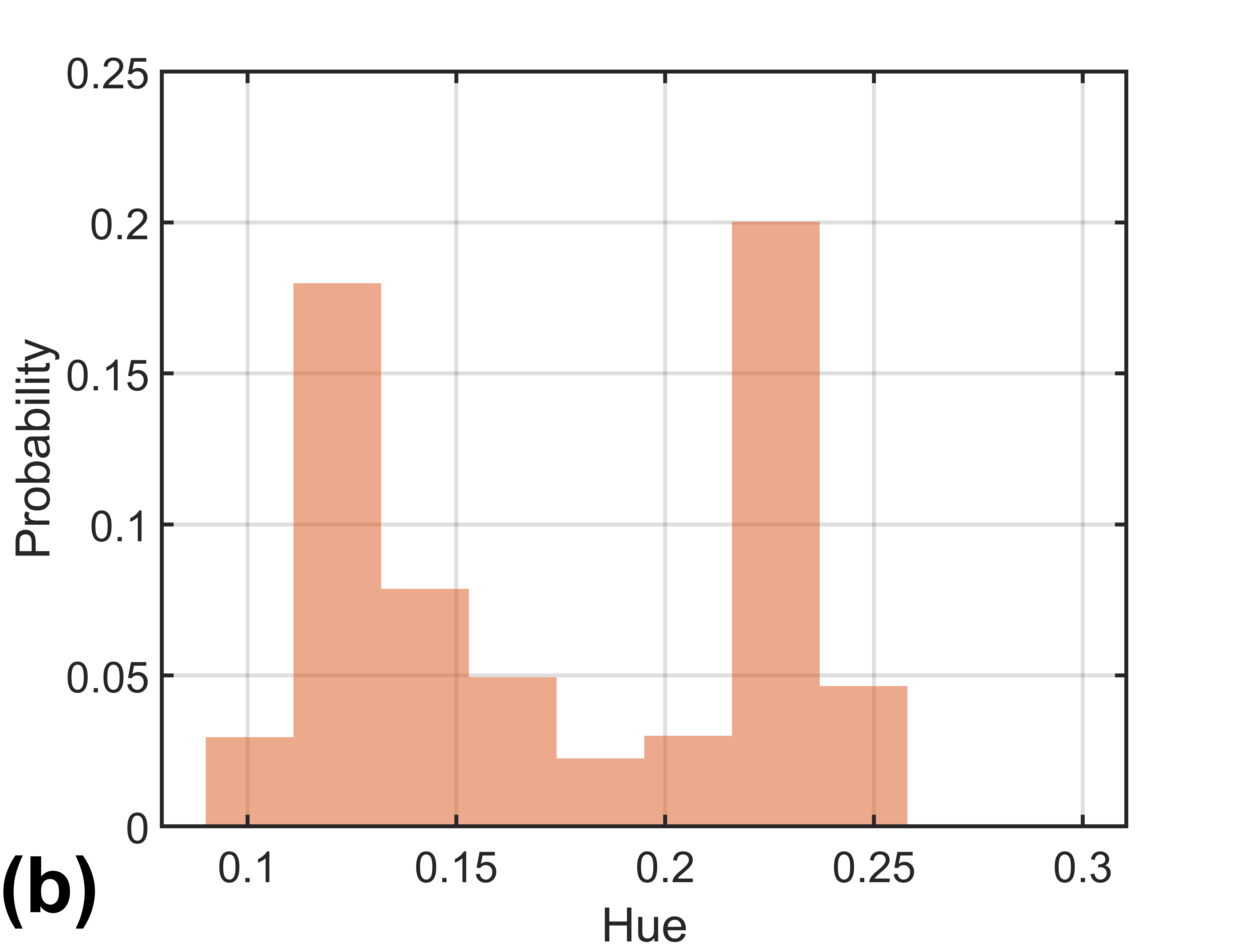}
        %\caption{Second}
    \end{subfigure} 

    \vspace{0.5cm}

 \begin{subfigure}[t]{0.45\textwidth}
 \centering
        \includegraphics[width=\textwidth]{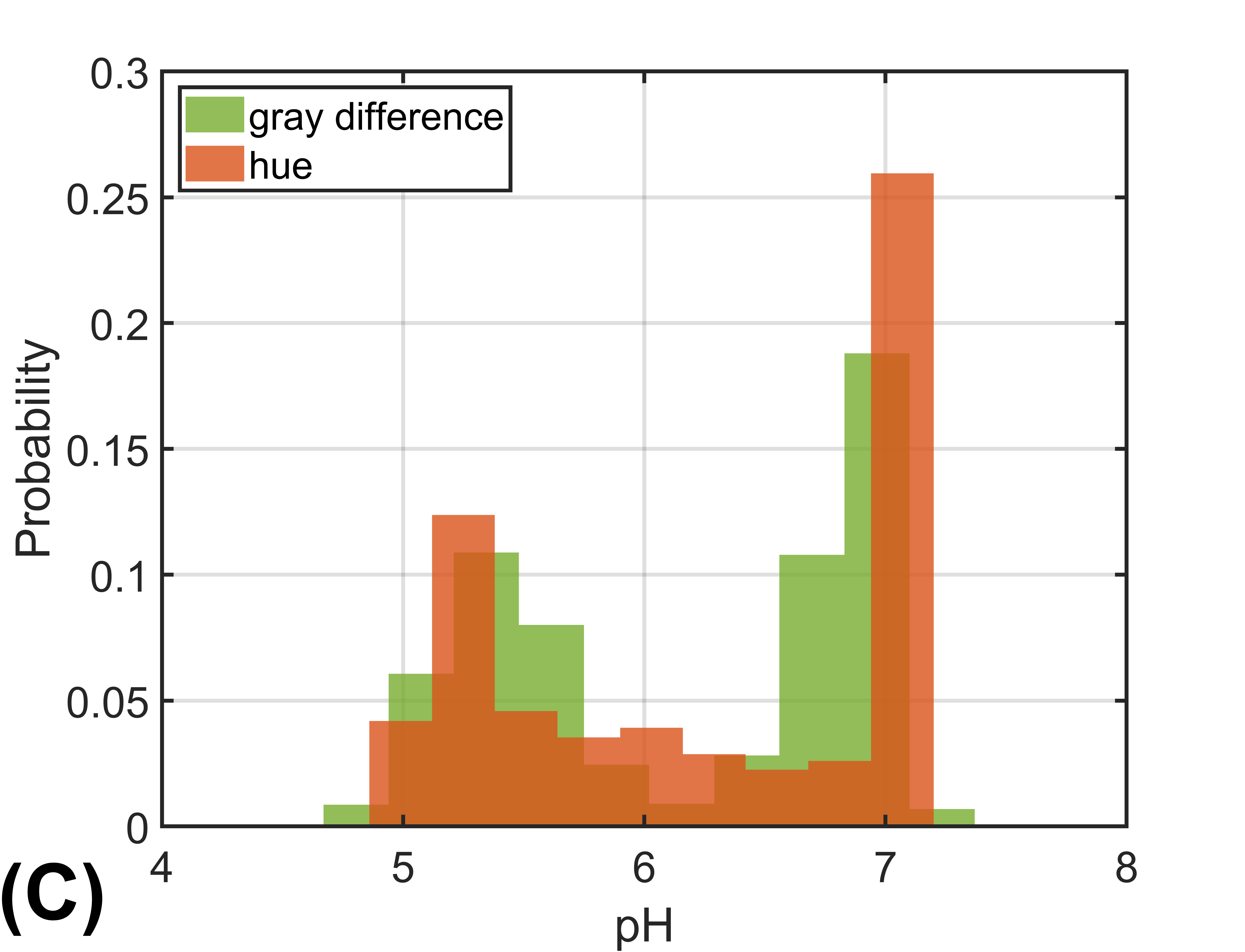}%{plots/concentration_map_polar.png}
        %\caption{Second}
    \end{subfigure}  
    \hfill
 \begin{subfigure}[t]{0.45\textwidth}
    \centering
        \includegraphics[width=\textwidth]{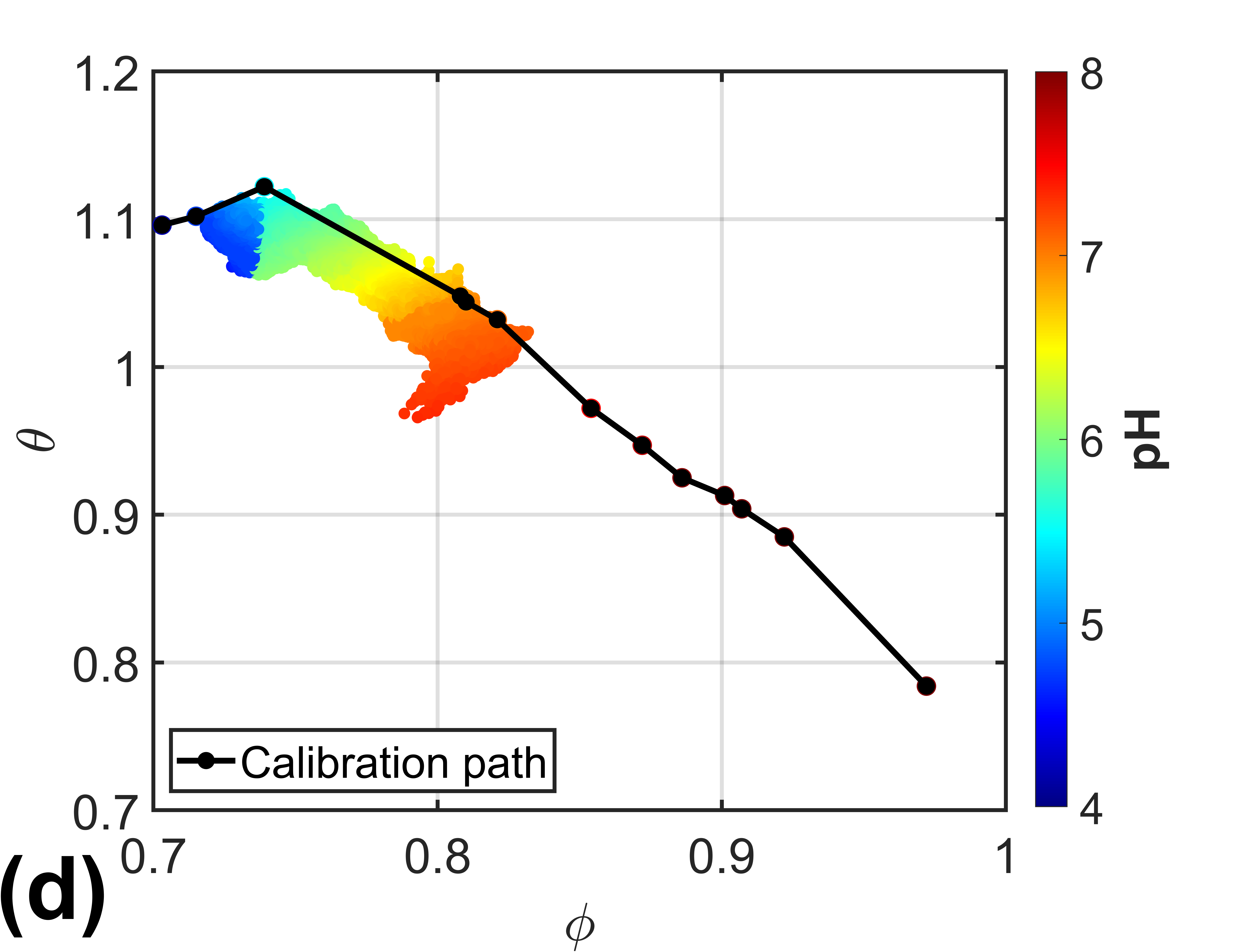}
        %\caption{Second}
    \end{subfigure}   
    \caption{(a): The histogram of the gray difference of the pixels in the square in Figure \ref{quanlitative_visulization}.(b): The histogram of the Hue of the pixels in the square. (c): The corresponding pH to the gray difference and Hue. (d): The distribution of $\mathbf{(\phi,\theta)}$ and pH of the pixels in the square.}
     \label{single plume}
   %\hfill
    \end{figure}

\begin{figure}[H]
    \centering
    % First figure
    \begin{subfigure}[t]{0.25\textwidth}
        \includegraphics[width=\textwidth]{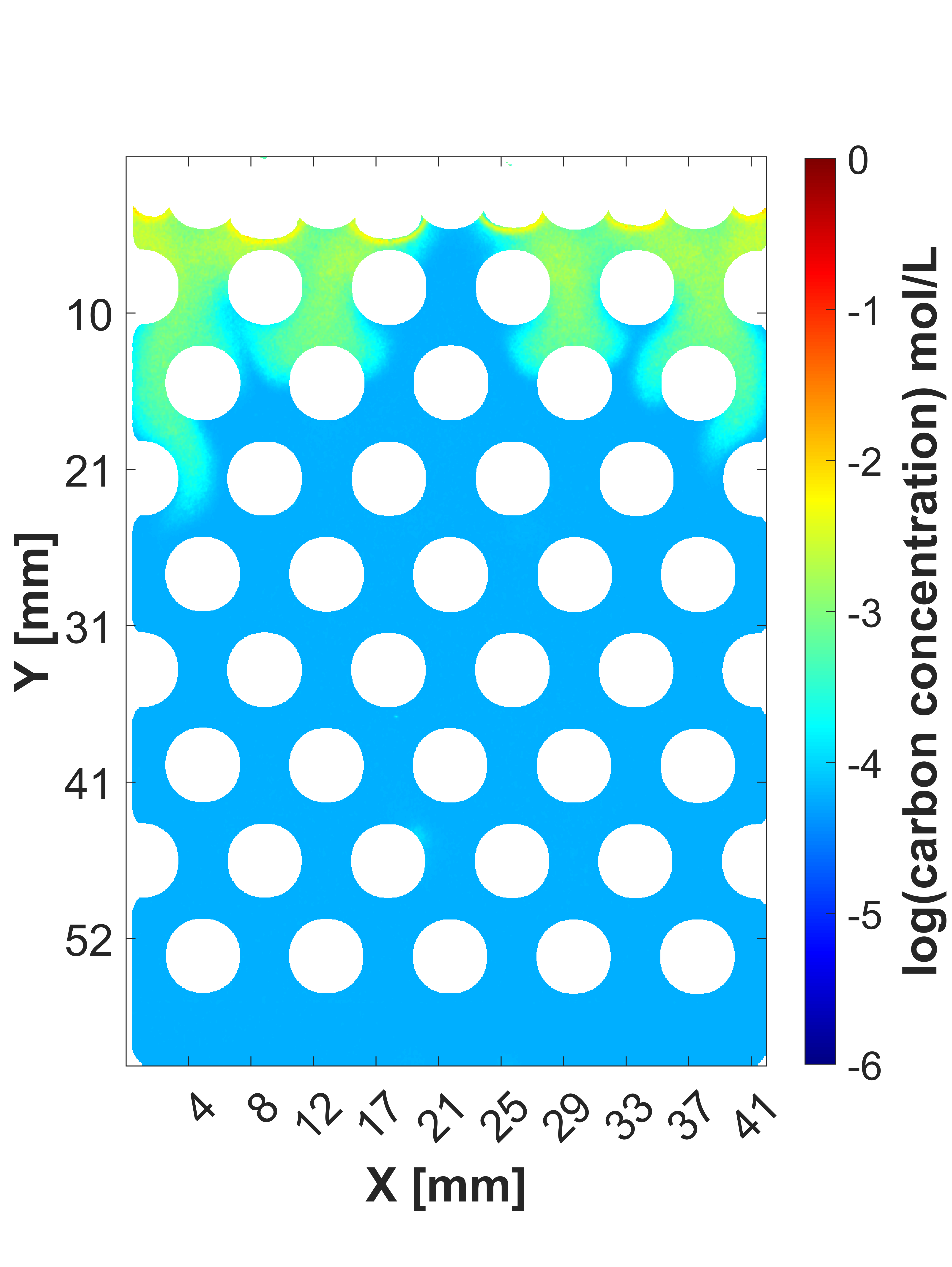}%{plots/concentration_map.png}
        %\caption{First}
        \label{fig:first}
    \end{subfigure}
    % Second figure
    \begin{subfigure}[t]{0.25\textwidth}
        \includegraphics[width=\textwidth]{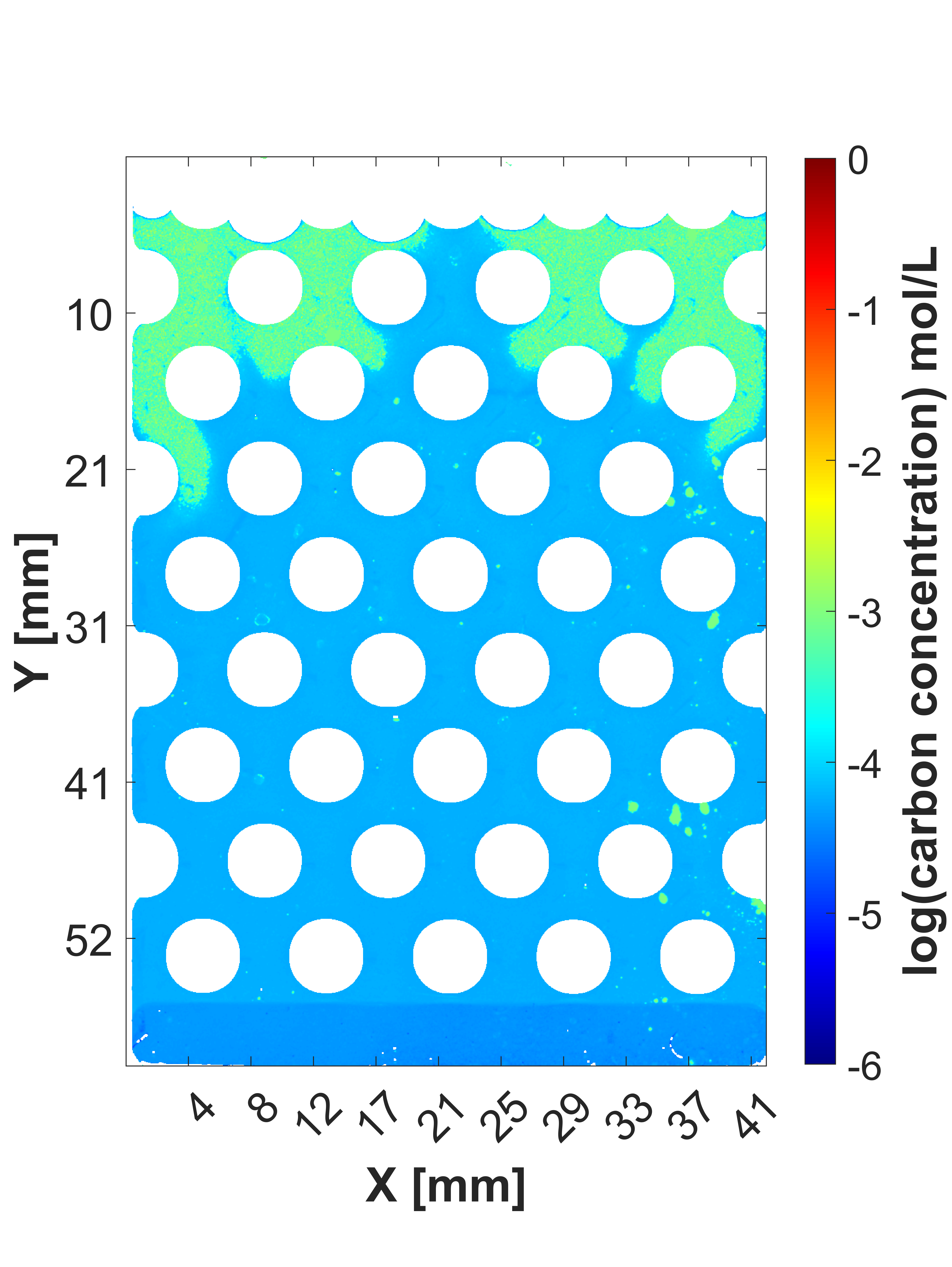}%{plots/concentration_map_gray.png}
        %\caption{Second}
    \end{subfigure}   
   \begin{subfigure}[t]{0.25\textwidth}
        \includegraphics[width=\textwidth]{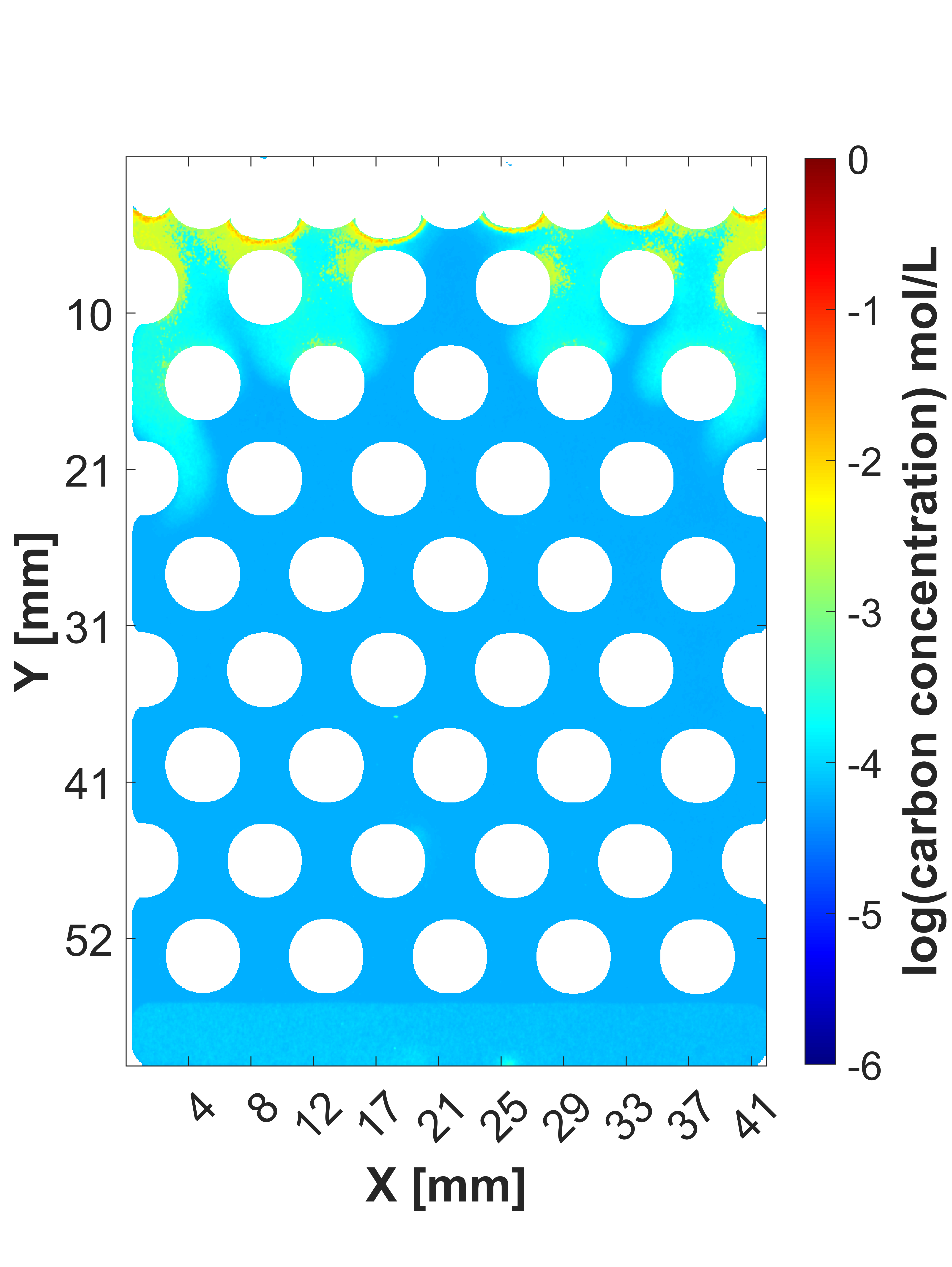}%{plots/concentration_map_polar.png}
        %\caption{Second}
    \end{subfigure}   
    \caption{The spatial concentration in the porous media after 10 minutes of carbon injection. From left to right are the carbon concentration map by Hue, gray difference, and $\mathbf{(\phi,\theta)}$ respectively.}
     \label{concentration in porous media}
    \end{figure}

The pH maps by the three methods are shown in Figure \ref {pH in porous media}. Their corresponding carbon concentration maps were computed by Equation \ref{carb-con} and were shown in Figure \ref {concentration in porous media}. %Those maps provided us with an intuitive understanding of the dynamics of pH and carbon concentration driven by density convection.
However, as previously stated, the non-monotonic calibration paths of the gray difference significantly impair the accuracy of pH interpolation within the 4-6 range. To further investigate the effect, the gray difference of one convective plume marked by the square shown in Figure \ref {quanlitative_visulization}(a) was analyzed. Since the region inevitably includes some carbon-free background fluid, the histograms in Figure \ref{single plume} have two peaks, characterizing the carbon plume and the carbon-free background fluid.  
Figure \ref {single plume}(a) demonstrates that the gray difference for most of the plume's pixels falls within this critical range, which causes errors in determining pH in the region.
%Moreover, pH interpolation is very susceptible to the shift of the gray difference, making the pH in the plume not very continuous.
Moreover, the gray-difference technique is significantly more sensitive to artifacts such as staining and background noise than the angular and Hue methods, as shown in Figures \ref {pH in porous media} and \ref{concentration in porous media}. This is likely due to its dependence on carbon-free background images, which were not used to compute Hue and $\mathbf{(\phi,\theta)}$. For example, Figure \ref{E:7.010} shows residual staining from previous experiments within the empty model. It carried over directly into the gray-difference calculation.

In addition, the pH of the plume by the angular technique was also analyzed. Figure \ref{single plume}(c) reveals a distinct pH discontinuity, or 'jump', in the vicinity of the maximum $\mathbf{(\phi,\theta)}$ of the calibration path. This sensitive zone makes determining the true pH in this range particularly challenging. However, as analyzed in the error analysis in Section \ref{Error}, the pH determinations remain highly precise, with a low standard deviation, when the measured pH is outside this sensitive zone. More details can be found in Table \ref{E:table}.
%In addition to the non-monotonic calibration path arising from the indicators' characteristics, this jump may also be caused by the interpolation method-projection. But this study doesn't focus on improving the mathematical interpolation; instead, it will mainly compare various experimental techniques. 

However, the Hue's interpolation to pH is unique and without a pH jump. The calibration map from Hue to pH is continuous and smooth. 
It is also worth noting that a critical limitation emerges when comparing the techniques' performance across varying fluid thicknesses. Both the gray-difference and angular methods produce inconsistent estimates for pH and carbon concentration in the bottom layer when fluid thickness varies, as demonstrated in Figures \ref{pH in porous media} and \ref{concentration in porous media}. In contrast, the Hue technique delivers highly consistent, thickness-independent results, a finding further demonstrated in Section \ref{thickness_sec}.

Overall, the Hue technique offers superior and more accurate insight into the carbon convection dynamics. Specifically, the pH and carbon-concentration distributions analyzed using the Hue technique exhibit clear concentration gradients across each finger, which aligns consistently with observations reported in previous studies\cite{zhang2024permeability}\cite{wang2020enhanced}\cite{islam2016reactive}\cite{rasmusson2017refractive}.

%Its pH and concentration patterns within the range are much different from those of the Hue technique. The pH and carbon concentration gradient across each finger is almost invisible.   However, the pH levels around the tips of the fingers are consistent with the other two methods.

%Although the patterns by the gray and hue methods show a large similarity, subtle differences can still be noticed in Figure.\ref{pH in porous media}.\ref{concentration in porous media}. It seems  $\mathbf{(\phi,\theta)}$ can measure a bit more details than what hue can do. That is probably because it utilizes two parameters to interpolate pH, which enhances its ability to capture more color variations.

\begin{figure} [H]
    \centering
    \includegraphics[width=1\linewidth]{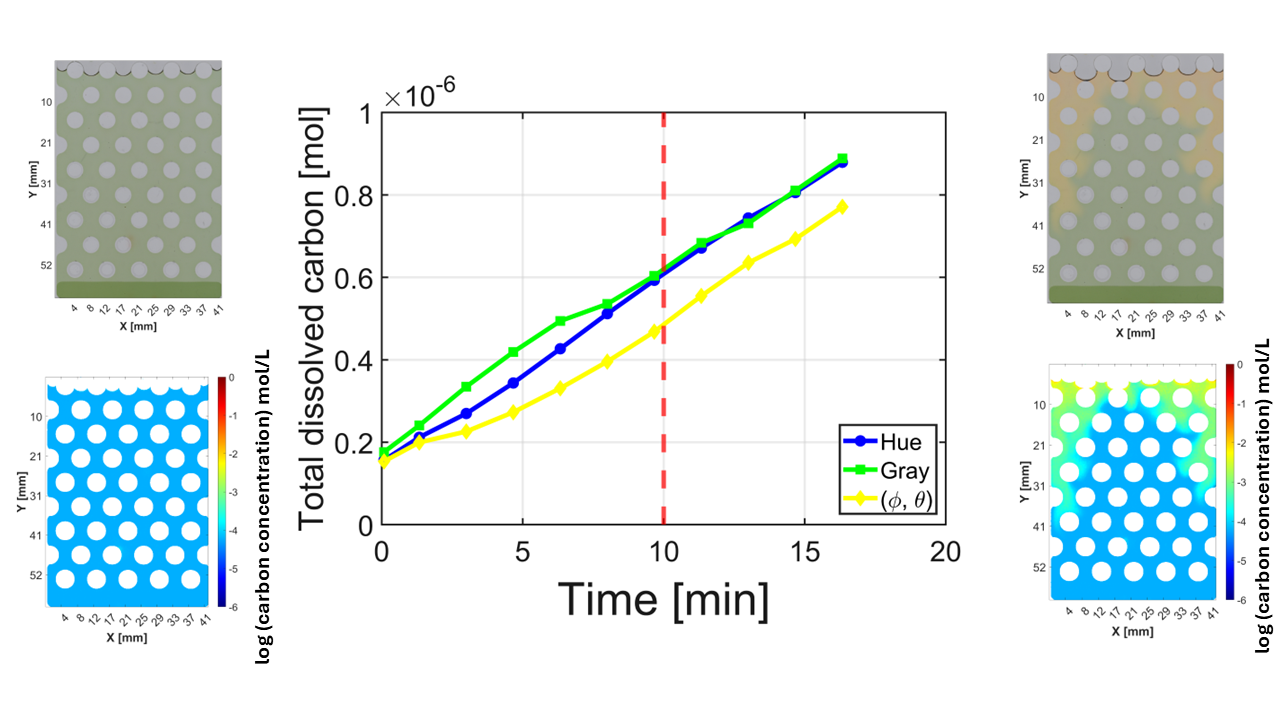}
    \caption{The comparison of total dissolved carbon by three methods. The dashed line represents the time of 10 minutes after carbon injection, corresponding to the time shown in Figure \ref{pH in porous media} and Figure \ref{concentration in porous media}. The left inset of the figure illustrates the initial state of the experiment ($t=0$). The upper panel displays the raw image, while the lower panel shows the corresponding spatial carbon concentration interpolated using the Hue technique. Similarly, the right inset depicts the final state of the plot ($t=16.3$ min).}
    \label{dissolved carbon}
\end{figure}

The total dissolved carbon estimated by the three techniques is shown in Figure \ref{dissolved carbon}. Overall, the results exhibit general agreement, but differences exist. 
This underestimation by $\mathbf{(\phi,\theta)}$-based  method is attributed to the fact that the majority of the plume's pixels fall within the previously discussed $\text{pH-jump}$ zone, which leads directly to an underestimation of carbon dissolution in that range. The gray-difference technique reports slightly higher dissolved carbon than the Hue technique at the beginning, but the discrepancy decreases over time, and the two converge to nearly the same value by the end. Figure \ref {single plume}(c) shows the pH distribution of the plume by Hue and gray-difference techniques. Although the pH map of the plume appears very different in Figure \ref{pH in porous media}, the two distributions are similar. This similarity validates the comparable total dissolved carbon estimates achieved after 10 minutes of carbon injection, as shown in Figure \ref{dissolved carbon}.

%The total dissolved carbon by the three techniques is present in Figure.\ref{dissolved carbon} where the hue-based method has the highest total dissolved carbon while the gray-based method has the lowest. As mentioned above, the pH measurement based on the gray-value method can not accurately determine the pH range from 4-6, leading to a higher pH measurement and thereby lower dissolved carbon compared to the other two methods. The discrepancies of the total carbon content between the methods of Hue and gray could be attributed to the measurement of carbon concentration close to the interface. 
    
\section{Conclusion}
Density-driven convection plays a critical role in geological carbon storage, as this process significantly enhances the dissolution rate of carbon into groundwater and influences the spatial and temporal distribution of $\text{pH}$ and carbon concentration within porous media. Accurate determination of these parameters is therefore vital for characterizing convection and the subsequent carbon mineralization processes.

In this study, we utilized a universal $\text{pH}$ indicator with a large measurable range ($\text{pH}$ 4–9.5) and high resolution.  We successfully conducted carbon injection experiments within quasi-2D porous media saturated with the indicator solution, where the color changes indicate variations in pH and carbon concentration. The experimental setup proved practical and easy to operate, using only standard digital cameras to capture the dynamics of these color variations. We then compared three common techniques—Hue, gray-difference, and $\mathbf{(\phi,\theta)}$ for quantifying the color data captured in the images. 

The comparative analysis demonstrates that the Hue technique is the most robust method. The quantification of color by Hue proved to be invariant to variations in camera settings, LED luminance, and, critically, fluid thickness. This stability makes the resulting calibration curve reliable across various experimental conditions.
Furthermore, the Hue technique has a monotonic calibration path for $\text{pH}$ interpolation. This characteristic enables fast computation and accurate $\text{pH}$ determination across the entire range, making it an excellent candidate for accurately mapping convection dynamics and determining $\text{pH}$ and carbon concentration in density-driven convection experiments.

In contrast, the $\mathbf{(\phi,\theta)}$-based method is less stable than the Hue technique, concerning variations in camera settings and LED luminance, and they exhibit much greater deviation when fluid thickness varies.  The calibration path for this technique is non-monotonic, resulting in inaccurate pH measurements near the maximum $\mathbf{(\phi,\theta)}$ in Figure \ref{projection method}(a). 
Error analysis demonstrates that this technique is still effective for determining $\text{pH}$ above 5.8. However, because it uses two variables ($\mathbf{(\phi,\theta)}$) for interpolation, it is more mathematically demanding, although this complexity makes it more accurate for interpolating $\text{pH}$ in a certain range. 

The gray techniques are sensitive to changes in the surrounding environment, such as camera settings, and LED luminance, which makes the calibration curve less applicable for use. Besides, the calibration path is non-monotonic in the range of 4-6, leading to erroneous interpolation in this critical zone. This makes the total dissoved carbon a bit higher than that of the Hue technique. While the error analysis demonstrates that it can still determine pH beyond the critical zone, it has reported the highest standard deviation among the three techniques, indicating the technique has more noise and errors. 

In summary, while all three techniques can determine $\text{pH}$ in a certain range, the Hue technique offers superior stability, universality, and computational simplicity, establishing it as the most reliable quantitative method for this class of experiment.

\section*{Acknowledgement}

We acknowledge the financial support from the Research Council of Norway through the PoreLab Center of Excellence (project number 262644) and the FlowConn Researcher Project for Young Talent (project number 324555). We also appreciate the support from the Njord Center, Faculty of Mathematics and Natural Sciences at the University of Oslo through the CO2Basalt project. %Additionally, we would like to express our gratitude to the input from the anonymous reviewers that significantly contributed to this work.

\section*{Authors Contributions}
YX designed the experiments, performed the data analysis, and contributed to the conceptualization and manuscript writing.  MM assisted with data analysis and revisions to the manuscript. EGF contributed to manuscript revisions. KJM participated in manuscript writing and revisions. All authors were involved in discussions related to the work. 

\appendix
\renewcommand{\theequation}{A.\arabic{equation}}
\renewcommand{\thefigure}{A.\arabic{figure}}
\setcounter{equation}{0}
\setcounter{figure}{0}
%\section*{Appendix}

%\section*{A.}

\newpage
\bibliographystyle{Frontiers-Vancouver} 
\bibliography{biblio}
\end{document}